\documentclass[11pt]{article}
\usepackage[margin=1in]{geometry}
\usepackage{lmodern}

\usepackage[T1]{fontenc}

\usepackage[utf8]{inputenc}
\usepackage[pdftex, pdftitle={Article}, pdfauthor={Author}]{hyperref} 
\usepackage[parfill]{parskip}
\usepackage{float} 
\usepackage{array}
\usepackage{multirow}
\usepackage{caption}

\usepackage{lineno}

\newcolumntype{C}{>{\centering\arraybackslash}p{1 cm}}
\newcolumntype{V}{>{\centering\arraybackslash}p{2 cm}}
\newcolumntype{B}{>{\centering\arraybackslash}p{2.8 cm}}
\usepackage{chemformula}
\usepackage{graphicx}
\usepackage{amsmath,amssymb,amsthm}
\usepackage[
  backend=biber,
  style=nature,
  citestyle=numeric-comp,
  sorting=none,
  autocite=superscript,
  defernumbers=true,
  ]{biblatex}

\usepackage{setspace} 
\usepackage{authblk}

\newcounter{extendedfigure}
\newenvironment{extendedfigure}
  {\renewcommand{\thefigure}{\theextendedfigure}\renewcommand{\figurename}{}\stepcounter{extendedfigure}
  \begin{figure}}
  {\end{figure}}

\usepackage{titlesec}
\titleformat{\part}
  {\normalfont\Huge\bfseries\centering} 
  {\partname\ \thepart}{20pt}{\Large}

\title{\fontsize{15}{16}\selectfont\textbf{Emergent frustrated magnetism in strain-patterned graphene}}
\date{}
\author[1]{Yu-Chiang Hsieh}
\author[2]{Wen-Han Kao}
\author[3]{Christophe De Beule}
\author[1]{Sheng-Zhu Ho}
\author[1]{Ru-Long Gou}
\author[1]{Bo-Nian Chen}
\author[1]{Kuan-Yu Chou}
\author[1]{Kuo-En Chang}
\author[1]{Chin-Chia Chang}
\author[4]{Ying-Mei Yang}
\author[4]{Ching-Hua Kao}
\author[1]{Hao-Chien Chiang}
\author[1]{Jyun-Lin Chen}
\author[1]{Sheng-Chin Ho}
\author[5]{Kenji~Watanabe}
\author[6]{Takashi~Taniguchi}
\author[1,4]{Ming-Hao Liu}
\author[1,4]{Ching-Hao Chang}
\author[1,4${\ast}$]{Yi-Chun Chen}
\author[7${\ast}$]{Ying-Jer Kao}
\author[1,4${\ast}$]{Tse-Ming Chen}

\affil[1]{Department of Physics, National Cheng Kung University, Tainan 701, Taiwan.}
\affil[2]{Department of Physics, University of Wisconsin-Madison, Madison, Wisconsin 53706, USA.}
\affil[3]{Department of Physics, University of Antwerp, Groenenborgerlaan 171, 2020 Antwerp, Belgium.}
\affil[4]{Center for Quantum Frontiers of Research \& Technology (QFort), National Cheng Kung University, Tainan 701, Taiwan.}
\affil[5]{Research Center for Electronic and Optical Materials, National Institute for Materials Science, Namiki 1-1, Tsukuba, 305-0044, Ibaraki, Japan.}
\affil[6]{Research Center for Materials Nanoarchitectonics, National Institute for Materials Science, Namiki 1-1, Tsukuba, 305-0044, Ibaraki, Japan.}
\affil[7]{Department of Physics, National Taiwan University, Taipei 10617, Taiwan.}
\begin{document}

\begin{refsegment}
\maketitle
\normalsize{$^\ast$To whom correspondence should be addressed: 
{ycchen93@mail.ncku.edu.tw}, 
{yjkao@phys.ntu.edu.tw},
{tmchen@phys.ncku.edu.tw}.}

\renewcommand{\abstractname}{}

\begin{abstract}

\textbf{Geometrically frustrated magnetism conventionally arises from pre-existing magnetic moments on lattices whose geometry prevents their interactions from being simultaneously satisfied, giving rise to highly degenerate states and rich collective behavior\autocite{bramwell_S01, balents_N2010, mengotti_NP11, sk_NRP20}. Creating such frustration in an intrinsically non-magnetic material presents a fundamentally different challenge, requiring both the magnetism and the competing interactions to emerge from correlated electrons. Here we show that this can be realized in graphene through lithographically programmable strain engineering. Patterning strain and the associated pseudo-magnetic field (PMF) into superlattices creates a correlated electronic system with flat bands and strong interactions. Transport measurements reveal interaction-driven insulating behavior, anisotropic magnetic hysteresis, and slow relaxation dynamics reminiscent of spin freezing, qualitatively captured by Monte Carlo simulations of competing magnetic moments on the PMF-defined ruby superlattice. Cryogenic magnetic force microscopy further reveals magnetic textures associated with the PMF landscape. These observations demonstrate frustrated magnetism emerging from correlated electrons in otherwise non-magnetic graphene. Our results establish lithographic strain engineering as a general and scalable route to flat bands and correlated states in van der Waals materials, providing a versatile platform for programmable quantum matter.}

\end{abstract}

The ability to engineer lattice structures and interlayer coupling in van der Waals (vdW) materials provides a powerful route to tailor electronic properties and access exotic quantum phases and functionalities. A prominent example is twisting adjacent vdW layers at the so-called magic angles, which reshapes the electronic structure into flat bands~\autocite{Bistritzer_PNAS11}. In such flat-band systems, suppressed kinetic energy and the enhanced density of states make Coulomb interactions dominant. These strong interactions, together with nontrivial band topology in some cases, have enabled strongly correlated quantum phases such as correlated insulators~\autocite{cao_N18}, unconventional superconductors~\autocite{cao_N18SC}, orbital Chern magnets~\autocite{sharpe_S19, serlin_S20}, fractional Chern insulators\autocite{xie_N21, cai_N23, park_N23}, and heavy fermions\autocite{song_PRL22,zhao_nature23}.

Beyond twist engineering, designer strain offers a potentially more scalable route to lithographically program band structure, topology and interactions in vdW materials. Uniform strain has proven effective for tuning electronic properties~\autocite{conley_NL13, wu_N14}, but richer possibilities emerge with spatially varying strain profiles, potentially enabling pseudo-magnetic fields~\autocite{guinea_NP10, levy_S10}, modulation of intra- and interlayer couplings\autocite{mccann_RPP13, ho_NE21}, and electronic band flattening~\autocite{mao_N20, milovan_PRB20, Christophe_PRB23, de_25PPRL}. To date, however, experimental studies of electronic properties under spatially varying strain have largely relied on local STM/STS measurements on individual bubbles or buckled nanostructures\autocite{levy_S10, mao_N20}, as clean and controllable strain superlattices in scalable device architectures have long been challenging to realize. This limitation was recently mitigated by lithographic patterning of atomically flat hexagonal boron nitride (hBN) substrates, which enables clean, deterministic, and programmable strain engineering in vdW materials~\autocite{hsieh_NL23}, and has enabled nontrivial quantum geometry and phenomena in corrugated bilayer graphene~\autocite{ho_NE21}. Nevertheless, achieving genuinely flat bands and, more importantly, interaction-driven correlated phases in vdW straintronic systems has remained elusive.

Here, we report the emergent frustrated magnetism in strain-engineered graphene by patterning a strain-induced pseudo-magnetic field (PMF) into a ruby superlattice. The resulting electronic structure hosts topological ultra-flat bands with a total bandwidth of only a few meV, enabling strong correlations to drive novel quantum phases. Low-temperature transport reveals interaction-driven insulating behavior and unusual magnetic hysteresis with slow relaxation dynamics reminiscent of spin freezing. Together with theory, these observations indicate that PMF-driven electron localization and strong interactions favour local valley polarization and the emergence of magnetic moments, whose competing interactions on the PMF-defined ruby lattice give rise to geometrical frustration. A minimal spin-ice-like model qualitatively captures the hysteresis and relaxation dynamics through magnetic-charge-defect fluctuations. Cryogenic magnetic force microscopy further reveals distinct magnetic textures, with spatial features suggestive of local geometrical constraints. All central results were reproduced in multiple devices.

These observations establish strain-patterned graphene as a distinct realization of geometrically frustrated magnetism. Here, the strain-induced PMF defines local plaquette constraints that motivate a spin-ice-inspired model, qualitatively capturing the observed magnetic textures and frustration-driven dynamics. The underlying mechanism, however, is fundamentally different from conventional spin ice: rather than relying on pre-existing moments supplied by magnetic ions~\autocite{bramwell_S01,caste_N08,fennell_S09,morris_S09} or nanomagnets~\autocite{mengotti_NP11,branford_S12,perrin_N16,sk_NRP20}, both the magnetic moments and their competing interactions are rooted in correlated electrons. This suggests a distinct form of frustrated electronic magnetism in an electrically tunable, fermionic 2D system.

\subsection*{Lithographic strain superlattice and electronic structure}
Figure~1 introduces our devices and shows how patterned strain drives graphene into flat bands with strong localization and interactions. Periodic strain profiles are engineered by transferring graphene onto hBN patterned with lithographically defined nanoholes arranged in a hexagonal superlattice (Fig.~1a; see Methods for device fabrication and Supplementary Information~S1 for Raman characterization of strain). In this work, we study strain-patterned bilayer and trilayer graphene with Bernal stacking, and here focus on bilayer devices for simplicity, as both give consistent results. The strain field around each nanohole varies slowly with respect to the graphene lattice, leaving the $K$ and $K'$ valleys decoupled, and its shear component couples to the electrons as pseudogauge field that generates a PMF~\autocite{guinea_NP10,fogler_PRL08}. For a single isolated nanohole, the local strain geometry and corresponding PMF for valley $K$ are illustrated schematically in Fig.\ 1b. Although the nanohole is effectively circularly symmetric for long-wavelength electrons, the PMF depends strongly on the local strain orientation relative to the graphene lattice~\autocite{verb_PRB15}. It vanishes when the radial direction aligns with the zigzag direction and reaches its maximum along the armchair direction. Thus, the PMF varies as $\sin(3\theta)$ with $\theta$ the angle between the zigzag direction and the $x$ axis. For a given valley, this yields a threefold-symmetric PMF with a sign-changing domain wall on a ring of maximal shear strain. 

Extending from a single nanohole to a device-scale hexagonal array of holes produces a periodic nanostructure with $C_{6v}$ symmetry (Fig.\ 1c). Without in-plane lattice relaxation, the resulting periodic PMF (Fig.\ 1d) reaches up to $60$~T near the hole edges. Moreover, the maxima in the PMF magnitude form an expanded honeycomb superlattice, specifically the ruby superlattice illustrated in Fig.\ 1e. To capture a more realistic case, we account for the fact that strained graphene can lower its elastic energy by in-plane atomic displacements, resulting in reduced strain and PMFs~\autocite{guinea_PRB08,phong_PRL22, Christophe_PNAS23} (Supplementary Information~S2).
Interestingly, elastic coupling of neighboring holes extends the finite PMF regions (Figs.\ 1f and 1g) depending on the orientation of the graphene lattice (Extended Data Fig.\ 1). The PMF amplitude is reduced by one order of magnitude to about $5$~T, still strong enough to localize electrons on the length scale relevant to our experiments. Importantly, the PMFs for the $K$ and $K'$ valleys (Figs.\ 1f and 1g) appear with opposite signs due to time-reversal symmetry, so a net magnetic response can emerge when this symmetry is broken.

Figure~1h shows the calculated band structures for the $K$ and $K'$ valleys for the PMF superlattice shown in Figs.\ 1f and 1g (Supplementary Information~S3). The lowest bands are strongly flattened, with an overall bandwidth below $6$~meV, and become ultraflat near the Dirac points, shifted to the center of the superlattice Brillouin zone (SBZ). Introducing a small inversion-symmetry breaking term, as naturally expected in our graphene/hBN strained devices~\autocite{arrighi_2023NC},
further reduces the entire bandwidth to below $1$~meV across the whole SBZ, while also simultaneously flattening the remote bands (Fig.~1i and 1j, with the latter providing a clearer 3D view of valley-$K$ bands). These flat bands remain robust over a broad range of lattice orientations and interlayer biases (Extended Data Fig.~1). From a complementary perspective, these ultraflat bands can be understood in real space as local pseudo-Landau levels. The strong PMFs result in energy quantization with localized wavefunctions following the strain-induced flux patterns. Such localization suppresses kinetic energy, allowing Coulomb interactions to dominate and favour interaction-driven valley polarization and associated time-reversal-symmetry breaking\autocite{abanin_2012PRL,ghaemi_2012PRL}. Beyond flatness, these bands are also topologically nontrivial: the lowest two conduction bands near charge neutrality carry valley Chern numbers $C=(C_K-C_{K'})/2=2$ and $-1$, respectively, and opposite for the valence band. 

\subsection*{Electronic signatures of strong interaction and frustrated magnetism}

We now move on to experimentally probe the strong electronic interactions in our system. Figure~2a shows the temperature-dependent resistivity of strain-patterned graphene, together with an unstrained region for direct comparison (inset of Fig.~2a). 
The resistivity of strain-patterned graphene increases significantly with decreasing temperature, indicative of insulating behavior, whereas the unstrained region exhibits graphene's typical weak semimetallic response. The temperature dependence of the strain-patterned graphene is further analyzed using Arrhenius and Efros–Shklovskii variable-range hopping (ES-VRH) models (Fig.~2b). Importantly, the data are well described by ES-VRH, indicating that the strong localization and electron-electron interactions dominate the transport and that the system is in a correlated regime~\autocite{tsigan_PRL02}. 

We next investigate the low-temperature magnetotransport of strain-patterned graphene. Figures~2c and 2d show clear hysteresis between $0.2$ and $4$~K under out-of-plane and in-plane magnetic fields, respectively. The observed hysteresis resembles frustration-driven responses in artificial spin ice~\autocite{branford_S12,park_PRB17} and metallic pyrochlores~\autocite{pearce_NC22,tian_NP16}, and remains relatively robust against carrier density (Extended Data Fig.~2).
Unlike conventional ferromagnets, where hysteretic magnetoresistance typically peaks near the coercive field due to domain reversal, our system exhibits an unusual negative-magnetoresistance-like response with the resistivity peaking near $B=0$, a behavior associated in frustrated magnets with the field-dependent density of magnetic-charge defects~\autocite{pearce_NC22}. More importantly, the hysteresis window, quantified by the magnetoresistivity asymmetry, shows a clear non-monotonic temperature dependence and peaks near $T=1$~K (Fig.~2e and Extended Data Fig.~3), similar to that reported in metallic frustrated systems~\autocite{kim_PRB18, ye_PRB17}. The magnetic hysteresis is progressively suppressed with increasing drive current (Supplementary Information~S4). It is important to note that no hysteresis is observed in either strain-patterned monolayer graphene, where the absence of interlayer coupling limits strain-induced band reconstruction, or bilayer graphene with one-dimensional corrugation, a strain pattern without frustrated geometry (Extended Data Figs.~4 and 5). These control experiments rule out strain or fabrication-induced defects alone as the origin of the observed hysteresis, and highlight the essential roles of strong electronic interactions and the frustrated geometrical constraints.

\subsection*{Frustration-induced freezing and magnetic-charge-defect dynamics}
To understand the microscopic origin of the hysteresis and its connection to magnetic frustration, we perform classical Monte Carlo simulations on the ruby lattice (Fig.~2f), which captures the effective magnetic geometry of our system. Antiferromagnetic Ising interactions on the triangular plaquettes impose local geometrical constraints, giving rise to an extensively degenerate manifold and spin-ice-like magnetic-charge configurations (Figs.~2g--i). The rectangular plaquettes further host mobile magnetic-charge defects, providing a useful picture for the field-dependent dynamics. Most importantly, the forward and backward field sweeps access two distinct low-field states: during the forward sweep through zero field, the system passes through a frustrated low-charge state (Fig.~2k, top), whereas during the backward sweep it becomes trapped in a metastable state with residual magnetic-charge defects (Fig.~2k, bottom). These field-history-dependent charge configurations give rise to the simulated hysteresis (Fig.~2j), and underlie the distinct relaxation dynamics discussed later. The simulations also reproduce the non-monotonic temperature dependence of the hysteresis window, with defect redistribution increasingly suppressed by spin freezing at the lowest temperatures. A quantitative comparison between the simulated and experimental results gives consistent energy scales and an effective magnetic moment of $\sim 9.0~\mu_B$, of similar magnitude to interaction-driven magnetic moments reported in twisted bilayer graphene~\autocite{Li2020_TBG,Tschirhart2021_TBG,Sharpe2021_TBG}. Further details of the microscopic charge-defect dynamics are provided in Supplementary Information~S5.

The hysteresis window is strongly sweep-rate dependent: slower sweeps suppress the magnetoresistivity asymmetry (Fig.~3a), consistent with frustration-driven dynamics. The two field-history-dependent states identified above also show distinctly different relaxation behavior: the metastable state is effectively frozen below $0.2$~K, whereas the frustrated state continues to relax and remains dependent on the sweep rate (Fig.~3b and 3c and Extended Data Fig.~6). This state-dependent relaxation provides clear evidence for spin-freezing dynamics in our system. Although Monte Carlo steps do not map directly onto physical time, simulations with different numbers of steps show good qualitative agreement with the experimental sweep-rate dependence of the hysteresis (Fig.~3d).

The magnetic hysteresis exhibits a clear anisotropy as the field is rotated within the graphene plane. The hysteresis window, quantified by the magnetoresistivity asymmetry, varies strongly with field angle and shows a twofold-like angular dependence (Fig.~3e and Extended Data Fig.~7). Such an anisotropic response is not unexpected in magnetotransport, as the fixed current direction introduces an additional reference axis, and the measured response depends on the relative orientations of the current, magnetic field, and magnetic configuration. Similar angular dependence of hysteretic magnetotransport has been reported in artificial spin-ice systems~\autocite{branford_S12,le_2017PRB}. Furthermore, the remanent resistivity measured as the field sweep passes through zero exhibits a distinct threefold-like angular anisotropy (Extended Data Fig.~7), indicating that the zero-field magnetic state retains a memory of the preceding field orientation and is constrained by the underlying magnetic geometry~\autocite{le_2017PRB}.

\subsection*{Real-space imaging of the magnetic response}

We next turn to scanning probe microscopy as a complementary spatial probe of the magnetic response. Figures~4a and 4b show the atomic force microscopy (AFM) topography and corresponding Kelvin probe force microscopy (KPFM) image of one of our devices. The graphene follows the patterned surface and remains continuous across the patterned region, and the KPFM reveals a periodic surface-potential modulation that follows the underlying superlattice and is consistent with the strain-induced deformation potential. Similar behaviour is reproduced in a separate device, whereas patterned hBN without graphene shows the opposite potential contrast (Extended Data Fig.~8), further supporting the structural integrity of the strained graphene.

Local magnetic moments associated with interaction-driven valley polarization are expected to emerge near the PMF maxima of the distorted ruby superlattice (Fig.~4c). In such a frustrated geometry, competing moments can generate complex real-space textures at low temperatures. We probe this magnetic response using cryogenic magnetic force microscopy (MFM). At $T=1.6$~K and $B=4$~T, MFM reveals a distinct and spatially complex magnetic texture in strain-patterned graphene (Fig.~4d). Nanohole positions are superimposed for comparison; possible topographic contributions are further examined in Supplementary Information~S6. 
Magnetostatic simulations based on the same ruby-lattice geometry reproduce the main spatial features of the observed texture (Figs.~4e and 4f), consistent with the PMF-defined frustrated geometry. The magnetic texture also rapidly disappears with increasing temperature (Fig.~4g), supporting its connection to the low-temperature correlated magnetic state. 

\subsection*{Discussion and outlook}

An important question is what form of frustrated magnetism is realized here. The observed magnetic hysteresis, metastability and spin-freezing dynamics are characteristic of frustrated magnets, but are not unique to a particular phase and can also occur in glassy magnets. Several observations, however, make a conventional disorder-driven spin glass less likely. The hysteresis shows a twofold-like anisotropy, while the remanent resistivity exhibits a distinct threefold-like anisotropy, indicating a geometry-dependent rather than disorder-dominated magnetic response. Moreover, no hysteresis is observed in strain-patterned monolayer graphene or bilayer graphene with one-dimensional strain, highlighting the importance of strong electron–electron interactions and frustrated geometry. Zero-field-cooled and field-cooled measurements also show nearly identical hysteresis, with no clear spin-glass-like irreversibility (Supplementary Information~S7). Instead, the observed magnetic response and dynamics are consistent with the ruby-lattice model. However, our experiments do not establish thermodynamic signatures of a canonical spin-ice phase, such as residual Pauling entropy, which are difficult to access in a micron-scale 2D electronic system because of its small active volume and magnetic signal. Moreover, the emergent magnetic moments and interactions in our electronic system may differ substantially from those of conventional spin ices. We therefore use the spin-ice-like model as an effective framework for understanding the geometrical constraints and magnetic-charge dynamics, rather than as a unique identification of the magnetic phase.

As a controllable design parameter, strain complements twistronics by offering geometric freedom beyond moir\'e periodicity and compatibility with wafer-scale fabrication. Lithography allows arbitrary strain landscapes to be defined directly in atomically thin vdW materials, creating flat bands and otherwise inaccessible quantum phases, such as the frustrated magnetism demonstrated here. In strain-patterned graphene, both the magnetic moments and their competing interactions emerge from correlated electrons. Unlike conventional frustrated magnets with relatively rigid local moments and interactions, these emergent moments and interactions may evolve with carrier density, strain, localization and electrical drive, opening access to a broader range of frustrated states. Such competition and tunability may favor more exotic phases characterized by strong quantum fluctuations, such as quantum spin liquids. Because PMF landscapes can be lithographically designed with diverse geometries beyond the ruby lattice, this approach provides a scalable route to flat bands, correlated phases and programmable quantum matter in vdW systems.

\newpage

\begin{figure}[H]
    \centering
    \includegraphics[width=1\textwidth]{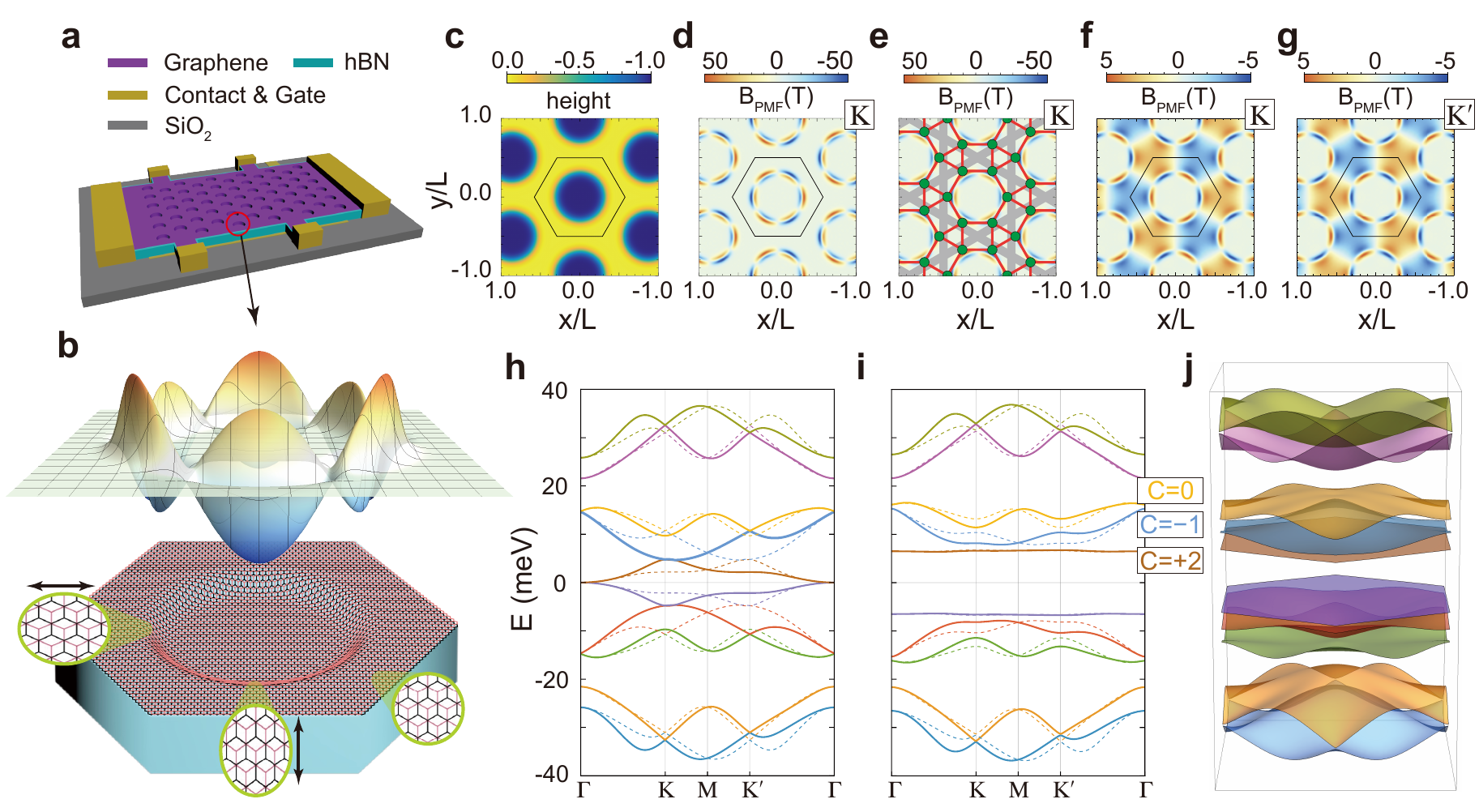}
    \caption{\textbf{Pseudo-magnetic field (PMF) superlattice and flat bands engineered in strain-patterned graphene.} 
    \textbf{a}, Schematic of strain-patterned graphene devices. Devices are fabricated by transferring graphene onto a lithographically patterned hBN substrate containing a hexagonal array of nanoholes, where the lattice deforms to define a strain superlattice. \textbf{b}, 3D visualization of the PMF landscape from nanohole-induced strain. Radial strain around each nanohole produces a PMF distribution that is maximized when aligned with graphene's armchair directions and vanishes along zigzag directions. The top panel shows the simulated PMF intensity in the $K$ valley, and the bottom panel illustrates the corresponding strain field, with zoom-ins highlighting lattice deformations near hole edges at two example orientations (ellipses), and pristine graphene lattices in unstrained flat regions for comparison (circles). \textbf{c}, \textbf{d}, Periodic height profile of the device surface landscape~(c), modeled from the AFM topography of our fabricated devices, and the corresponding simulated PMF distribution~(d), with the black hexagon indicating the unit cell. Graphene's armchair direction is aligned along the $x$-axis. \textbf{e}, Same as~(d), but with visual guides highlighting the PMF superlattice. Green dots mark the PMF maxima, which form a ruby lattice (red lines). The ruby lattice can be regarded as an expanded honeycomb with each vertex replaced by a triangular motif, or equivalently as a kagome-like geometry with PMFs sitting on kagome bonds (gray lines). \textbf{f}, \textbf{g}, Simulated PMF distributions including lattice relaxation effects for the $K$ and $K'$ valleys, respectively. In-plane atomic deformations reduce the strain overall, lowering the elastic energy, but extend it and the associated PMFs between the nanoholes. The $K$- and $K'$-valley PMFs exhibit opposite signs due to time-reversal symmetry. \textbf{h}, Calculated electronic band structures for the $K$ (solid) and $K'$ (dashed) valleys based on the strain superlattices in (f) and (g). The lowest bands exhibit pronounced flattening, with a total bandwidth below $6$~meV, and are particularly flat near charge neutrality. \textbf{i}, \textbf{j}, Same as (h) but with a small inversion-symmetry-breaking interlayer bias ($13$ meV), which reduces the total bandwidth to below $1$ meV and yields an ultra-flat bands across the entire Brillouin zone (i). A 3D view illustrates this band flatness (j). These bands are also topologically nontrivial, with the lowest three $K$-valley conduction bands carrying valley Chern numbers of $C=2$, $-1$, and $0$.
        }
    \label{fig1}
\end{figure}

\newpage

\begin{figure}[H]
    \centering
    \includegraphics[width=0.92\textwidth]{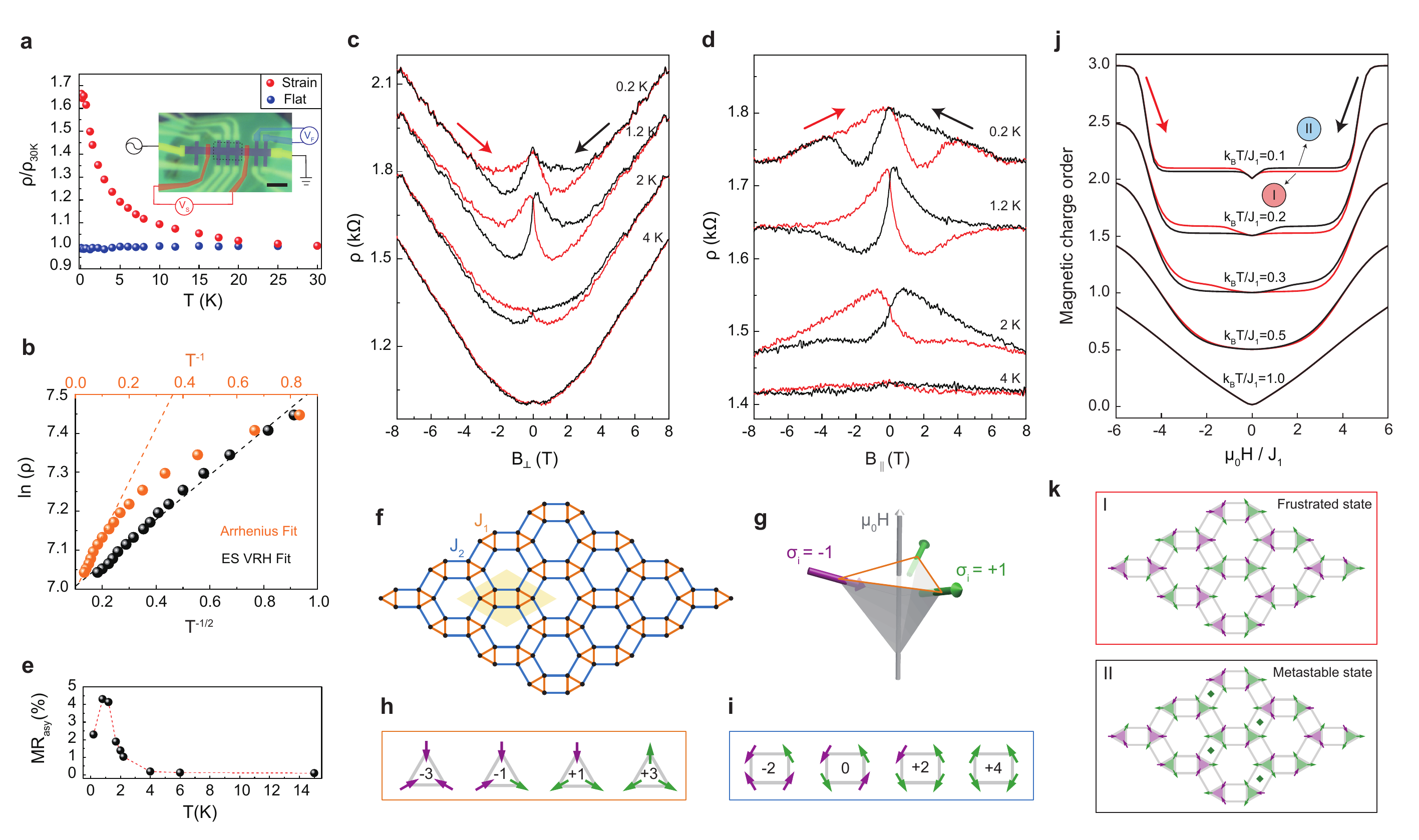}
    \caption{\textbf{Emergent strongly correlated frustrated behavior revealed by transport and simulation.} 
    \textbf{a}, Normalized resistivity ($\rho/\rho_{\mathrm{30K}}$) as a function of temperature, measured in the same trilayer graphene device from strain-patterned (red dots) and unstrained flat (blue dots) regions. The strain-patterned graphene exhibits a pronounced insulating upturn, whereas the unstrained region displays weakly metallic behavior. Inset: Optical micrograph of the device, with the dashed rectangle indicating the strain-patterned region (scale bar, 5~$\mu$m). \textbf{b}, Fits of the strain-patterned region's resistivity to Arrhenius activation and Efros–Shklovskii variable-range hopping (ES-VRH) models. The good agreement with ES-VRH, together with the deviation from Arrhenius, indicates that strong Coulomb interactions and interaction-driven localization underlie the insulating behavior. \textbf{c}, \textbf{d}, Magnetotransport hysteresis of strain-patterned graphene under out-of-plane (c) and in-plane (d) magnetic fields, measured with a driving current of $10$~nA and a sweep rate of $0.06$~T/min. Red and black lines denote forward and backward sweeps, respectively, with the observed asymmetry reminiscent of frustration-driven responses. Traces are offset for clarity. \textbf{e}, Temperature dependence of the hysteresis window, quantified by magnetoresistivity asymmetry. The window size exhibits a non-monotonic evolution with a maximum near $T=1$~K. \textbf{f}, \textbf{g}, Schematic of the ruby-lattice model~(f) and a zoom-in of one triangle defining the canted spin orientation~(g) used for the simulations. $J_1$ and $J_2$ denote intra- and inter-triangle couplings. \textbf{h}, \textbf{i}, Representative spin configurations on triangular (h) and rectangular (i) plaquettes illustrating local magnetic constraints of the ruby lattice. The number inside each plaquette indicates $Q = \sum_i \sigma_i$, a proxy for effective magnetic charge defect. The local constrained configurations (2--1 on triangles, 2--2 on rectangles) are charge-balanced, whereas departures from these configurations represent magnetic-charge defects. \textbf{j}, Monte Carlo simulations of triangular magnetic charge order (MCO) versus out-of-plane magnetic field $\mu_0 H / J_1$ for various temperature $k_BT/J_1$. Hysteresis emerges at low temperature, mirroring the magnetotransport behavior observed in experiments. Traces are offset for clarity. \textbf{k}, Magnetic configurations from simulations at $\mu_0 H/J_1 = 2$, illustrating (I) a frustrated state with zero magnetic charge and (II) a metastable state with finite charges, linked to the forward- and backward-sweep points indicated in panel (j). The green dots highlight the $Q = +2$ effective magnetic charge within the rectangular plaquette.
    }
    \label{fig2}
\end{figure}
\newpage

\begin{figure}[H]
    \centering
    \includegraphics[width=0.9\textwidth]{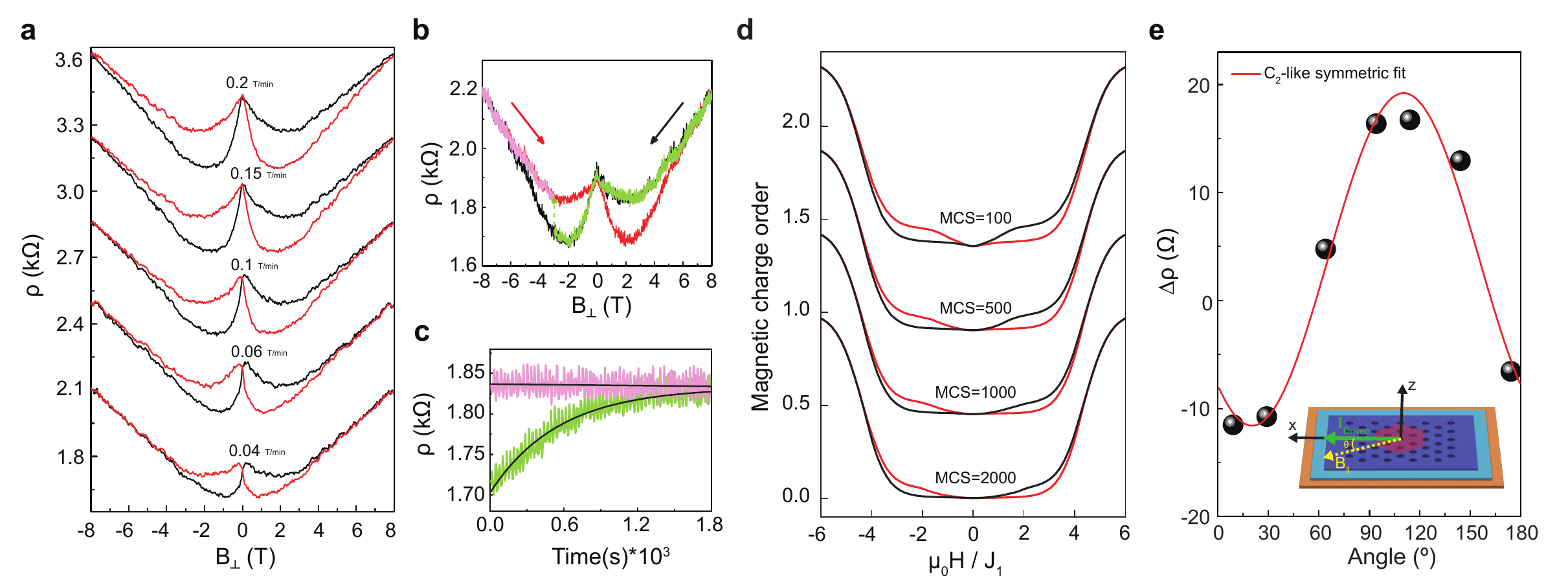}
    \caption{\textbf{Dynamic signatures of magnetic frustration in strain-patterned graphene.} 
    \textbf{a}, Sweep-rate dependence of magnetic hysteresis at $T=1.2$~K, showing non-equilibrium dynamics. \textbf{b}, Relaxation measurements of the magnetoresistivity at $200$~mK, measured with a sweep rate of $0.2$~T/min. Red and black lines denote the forward and backward field sweeps, respectively. Pink and green lines indicate the field-sweep protocols used to prepare the metastable and frustrated states, respectively, before holding the magnetic field at $-3$~T.  \textbf{c},  Time evolution of the resistivity for the metastable and frustrated states at $-3$~T, corresponding to the conditions shown in~(b). The metastable state remains highly stable, with the resistivity unchanged over the measured time, while the black lines highlight the relaxation trends. \textbf{d}, Monte Carlo simulations of triangular MCO at $k_BT/J_1=0.2$ under different sweep-rate conditions, modeled via the number of Monte Carlo steps (MCS). Slower sweeps (larger MCS) reproduce the experimental trends, reflecting the spin-ice frustration dynamics. \textbf{e}, Angle-dependent resistivity difference between the forward and backward branches at $0.8$~T, measured in a separate strain-patterned graphene device (black dots). The red curve represents a fit to a twofold-symmetric cosine function over the experimentally accessible angular range. The inset illustrates the anisotropic magnetotransport measurement geometry and defines the angle between the strain-pattern orientation and the in-plane magnetic field. The red hexagon indicates the orientation of the hexagonal hole array, with the $x$-axis defined along the graphene zigzag direction and the driving current applied parallel to the $x$-axis.}

    \label{fig3}
\end{figure}

\newpage

\begin{figure}[H]
    \centering
    \includegraphics[width=0.75\textwidth]{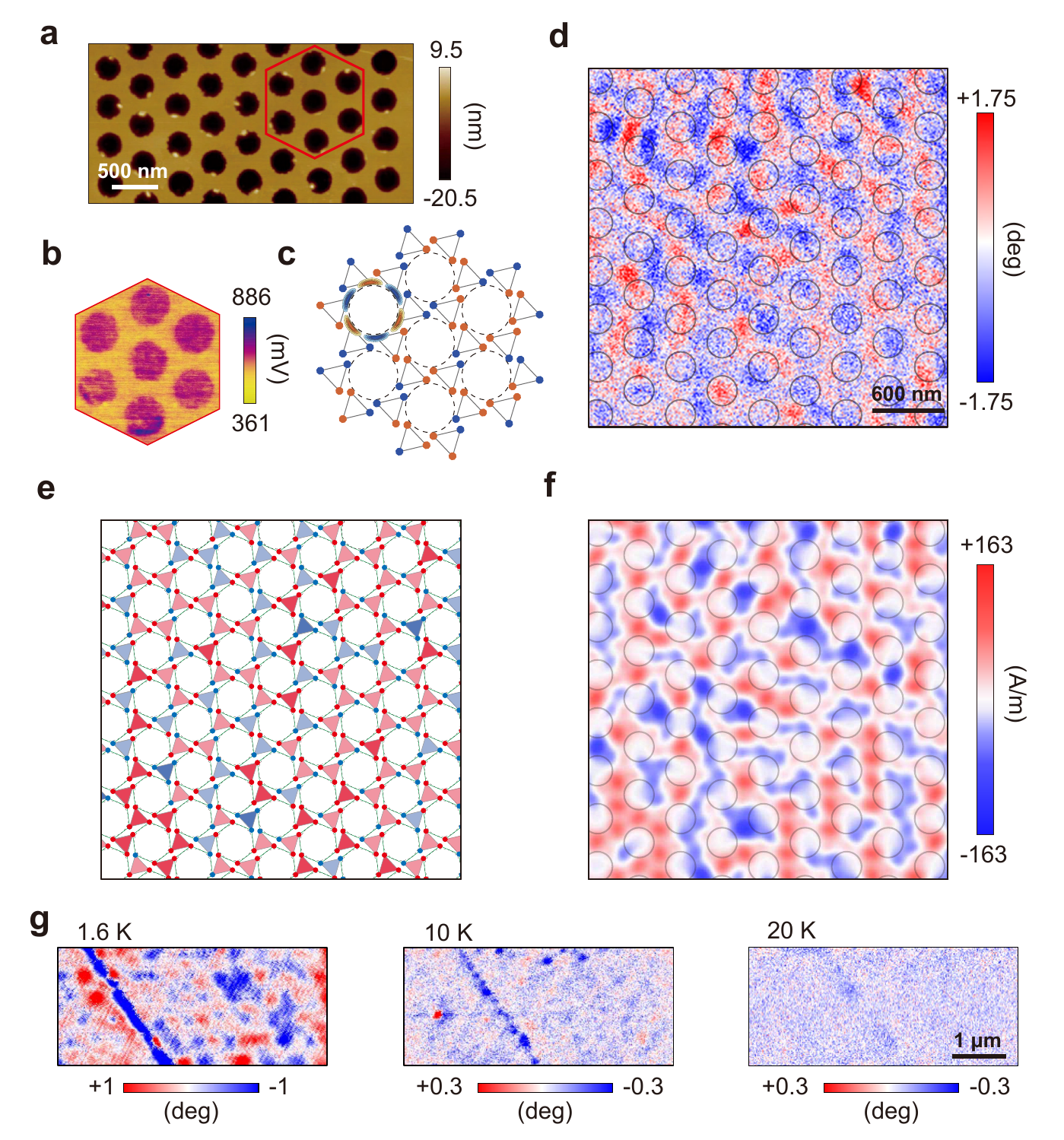}
    \caption{\textbf{Imaging the geometrically frustrated magnetism.} 
    \textbf{a}, Atomic force microscopy (AFM) topography of a strain-patterned graphene device. \textbf{b}, Kelvin probe force microscopy (KPFM) map of the surface potential from the region marked by the red hexagon in~(a). Higher surface potential areas coincide with strong PMFs, consistent with the strain-induced deformation potential. \textbf{c}, Schematic illustration of the PMF configuration, forming an expanded honeycomb (distorted ruby) superlattice. The graphene zigzag axis is rotated by $7^\circ$ relative to the $x$-axis, with the orientation determined using the long-edge alignment method for this device. Blue and orange dots mark PMF maxima with opposite magnetization directions under locally broken time-reversal symmetry, arranged according to the ice rule. \textbf{d}, Cryogenic magnetic force microscopy (MFM) image taken at $1.6$~K under an out-of-plane field of $B = 4$~T, with the nanohole positions determined from the corresponding surface topography superimposed for comparison. The MFM contrast reveals a complex magnetic texture in strain-patterned graphene. \textbf{e}, \textbf{f}, COMSOL Multiphysics simulation of vertically arranged magnetic dipoles and the resulting stray-field distribution. Following the distorted ruby geometry in~(c), the simulation reproduces the main spatial features of the experimentally observed MFM texture in~(d). The circular outlines in (d) and (f) indicate the nanohole positions. Panels (d-f) have the same spatial dimensions, allowing direct comparison between experiment and simulation. \textbf{g}, MFM images of a separate device at different temperatures, showing rapid suppression of the magnetic response with increasing temperature. All images in~(g) share the same scale.}
    \label{fig4}
\end{figure}

\newpage

\section*{Methods}
\subsubsection*{Device fabrication}

Strain-patterned graphene devices were fabricated using a standard polymer-assisted dry-transfer technique. Hexagonal boron nitride (hBN) and graphene flakes were mechanically exfoliated onto separate SiO$_2$/Si substrates. Selected hBN flakes were transferred onto pre-exfoliated graphite on the target substrate, which served as a bottom local gate. The heterostructure was annealed in high vacuum at 500~$^\circ$C for 3 hours to remove surface residues and ensure a clean interface.

Periodic nanoholes were subsequently defined in the hBN using electron-beam lithography and inductively coupled plasma reactive ion etching (ICP-RIE) etching. A technique employing cyclic-purge ICP-RIE was adopted, where alternating etch and vacuum-purge intervals were repeated to maintain uniform etching and produce atomic-scale flatness in the etched regions\autocite{hsieh_NL23}. A subsequent high-vacuum annealing step at 300~$^\circ$C removed resist residues and completed the substrate preparation. Graphene was then dry-transferred onto the pre-patterned hBN to form strain-patterned graphene, followed by a final anneal at 300~$^\circ$C to improve cleanliness and facilitate strain conformation to the underlying surface topography.

Electrical contacts are fabricated using a dual-resist process (copolymer $+$ PMMA A4) and metalized by electron-beam evaporation of a $5$~nm Cr adhesion layer and $\sim70$~nm Pd/Au $(1:1)$ alloy. For the devices used for transport measurements, additional Hall-bar geometries were defined and fabricated by electron-beam lithography and plasma etching, followed by polymer-cleaning to remove residual resist.

\subsubsection*{Raman spectroscopy}
Raman spectra were acquired using a micro-Raman system with a 532-nm continuous-wave solid-state laser as the excitation source. The laser was focused onto the sample through a $100\times$ objective lens with a numerical aperture of 0.95. The incident laser power at the sample was maintained below 10~mW using an adjustable neutral-density filter to minimize local laser heating. The spectra were collected in a confocal backscattering geometry under ambient conditions using an iHR550 spectrometer (Horiba Jobin Yvon).

\subsubsection*{Kelvin probe force microscopy measurement}
Kelvin probe force microscopy (KPFM) measurements were carried out using the sideband mode on a commercial scanning probe microscopy platform (Multimode 8, Bruker). Topography and surface potential are simultaneously acquired during the scans. A Pt/Ir-coated conductive probe (PPP-EFM, NANOSENSORS) with a nominal spring constant of $2.8$~N/m was employed and driven near its mechanical resonance frequency ($\sim75$~kHz) for sensitive surface detection. To enable sideband KPFM operation, an additional AC modulation voltage of $6$~V at $2$~kHz is superimposed on the DC bias, generating electrostatic sidebands at approximately 75~$\pm 2$~kHz. These sidebands are used for the detection and demodulation of the electrostatic force gradient, allowing for spatially resolved surface potential mapping with enhanced sensitivity and reduced crosstalk from topographic signals.

\subsubsection*{Cryogenic magnetic force microscopy measurement}
Magnetic force microscopy (MFM) measurements were performed in a cryogenic environment down to $1.6$~K to probe interaction-driven, geometrically frustrated magnetism in strain-patterned graphene. The experiments were conducted using a commercial cryogenic scanning probe microscope system (attoAFM I, Attocube) integrated with a closed-cycle cryostat (attoDRY 2100, Attocube) equipped with a $9$~T superconducting magnet. The system allows precise temperature control under applied magnetic fields. 
 In our experiments, we employed the lift-height mode configuration, where the MFM tip is maintained at a constant height above the sample surface. This effectively suppresses short-range surface forces, allowing long-range magnetic interactions to dominate the contribution. A cobalt (Co)-coated tip (NANOWORLD MFMR) with a nominal spring constant of $2.8$~N/m is used in the measurements, driven with an AC voltage of $\sim0.5$~V at a resonance frequency of $\sim75$~kHz for tapping operation.

\subsubsection*{COMSOL modeling of stray-field distribution}

Finite-element magnetostatic simulations were performed using COMSOL Multiphysics 6.2 with the AC/DC Module (\textit{Magnetic Fields, No Currents}) to model the stray-field distribution of nanoscale magnetic bars arranged in a distorted ruby lattice. 

The system was governed by the magnetostatic equation
\[
\nabla \cdot \left( \mu_0 \mu_r \mathbf{H} + \mathbf{B}_r \right) = 0,
\]
where $\mu_0$ is the vacuum permeability, $\mu_r$ is the relative permeability, and $\mathbf{B}_r$ is the remanent flux density. In the absence of free currents, the magnetic field can be expressed as
\[
\mathbf{H} = -\nabla \phi ,
\]
with $\phi$ denoting the magnetic scalar potential.

The simulated magnetic texture consisted of a distorted ruby array of cylindrical magnetic bars (20~nm radius, 10~nm height) embedded in a nonmagnetic air matrix ( $L_x ={5090}$~nm, $L_y = {5468}$~nm, $L_z ={2000}$~nm).
Each magnetic bar was assigned a relative permeability of 1.05 and a remanent flux density of $0.1$~T to approximate the experimental magnetic response.  
To approximate the open-boundary condition, the outer surfaces of the simulation domain were constrained by $\mathbf{n}\cdot\mathbf{B}=0$, and the overall domain size was extended sufficiently large to capture the natural decay of stray fields and avoid artificial boundary reflections.
Magnetic bars were positioned in the $xy$-plane at the mid-plane of the simulation space ($z = 1000$~nm), arranged in a distorted ruby lattice (or expanded honeycomb) pattern corresponding to the locations where the pseudo-magnetic field (PMF) reaches its maximum around each hole, as illustrated in the main text and Fig.~4e. The computational mesh employed a tetrahedral grid with adaptive refinement to a minimum element size of $\approx {10}$~nm near the magnetic–air interfaces, using quadratic Lagrange basis functions. The stationary solver was used with a relative tolerance of $10^{-6}$ to ensure convergence.

Post-processing included extraction of the vertical component of magnetic field to evaluate spatial variations and field confinement. The simulated stray-field distribution was sampled on a plane $240$~nm above the magnetic array and compared with experimental MFM images, confirming that the observed magnetic contrasts arise from geometrically frustrated spin configurations.

\subsubsection*{Electrical transport measurements}

Transport measurements were performed in a dilution refrigerator (Triton 200) with a base temperature of 100~mK and an integrated 8~T superconducting magnet. Standard low-frequency lock-in techniques were employed, applying an AC excitation current at 77~Hz through the sample and measuring the corresponding four-terminal voltage using a lock-in amplifier (SR830). The gate voltage, and thus the carrier density, was controlled by a DC source meter (Keithley 2400). Unless otherwise specified, all magnetoresistance measurements were carried out with a driving current of 10~nA and a gate voltage of $-3$~V, while sweeping the magnetic field at a rate of 0.06~T~min$^{-1}$. Angle-dependent magnetotransport measurements were performed using a TeslatronPT cryogenic system, with the device mounted on an attocube ANRv220 rotator to vary the angle between the in-plane magnetic field and the strain-pattern orientation. The main transport characteristics were reproduced across multiple devices, giving consistent results (Extended Data Fig.~9).

\printbibliography[segment=\therefsegment, heading=bibliography, title={References}]

\end{refsegment}

\newpage

\section*{Acknowledgements}

We thank L. Covaci, W.-H. Kang, Y.-D. Liou, and J. I.-J. Wang for insightful discussions and/or technical supports. This research was supported in part by the National Science and Technology Council in Taiwan (Grant Numbers 114-2123-M-006-002, 113-2112-M-002-033-MY3, 113-2112-M-006-025-MY3, 113-2123-M-006-002, 110-2124-M-006-007-MY3), by the Higher Education Sprout Project, Ministry of Education to the Headquarters of University Advancement at the National Cheng Kung University (NCKU). W.-H.K. acknowledges support from the Office of the Vice Chancellor for Research and Graduate Education at the University of Wisconsin–Madison with funding from the Wisconsin Alumni Research Foundation. K.W. and T.T. acknowledge support from the JSPS KAKENHI (Grant Numbers 21H05233 and 23H02052), the CREST (JPMJCR24A5), JST and World Premier International Research Center Initiative (WPI), MEXT, Japan. M.-H.L. acknowledges support from the National Science and Technology Council in Taiwan (Grant numbers 114-2112-M-006-029-MY3 and 112-2112-M-006-019-MY3).

\section*{Author Contributions}
Y.-C.H. and T.-M.C. conceived and designed the experiments. Y.-C.H. and R.-L.G. fabricated the devices with contributions from K.-Y.C., K.-E.C., Y.-M.Y., C.-H.K., H.-C.C., J.-L.C., and S.-C.H. W.-H.K. conducted the Monte Carlo simulations. C.D.B. conducted the theoretical calculations of the strain superlattice and its associated band structures, with support from M.-H.L. and C.-H.C. S.-Z.H. and B.-N.C. carried out the SPM measurements. C.-C.C. performed the magnetostatic simulations. Y.-C.H. conducted the transport measurements with contributions from R.-L.G., K.-Y.C., and K.-E.C. K.W. and T.T. grew the hBN crystals. Y.-C.H., W.-H.K., C.D.B, Y.-J.K., and T.-M.C. wrote the manuscript with inputs from all authors. Y.-C.H. and T.-M.C. coordinated the project. Y.-C.C., Y.-J.K., and T.-M.C. supervised the project.

\section*{Competing Interests Statement}
The authors declare no competing interests.

\section*{Data Availability}
The data that support the findings of this study are available from the corresponding author upon reasonable
request.

\newpage
\section*{Extended Data}

\begin{extendedfigure}[H]
    \centering
    \includegraphics[width=0.9\textwidth]{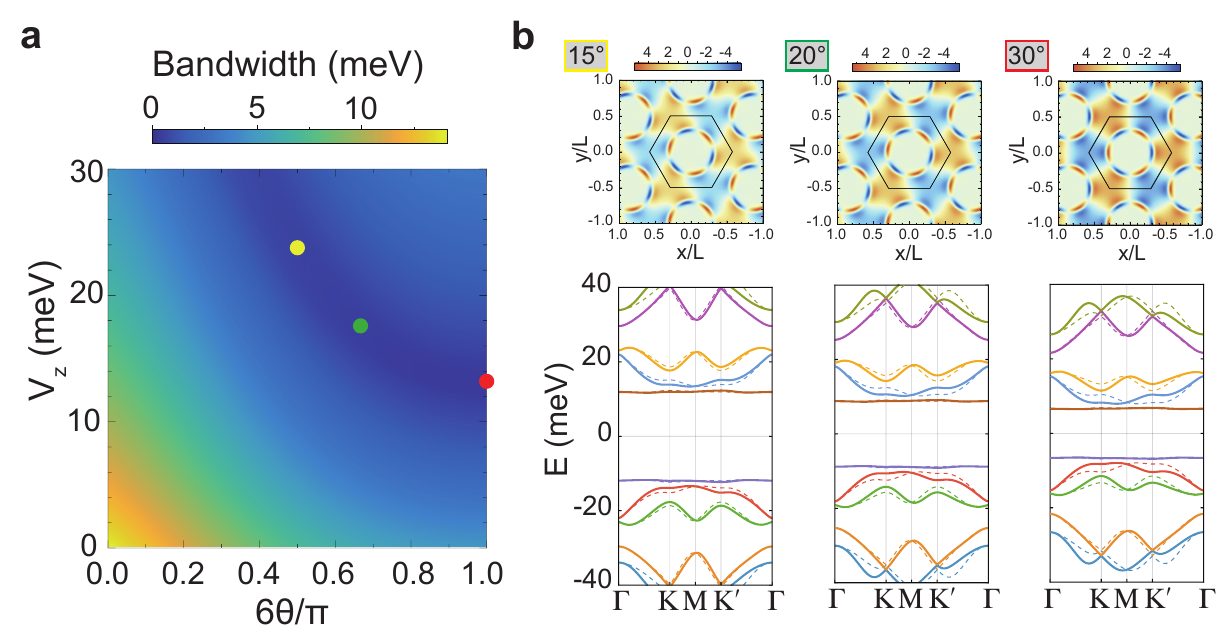}
    \caption{\textbf{Robustness of topological flat bands under varying lattice orientations.}
    \textbf{a}, Bandwidth of the lowest-energy bands as a function of interlayer bias $V_z$ and relative angle $\theta$. The distribution of the PMF depends on the relative angle ($\theta$) between the graphene lattice and the periodic corrugation, enabling different expanded honeycomb superlattices and thereby modifying the band structure. As defined in this study, $\theta$ is characterized by the angle between the zigzag direction of the graphene lattice and the $x$-axis. We systematically investigate how varying $\theta$ influences the resulting electronic band structures by extracting the bandwidth of the lowest bands as a function of the interlayer bias $V_z$ and $\theta$. The flat bands remain broadly robust under most conditions, with bandwidths below $5$~meV readily achievable using a modest interlayer bias ($V_z<30$ meV) across all orientations. \textbf{b}, Distributions of the PMF at $\theta = 15^\circ$, $20^\circ$, and $30^\circ$ are displayed in the upper insets and the corresponding electronic band structures at the parameters (colored dots in (a)) where ultra-flat bands emerge (lower insets). In addition to the $30^\circ$ case, which corresponds to the standard ruby lattice (see Fig. 1e), the $15^\circ$ and $20^\circ$ also form expanded honeycomb patterns, specifically, distorted ruby lattices. The $K$- and $ K'$-valleys are represented by solid and dashed lines, respectively. The strained pattern follows the same design as described in main text.
        }
	\label{EXT1} 
    
\end{extendedfigure}

\begin{extendedfigure}[H]
    \centering
    \includegraphics[width=0.9\textwidth]{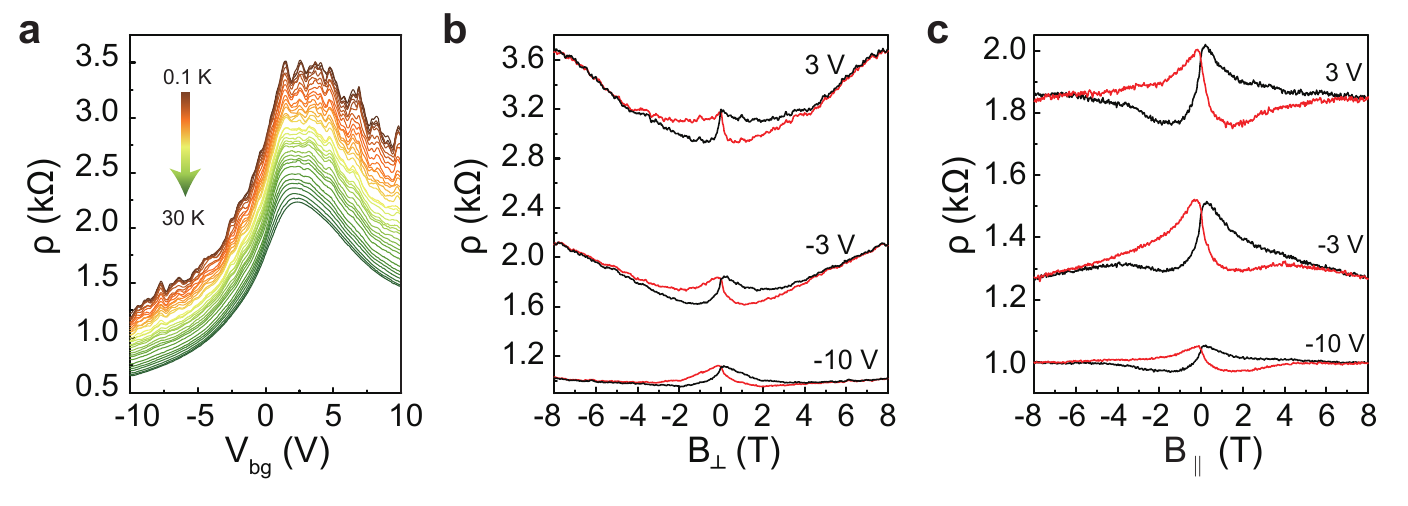}
    \caption{\textbf{Carrier-density dependence of frustrated magnetism.} 
    \textbf{a}, Resistivity as a function of back-gate voltage measured at various temperatures ranging from $0.1$ to $30$~K under zero magnetic field. Different colors denote the temperatures indicated by the arrows. \textbf{b}, \textbf{c}, Magnetic hysteresis obtained at different back-gate voltages under out-of-plane (b) and in-plane (c) magnetic fields, respectively. Our strain-patterned graphene exhibits both the Efros–Shklovskii variable-range-hopping (ES–VRH)-type interaction-driven resistivity enhancement (a) and frustrated magnetism (b,c) across a relatively broad range of carrier concentrations. The characteristic hysteresis loop, indicative of the frustrated phase, remains clearly visible when the system is tuned close to charge neutrality. As the gate voltage is shifted away from neutrality (e.g. $V_g=-10$~V), the hysteresis amplitude gradually decreases, reflecting a weakening of electron–electron interactions. Nevertheless, the overall hysteretic behavior persists, demonstrating that the emergent magnetic order is relatively robust against moderate carrier doping. The observed robustness indicates that the frustrated magnetism in strain-patterned graphene is largely governed by the strong pseudo-magnetic fields and associated strong electron localization, making the correlated state less sensitive to variations in carrier density.
        }
	\label{EXT2}
\end{extendedfigure}

\begin{extendedfigure}[H]
    \centering
    \includegraphics[width=0.9\textwidth]{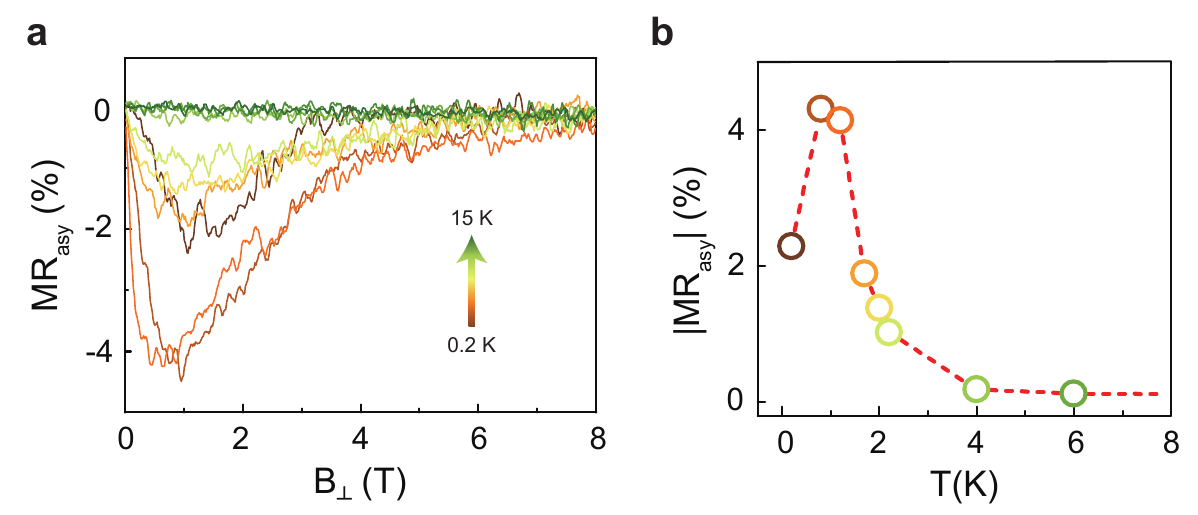}
    \caption{\textbf{Asymmetric analysis of magnetoresistance hysteresis.} 
    \textbf{a}, Asymmetric component of the forward sweep MR curve at various temperatures, extracted the asymmetric component of the normalized magnetoresistance (MR) using $\mathrm{MR}_{\mathrm{asy}}(B)=\frac{\rho(B)-\rho(-B)}{2\rho(0)}$ where \(\rho(B)\) is the longitudinal resistivity measured under a field \(B\). This approach removes the dominant symmetric MR background and isolates the asymmetric response, providing a cleaner probe to the non-equilibrium, geometrically frustrating magnetic excitations. The color coding indicates the temperature. \textbf{b}, Quantifying the maximum value of \(\mathrm{MR}_{\mathrm{asy}}(B)\) as a measure of the hysteresis window, we uncover a non-monotonic evolution with temperature. The maximum occurs around $1$~K and decreases at both higher and lower temperatures. For clarity, (b) presents the same data as Fig.~2e, with color coding corresponding to the hysteresis loops shown in (a).
        }
	\label{EXT3}
\end{extendedfigure}

\begin{extendedfigure}[H]
    \centering
    \includegraphics[width=0.9\textwidth]{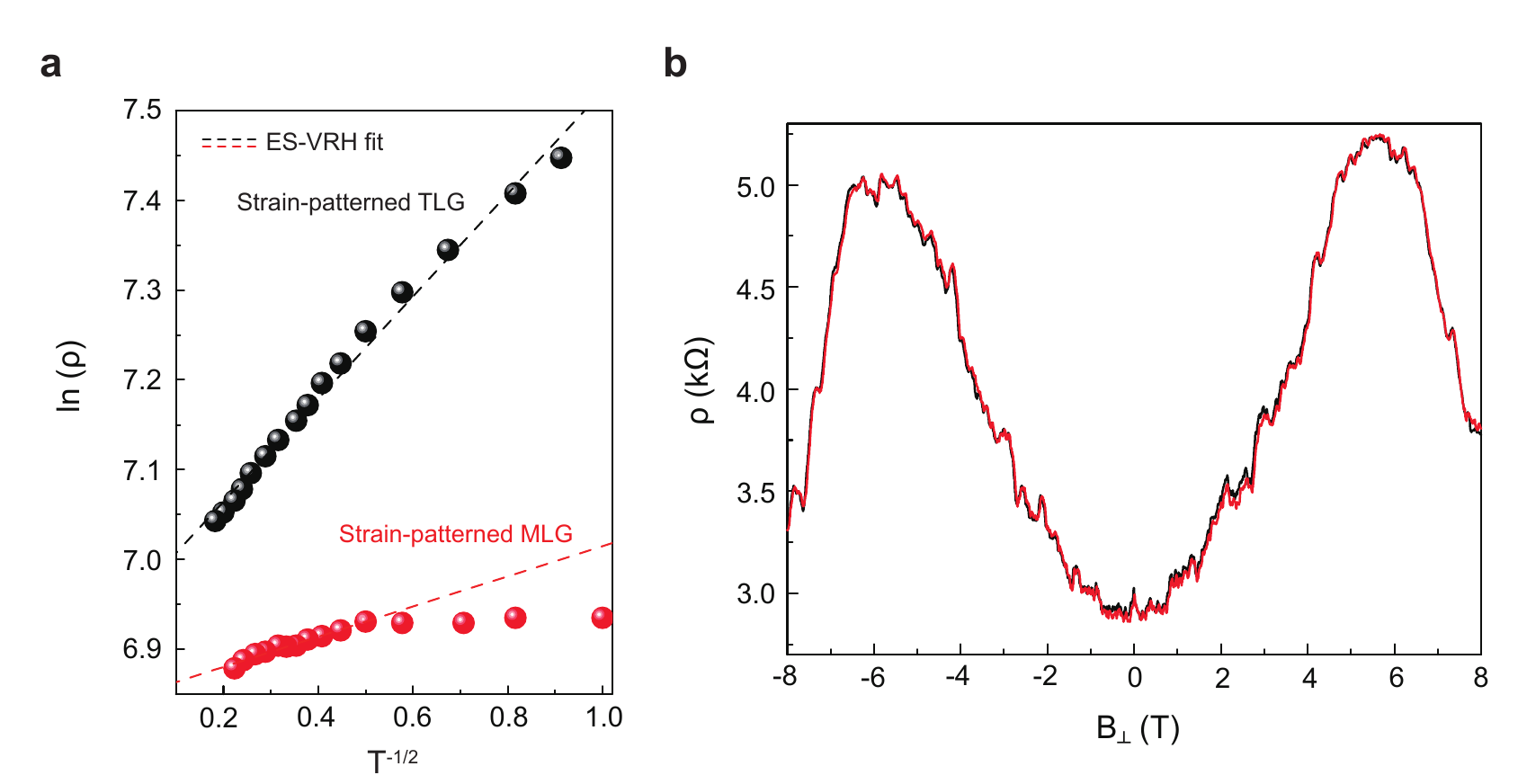}
    \caption{\textbf{Absence of interaction-driven insulating behavior and magnetic hysteresis in strain-patterned monolayer graphene.} \textbf{a}, Fits of the resistivity of strain-patterned monolayer and trilayer graphene devices to the Efros--Shklovskii variable-range hopping (ES-VRH) model. The strain-patterned monolayer device does not fit the ES-VRH model well, in contrast to the trilayer device, indicating substantially weaker interaction-driven localization. The strain-patterned trilayer graphene data are the same as those shown in Fig.~2 of the main text. \textbf{b}, Magnetotransport of the strain-patterned monolayer graphene device under an out-of-plane magnetic field, measured at $0.2$~K with a driving current of $100$~nA and a sweep rate of $0.2$~T/min. Red and black lines denote forward and backward sweeps, respectively. The two sweep directions nearly overlap, with no pronounced negative magnetoresistance or hysteresis.
        }
	\label{EXT4}
\end{extendedfigure}

\begin{extendedfigure}[H]
    \centering
    \includegraphics[width=0.9\textwidth]{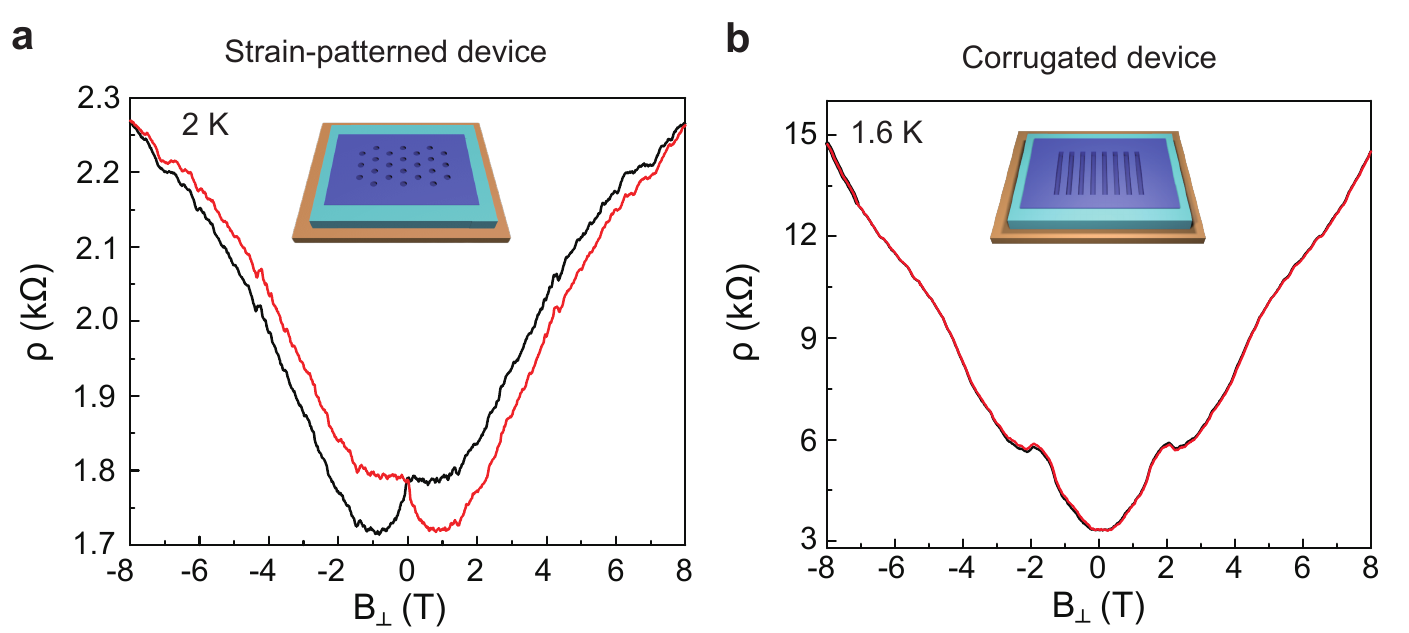}
    \caption{\textbf{Magnetotransport comparison between frustrated and non-frustrated PMF geometries.} 
    \textbf{a,b}, Magnetotransport measurements of strain-engineered graphene devices containing a hexagonal nanohole array (\textbf{a}) and a one-dimensional (1D) corrugation pattern (\textbf{b}). In the hexagonal-nanohole device, the spatial arrangement of the pseudomagnetic field (PMF) forms a geometrically frustrated pattern corresponding to the ruby lattice. In contrast, the corrugated device produces a simple 1D periodic PMF pattern without geometrical frustration. Pronounced magnetic hysteresis is observed only in the hexagonal-nanohole device, indicating that the magnetic response is closely associated with the frustrated PMF geometry rather than being a generic consequence of strain engineering or device fabrication. 
        }
        
	\label{EXT5} 

\end{extendedfigure}

\begin{extendedfigure}[H]
    \centering
    \includegraphics[width=0.9\textwidth]{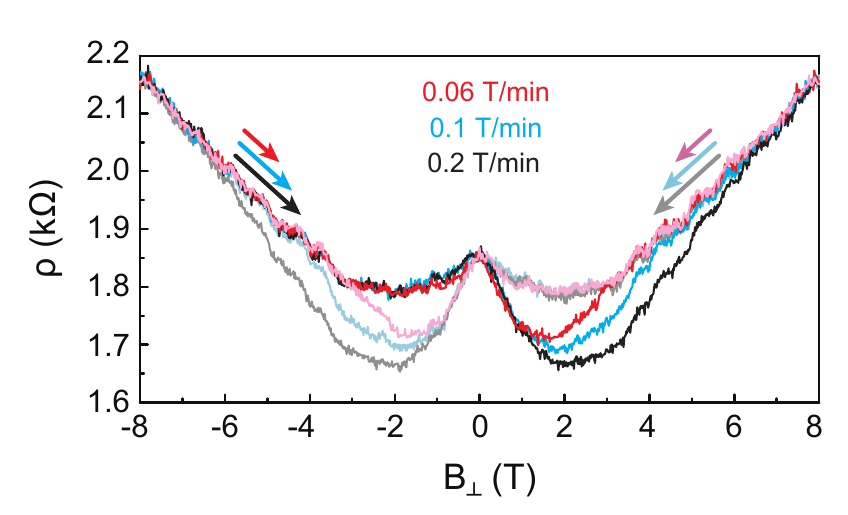}
    \caption{\textbf{Sweep rate dependent hysteresis behavior at $T=0.1$~K.} 
    Comparison of the magnetic hysteresis loops measured at $T=0.1$~K under various magnetic-field sweep rates to further examine the frustration-driven dynamics, with different colors representing the respective sweep conditions. Consistent with the trend observed at $1.2$~K (Fig.~3a), the hysteresis window becomes narrower as the sweep rate decreases. However, the frustrated and metastable states exhibit completely different relaxation behaviors. When the magnetic field is swept from zero toward high field, corresponding to the frustrated state, the magnetoresistivity strongly depends on the sweep rate, indicating frustration-driven dynamics of magnetic charge defect excitations toward quasi-equilibrium states. In contrast, when the magnetic field is swept back from high field to zero, corresponding to the metastable state, the magnetoresistivity remains essentially unchanged, suggesting that the system is dynamically frozen in the metastable configuration. These results indicate that the mechanisms of local magnetic configurations and relaxation are distinct in the frustrated and metastable regimes, reflecting intrinsic frustration-driven relaxation in our systems.
        }
	\label{EXT6}
\end{extendedfigure}

\begin{extendedfigure}[H]
    \centering
    \includegraphics[width=0.9\textwidth]{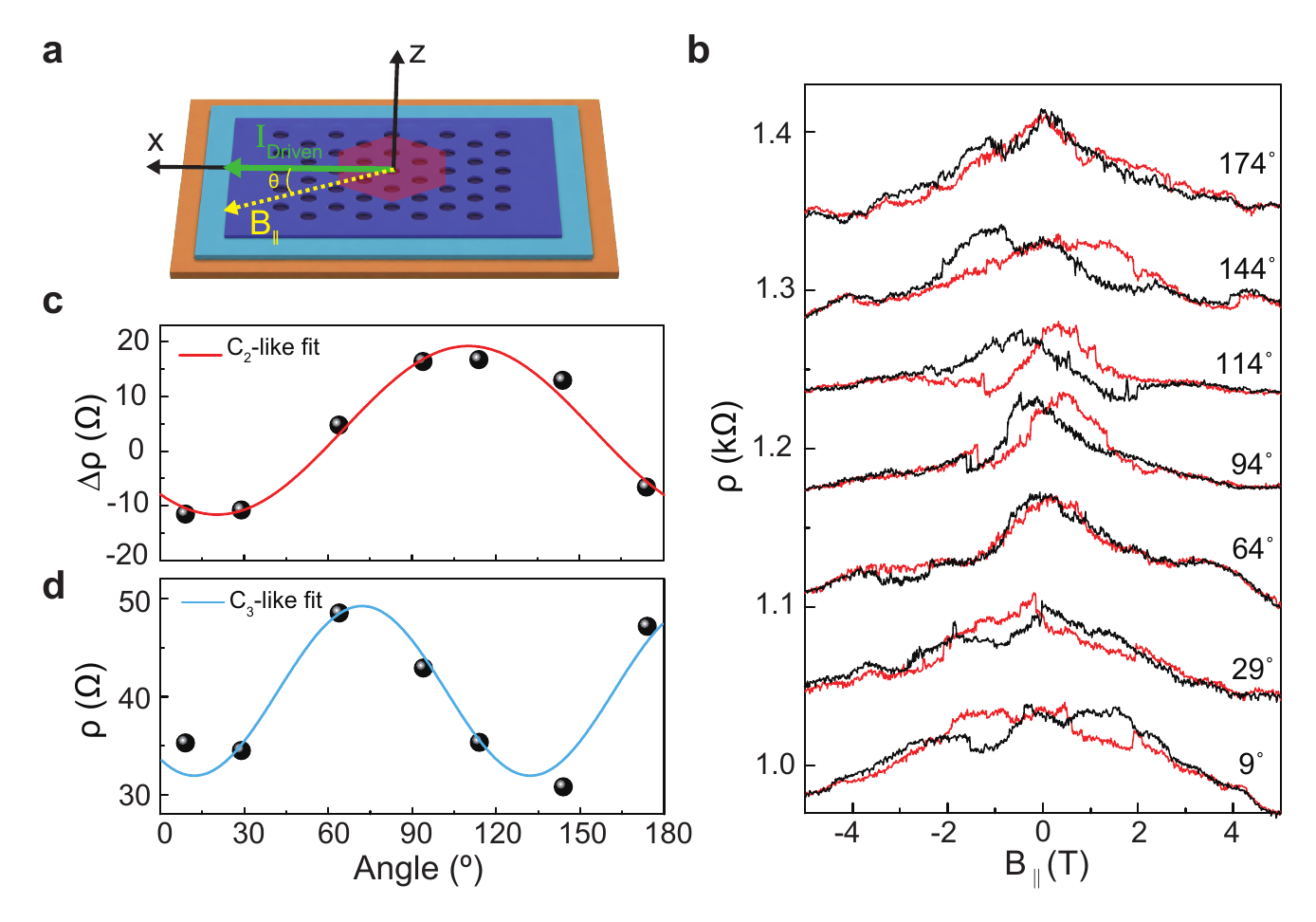}
    \caption{\textbf{Magnetotransport anisotropy.} \textbf{a}, Schematic illustration of the angle-dependent magnetotransport measurement, defining the angle between the strain-pattern orientation and the in-plane magnetic field. The red hexagon highlights the orientation of the hexagonal hole array, with the $x$-axis defined along the zigzag direction and the driving current applied parallel to the $x$-axis. \textbf{b}, Angle-dependent magnetic hysteresis measured in an independently fabricated strain-patterned device at $5$~K, with a driving current of $50$~nA and a sweep rate of $0.2$~T/min. The hysteresis window evolves with the in-plane magnetic-field angle. \textbf{c}, Angle-dependent resistivity difference between the forward and backward branches at $0.8$~T, denoted by the black dots. The red curve represents a fit to a twofold-symmetric cosine function. \textbf{d}, Angle-dependent zero-field remanent resistivity extracted from the forward magnetoresistance sweep at $B=0$. The blue curve represents a fit to a threefold-symmetric cosine function.
        }
	\label{EXT7}
\end{extendedfigure}

\begin{extendedfigure}[H]
    \centering
    \includegraphics[width=0.7\textwidth]{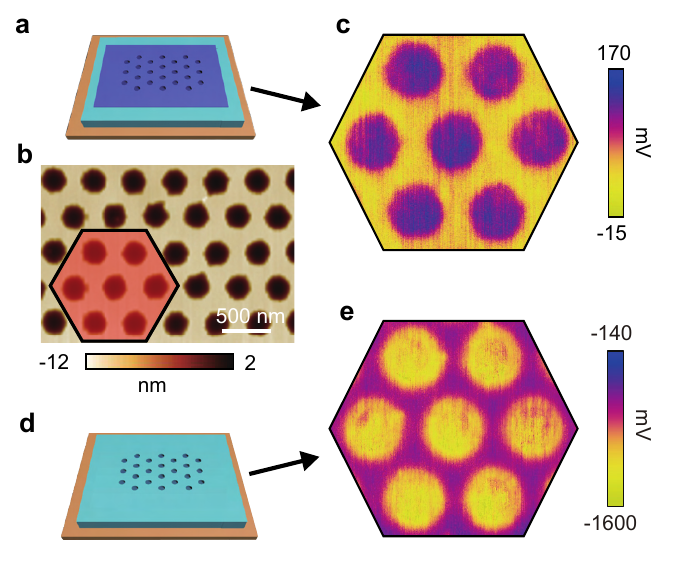}
    \caption{\textbf{AFM and KPFM characterization of strain-patterned graphene and bare patterned hBN.} \textbf{a},\textbf{b}, Schematic illustration of the strain-patterned graphene device (a) and the corresponding AFM topography (b). \textbf{c}, KPFM surface-potential map acquired from the region outlined by the red hexagon in b. The potential inside the nanoholes is higher than that in the surrounding flat regions. \textbf{d},\textbf{e}, Schematic illustration of the bare patterned hBN substrate (d) and the corresponding KPFM surface-potential map (e). In contrast to the graphene-covered device, the bare patterned hBN exhibits a pronounced potential modulation of the opposite sign, with the potential inside the nanoholes substantially lower than that in the surrounding regions.
        }
	\label{EXT8}
\end{extendedfigure}

\begin{extendedfigure}[H]
    \centering
    \includegraphics[width=0.9\textwidth]{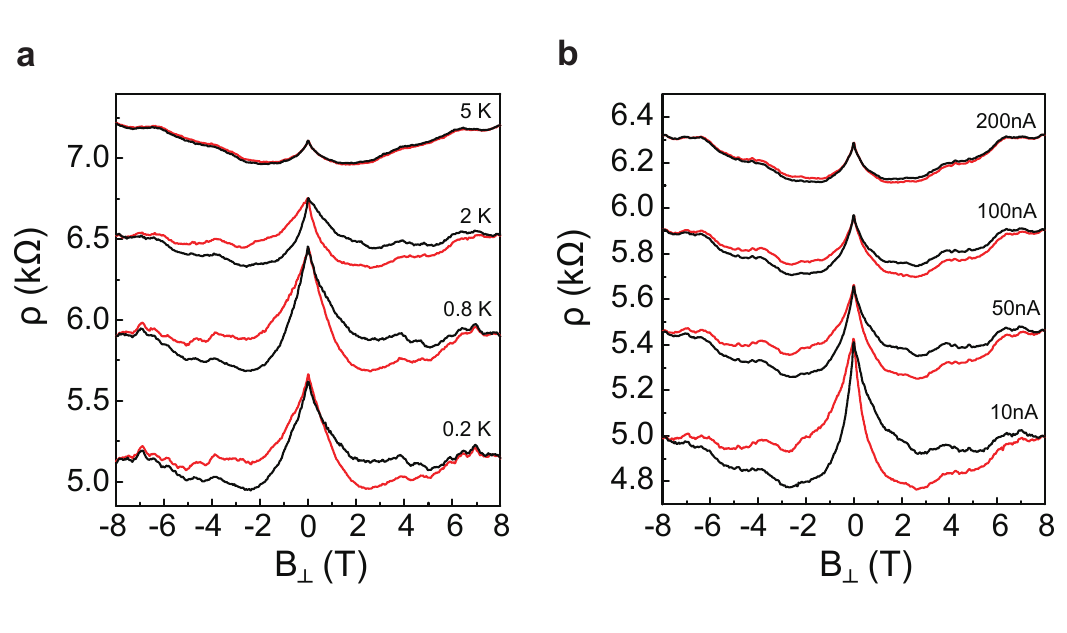}
    \caption{\textbf{Reproduce frustrated magnetism from another representative device.} 
    \textbf{a}, Frustration-driven hysteresis under an out-of-plane magnetic field at various temperatures, showing a non-monotonic evolution of the hysteresis window. \textbf{b}, Current dependence of magnetic hysteresis at $T=1.5$~K, where increasing current gradually reduces and ultimately eliminates the hysteresis window. The consistent transport characteristics indicate that the observed frustration-driven hysteresis is reproducible across samples and that the strain-engineering technique yields reliable device behavior. Traces are vertically offset for clarity. All measurements were performed at $0$~V gate voltage with a field sweep rate of $0.15$~T/min; the driving current in (a) is $10$~nA.
        }
        
	\label{EXT10} 

\end{extendedfigure}

\newpage
\part*{Supplementary Information}
\begin{refsegment}

\appendix
\setcounter{subsubsection}{0}
\renewcommand{\thesubsubsection}{S\arabic{subsubsection}}

\setcounter{figure}{0}
\renewcommand{\thefigure}{S\arabic{figure}}
\renewcommand{\figurename}{Figure}

\subsubsection{Raman characterization of strain-patterned trilayer graphene}
We provide Raman measurements of the strain-patterned trilayer graphene device used in the transport measurements presented in the main text, allowing us to estimate the strain magnitude. Compared with the flat region, both the G and 2D bands in the strained region exhibit redshifts and peak broadening, indicating nonuniform tensile strain, as shown in Figs.~S1a and b. The $D'$ band near 1618~cm$^{-1}$ is extremely weak, suggesting that the strain patterning does not introduce substantial defect scattering and that the graphene retains good quality.

To estimate the strain magnitude, we fitted the G bands measured in both the flat and strained regions. The G band in the flat region is well described by a single component centered at 1582.0~cm$^{-1}$ (Fig.~S1c), whereas the broadened G band in the strained region requires two components, with a dominant peak at 1580.2~cm$^{-1}$ and a lower-frequency component at 1576.6~cm$^{-1}$ (Fig.~S1e). This two-component structure is expected for our highly nonuniform strain profile, since the Raman spot averages over regions with substantially different strain. The lower-frequency component, which reflects the more strongly strained regions, is redshifted by approximately 5.4~cm$^{-1}$ relative to the flat region. Using the previously reported uniaxial-strain calibration for trilayer graphene, in which a G-band redshift of approximately 12~cm$^{-1}$ corresponds to 1\% tensile strain, this corresponds to an effective tensile strain of approximately 0.45\%. Because the strain in our device is neither uniform nor purely uniaxial, and the Raman spot is considerably larger than an individual nanohole, this value should be regarded as an experimental estimate of the strain scale rather than the maximum local strain.

The 2D band of trilayer graphene consists of multiple overlapping double-resonance Raman components, making a reliable quantitative extraction of the strain more difficult. Nevertheless, the strained region exhibits an overall redshift and broadening of the 2D band, and the fitted resonant components shift systematically to lower Raman frequencies compared with the flat region (Figs.~S1d and f). This provides further evidence for tensile strain, while the broadening and modified lineshape are consistent with the spatially nonuniform strain distribution and the associated modification of the electronic and phonon resonance conditions.

The Raman measurements therefore provide an independent experimental estimate of a sub-percent strain scale in the patterned region. Considering the spatial averaging of the Raman measurement and the strongly nonuniform, non-uniaxial strain distribution around the nanoholes, the experimentally estimated value of ~0.45\% is consistent with the local strain of order 1\% obtained from the relaxed theoretical calculations.

\begin{figure}
    \centering
    \includegraphics[width=1\linewidth]{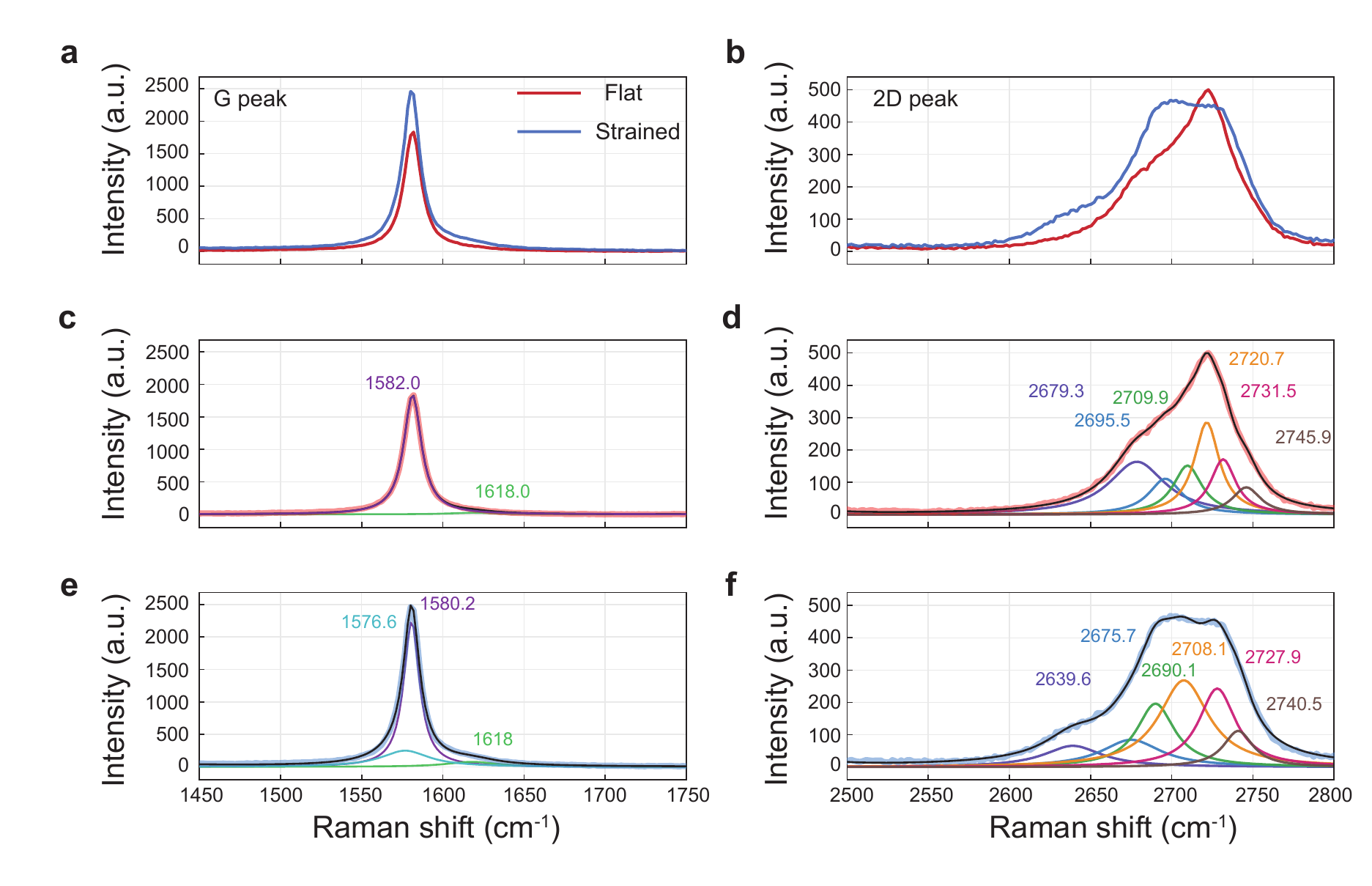}
    \caption{\textbf{Raman characterization of strain-patterned trilayer graphene.} \textbf{a},\textbf{b}, Raman spectra of (a) the G band and (b) the 2D band in the flat (red line) and strained (blue line) regions of strain-patterned trilayer graphene. The redshift and broadening of both bands in the strained region are consistent with nonuniform tensile strain. \textbf{c}-\textbf{f}, Spectral decomposition of the G and 2D bands measured in the flat region (c,d) and strained region (e,f), respectively. The G band in the strained region consists of a dominant peak (purple line) and a weaker lower-frequency component (cyan line), both redshifted relative to the G band measured in the flat region. This spectral structure reflects the spatially nonuniform tensile strain within the probed area, with the lower-frequency component arising from more strongly strained regions. An extremely weak $D'$ band (green) near 1618~cm$^{-1}$ indicates that the strain-patterned graphene retains good quality with little defect scattering. For trilayer graphene, the 2D band (d,f) was fitted with six resonant components, which shift systematically to lower Raman frequencies in the strained region. In panels c-f, the red and blue curves show the experimental Raman spectra measured in the flat and strained regions, respectively. The thin colored curves show the individual fitted components, while the black curves represent the overall fits obtained from their sum.}
    \label{figS1}
\end{figure}

\newpage

\subsubsection{Calculations of strain-induced pseudomagnetic fields}

{\normalsize \textit{S2.1 Pseudogauge field for strained graphene}}\\
\noindent Elastic strain couples to the low-energy electrons in graphene as a pseudogauge field \autocite{PhysRevLett.78.1932,KATSNELSON20073}:
\begin{equation} \label{eq:A}
    \mathbf A(\mathbf r) = \frac{\hbar}{e} \frac{\beta}{2a_0} R(3\theta) \begin{pmatrix} u_{yy} - u_{xx} \\ u_{xy} + u_{yx} \end{pmatrix} + \mathcal O(u_{ij}^2),
\end{equation}
with $\mathbf r = (x,y)$ and where $\theta$ is the angle between the zigzag direction of the graphene lattice (in the $xy$ plane) and the $x$ axis, and $R(3\theta)$ is a $2\times2$ rotation matrix for a counterclockwise rotation about $z$ by an angle $3\theta$. Here $\beta \approx 3$ is the electron Gr\"uneisen parameter and $a_0 \approx 0.142$ nm is the nearest-neighbor distance of pristine graphene. In the nearest-neighbor and central-force approximation, we have $\beta = a_0t'(a_0)/t_0$ with $t(|\mathbf d|)$ the hopping amplitude between coplanar carbon $p_z$ orbitals separated by a bond vector $\mathbf d$ and $t_0 \approx 3$ eV the nearest-neighbor hopping amplitude for pristine graphene. We also introduced the strain tensor for a two-dimensional (2D) elastic membrane embedded in 3D:
\begin{equation} \label{eq:straintensor}
    u_{ij}(\mathbf r) = \frac{1}{2} \left( \frac{\partial u_j}{\partial r_i} + \frac{\partial u_i}{\partial r_j} + \frac{\partial h}{\partial r_i} \frac{\partial h}{\partial r_j} \right),
\end{equation}
with $i,j = x,y$. The strain tensor is a symmetric rank-2 tensor ($u_{ij} = u_{ji}$) which decomposes in a scalar (trace) and deviatoric (traceless) part. The pseudogauge field corresponds to the latter which transforms as a $d$-wave (quadrupolar) object. However, it becomes equivalent to a vector when restricted to valley-preserving symmetries (the little group $D_{3h}$ at $K/K'$). Hence, it is allowed to enter the Hamiltonian in the same form as a vector potential (minimal coupling).

A periodic height modulation can be expanded as a Fourier series:
\begin{equation} \label{eq:hFourier1}
    h(\mathbf r) = \sum_{\mathbf g} h_{\mathbf g} e^{i\mathbf g\cdot \mathbf r},
\end{equation}
where the sum runs over reciprocal lattice vectors $\mathbf g$ of the periodic strain superlattice induced by the height profile. In particular, we consider a buckling pattern with $C_{6v} = \langle \mathcal C_{6z}, \mathcal M_y \rangle$ symmetry. Here the angle brackets indicate the group is generated by successive applications of the given sequence of operations. Hence, the height profile satisfies:
\begin{equation}
    h(\mathcal C_{6z}\mathbf r) = h(\mathbf r), \qquad h(x,y) = h(x,-y).
\end{equation}
In this case, we can simplify Eq.\ \eqref{eq:hFourier1} to
\begin{equation}
    h(\mathbf r) = \bar h + 2 \sum_{m=1}^\infty \sum_{n=0}^{m-1} h_{mn} \sum_{i=0}^2 \cos(\mathcal C_{3z}^i \mathbf g_{mn} \cdot\mathbf r),
\end{equation}
with $\mathcal C_{3z}$ a threefold rotation about the $z$ axis and
\begin{equation} \label{eq:hFourier2}
    \mathbf g_{mn} = m \mathbf g_1 + n \mathbf g_2, \qquad \mathbf g_2 = \mathcal C_{3z} \mathbf g_1,
\end{equation}
where $m = 1,2,\ldots$ labels the reciprocal shell and $n = 0,\ldots,m-1$ labels the reciprocal stars for a given shell. Here the $m$th shell contains $m$ stars that lie on the edge of a hexagon with radius $4\pi m/(\sqrt{3} L)$ with $L$ the period of the periodic strain field. Note that we define each star as six reciprocal vectors that are closed under $\mathcal C_{6z}$. In addition, stars related by $\mathcal M_y~(y \mapsto -y)$ have equal Fourier components. Taking $\mathbf g_1 = 4\pi \hat x / (\sqrt{3} L)$, we have $h_{m,n} = h_{m,m-n}$.

It is interesting to note how the orientation of the graphene with respect to the strain field changes the pseudomagnetic field (PMF). To this end, we define $\mathbf A_0(\mathbf r)$ as the pseudogauge field for the orientation where $x \parallel$ zigzag ($\theta = 0$):
\begin{equation}
    B(\mathbf r) = \nabla \times R(3\theta) \mathbf A_0 = \cos(3\theta) \nabla \times \mathbf A_0 + \sin(3\theta) \nabla \cdot \mathbf A_0.
\end{equation}
One therefore has to be careful to take into account the full $\mathbf A_0$ and only allow for gauge transformations ($\mathbf A \rightarrow \mathbf A + \nabla f$) after taking the orientation into account.\\

\noindent{\normalsize \textit{S2.2 Quenched strain}}\\
\noindent In the presence of an out-of-plane displacement field, graphene will relax elastically, producing in-plane displacements that lower the total elastic energy. We first consider the case without lattice relaxation, which is referred to as \textit{quenched strain}. In this case, there are no in-plane displacement fields by definition, and
\begin{equation}
    u_{ij} = \frac{1}{2} ( \partial_i h ) ( \partial_j h ) \equiv \frac{f_{ij}}{2}.
\end{equation}
The quenched pseudogauge field becomes
\begin{equation} \label{eq:Aq}
    \mathbf A_\text{quenched}(\mathbf r) = \frac{\hbar}{e} \frac{\beta}{4a_0} R(3\theta) \begin{pmatrix} f_{yy} - f_{xx} \\ 2f_{xy} \end{pmatrix}.
\end{equation}

\paragraph{Single hole}

\begin{figure}
    \centering
    \includegraphics[width=.8\linewidth]{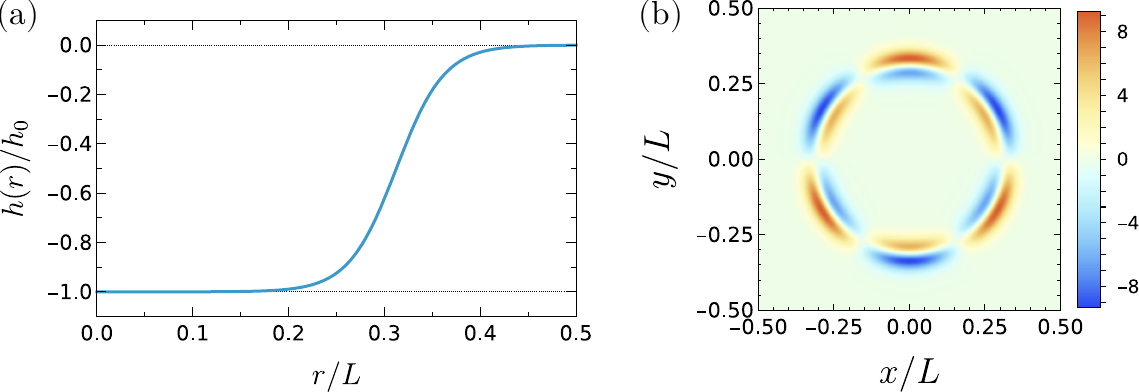}
    \caption{\textbf{Height profile and pseudomagnetic field of a (quenched) nanohole.} (a) Radial height profile of a single hole for $w/L = 0.05$ and $r_0/L = 0.3125$. (b) Corresponding PMF in units $10^6 (h_0^2 / L^3)$ tesla nm  for $\beta = 3$ and $a_0 = 0.142$ nm, in the absence of in-plane lattice relaxation. Here, we take $x \parallel$ zigzag direction of the graphene lattice. Changing the orientation of the graphene lattice, i.e.\ the angle $\theta$ between the zigzag direction and the $x$ axis, rotates the PMF profile by $3\theta$.}
    \label{fig:singlehole}
\end{figure}

Let us start by considering graphene subject to a single patterned nanohole. We model the corresponding height profile as
\begin{equation}
    h_\text{single}(x,y) = \frac{h_0}{2} \left[ \tanh \left( \frac{\sqrt{x^2+y^2} - r_0}{w} \right) - 1 \right],
\end{equation}
where $r_0$ gives the hole radius and $w$ determines the extremal radial slope $h_0/(2w)$ at $r = r_0$. Hence, the corresponding quenched strain tensor reaches a maximum at $r = r_0$. This height profile is illustrated in Fig.\ \ref{fig:singlehole}(a). The corresponding pseudomagnetic field (PMF) $B(\mathbf r) = \partial_x A_y - \partial_y A_x$ obtained with Eq.\ \eqref{eq:Aq} is shown in Fig.\ \ref{fig:singlehole}(b). We see that the PMF vanishes along the zigzag direction $\theta = 0$ mod $\pi/3$. This follows from the fact that the pseudogauge field is purely radial for the zigzag direction, and therefore a pure gauge. Indeed, in polar coordinates $(r,\varphi)$ with $r = \sqrt{x^2+y^2}$ and $\varphi$ the angle between $\mathbf r$ and the $x$ axis,
\begin{equation}
    \nabla \times \mathbf A(r) = \hat z \, \frac{1}{r} \frac{d}{dr} r A_\varphi, 
\end{equation}
where we used that the hole is circularly symmetric $h=h(r)$. We thus obtain a PMF with threefold rotation symmetry, consistent with the valley-preserving symmetries of graphene. Crystal symmetries that exchange valleys reverse the sign of the PMF, as expected from time-reversal symmetry. Furthermore, the PMF features a domain wall where it changes sign, centered at $r=r_0$ where the strain and thus the pseudogauge field is extremal. In a semiclassical picture, such a domain wall can trap an electron in a pseudomagnetic snake-like orbit. One can thus view the single nanohole as a leaky (i.e.\ when the radial direction aligns with the zigzag direction) strain-induced quantum dot where electrons are confined by pseudomagnetic fields.

\paragraph{Periodic height profile}

\begin{figure}
    \centering
    \includegraphics[width=.95\linewidth]{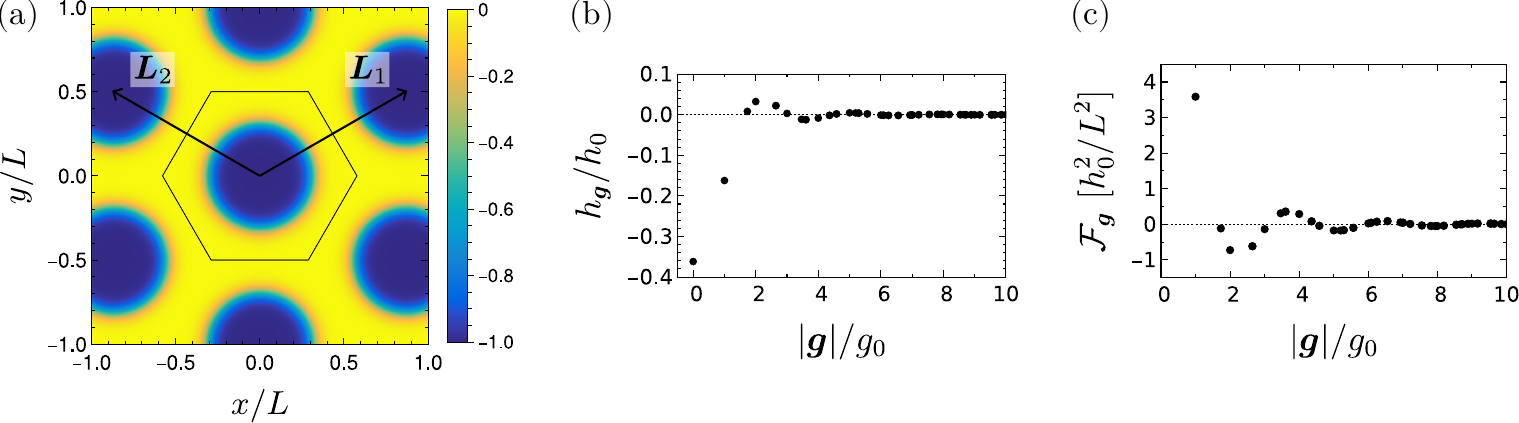}
    \caption{\textbf{Periodic array of nanoholes}. (a) Periodic height profile $h(\mathbf r)$ in units $h_0$ where the black hexagon is the unit cell of the strain superlattice. (b) Fourier components corresponding to (a) as a function of $|\mathbf g|/g_0$ with $g_0 = 4\pi/(\sqrt{3}L)$. (c) Fourier components $\mathcal F_{\mathbf g}$ which are related to the relaxed strain tensor in units $h_0^2/L^2$. Shown here for: $w/L = 0.05$ and $r_0/L = 0.3125$.}
    \label{fig:h}
\end{figure}

To obtain a periodic lattice of holes with lattice constant $L$ we consider the periodic extension of a single hole. Here, we ensure that the single hole is contained completely in the unit cell. That is, we take a hexagonal unit cell centered at the origin such that $h_\text{single}(r)$ is exponentially small at the borders of the unit cell. We then have
\begin{equation}
    h_{\mathbf g} = \frac{1}{A_\text{c}} \int_\text{symm.\ cell} d^2\mathbf r \, h_\text{single}(\mathbf r) e^{-i\mathbf g\cdot \mathbf r},
\end{equation}
with $A_\text{c} = \sqrt{3}L^2/2$ the unit cell area. Note that one has to integrate over a unit cell that respects the symmetry of the height profile. This is because $h_\text{single}(r)$ is not actually periodic. The results are shown in Fig.\ \ref{fig:h} for $w/L = 0.05$ and $r_0/L = 0.3125$. These values are obtained from fitting to the experiment. The Fourier components are real because of $\mathcal C_{2z}$ symmetry, and decay exponentially with increasing $|\mathbf g|$. The number of significant Fourier components increases with $w/L$ as expected.\\

\noindent{\normalsize \textit{S2.3 Annealed strain}}\\
\noindent The buckled graphene lattice can relax by displacing atoms in the $xy$ plane, thereby lowering its elastic energy. Hence, we introduce the in-plane displacement field
\begin{equation}
    \mathbf u(\mathbf r) = \sum_{\mathbf g} \mathbf u_{\mathbf g} e^{i\mathbf g\cdot \mathbf r}.
\end{equation}
Long-wavelength acoustic displacements in graphene can be modeled using the continuum theory of elasticity \autocite{landau_theory_1986}. Here, one views the graphene layers as an elastically isotropic membrane. In particular, the elastic potential energy \autocite{nelson_david_statistical_2004} and the coupling with the substrate can be modeled as \autocite{guinea_PRB08}
\begin{align}
    H_\text{elas} & = \frac{1}{2} \int d^2 \mathbf r \left[ \lambda u_{ii} u_{ii} + 2 \mu u_{ij} u_{ji} + \kappa ( \nabla^2 h )^2 \right], \label{eq:Helas} \\
    H_\text{sub} & = \frac{\gamma}{2} \int d^2 \mathbf r \left[ h(\mathbf r) - h_\text{sub}(\mathbf r) - h_\text{eq} \right]^2, \label{eq:Hsub} 
\end{align}
where $\lambda$ and $\mu$ are in-plane Lam\'e constants, $\kappa$ is the out-of-plane bending rigidity, and $\gamma$ controls the interaction with the substrate. Equivalently, $\lambda+\mu$ and $\mu$ are the bulk and shear modulus, respectively. In Eq.\ \eqref{eq:Helas} summation over repeated indices is implied. We further assume that the graphene is pinned to the substrate such that $h(\mathbf r) - h_\text{sub}(\mathbf r) = h_\text{eq}$. This is justified in the limit $L \gg \left( \kappa / \gamma \right)^{1/4} \approx 1 \; \text{nm}$ where the numerical value is for graphene on SiO$_2$ \autocite{guinea_PRB08,sabio_electrostatic_2008}. In this case, the interaction with the substrate dominates over the curvature term [last term in the integrand of Eq.\ \eqref{eq:Helas}] since the latter scales as $\kappa h^2 / L^4$ while the former scales as $\gamma h^2$ (inside the integral). In this limit, $H_\text{elas}$ is a functional of the in-plane displacement field only. Our goal is now to obtain the static zero-temperature ground-state configuration for the in-plane displacement field $\mathbf u(\mathbf r)$ given a fixed height profile $h(\mathbf r)$ under periodic boundary conditions.

Under these assumptions, the elastic energy density can be written as
\begin{equation}
    \mathcal H_\text{elas} = \frac{1}{A_\text{c}} \int_\text{cell} d^2 \mathbf r \left[ \left( \frac{\lambda}{2} + \mu \right) \left( u_{xx}^2 + u_{yy}^2 \right) + \lambda u_{xx} u_{yy} + 2 \mu u_{xy}^2 \right] + \text{constant},
\end{equation}
where $A_\text{c}$ is the area of the supercell defined by the periodic height modulation. For the triangular height profile, we have $A_\text{c} = \sqrt{3} L^2 / 2$. For convenience, we define the symmetric matrix
\begin{equation}
    f_{ij}(\mathbf r) \equiv \left[ \partial_i h(\mathbf r) \right] \left[ \partial_j h(\mathbf r) \right] = \sum_{\mathbf g} f_{ij\mathbf g} \, e^{i \mathbf g \cdot \mathbf r},
\end{equation}
where
\begin{equation}
    f_{ij\mathbf g} = \sum_{\mathbf g'} h_{\mathbf g-\mathbf g'} h_{\mathbf g'}  g_i' (g_j' - g_j),
\end{equation}
with $f_{ij,-\mathbf g} = f_{ij\mathbf g}^*$. The strain tensor, defined in Eq.\ \eqref{eq:straintensor}, thus becomes ($i,j = x,y$)
\begin{equation}
    u_{ij}(\mathbf r) = \frac{1}{2} \sum_{\mathbf g} \left[ i \left( g_i u_{j\mathbf g} + g_j u_{i\mathbf g} \right) + f_{ij\mathbf g} \right] e^{i \mathbf g \cdot \mathbf r},
\end{equation}
where we set $u_{i\mathbf 0} = 0$ since this amounts to a uniform translation. Plugging the Fourier expansions into the energy density $\mathcal H_{\text{elas}}$, and thus enforcing periodic boundary conditions, gives \autocite{phong_PRL22}
\begin{align}
     \frac{1}{A_\text{c}} \int_\text{cell} d^2 \mathbf r \, u_{ii}^2 & = \frac{1}{A_\text{c}} \sum_{\mathbf g,\mathbf g'} \int d^2 \mathbf r \left( i g_i u_{i\mathbf g} + \frac{f_{ii \mathbf g}}{2} \right) \left( i g_i' u_{i\mathbf g'} + \frac{f_{ii \mathbf g'}}{2} \right) e^{i ( \mathbf g + \mathbf g' ) \cdot \mathbf r} \\
     & = \sum_{\mathbf g} \left| i g_i u_{i\mathbf g} + \frac{f_{ii \mathbf g}}{2} \right|^2, \\
     \frac{1}{A_\text{c}} \int_\text{cell} d^2 \mathbf r \, u_{xx} u_{yy} & = \sum_{\mathbf g} \left( i g_x u_{x\mathbf g} + \frac{f_{xx \mathbf g}}{2} \right) \left( -i g_y u_{y\mathbf g}^* + \frac{f_{yy \mathbf g}^*}{2} \right) \\
     & = \frac{1}{2} \sum_{\mathbf g} \left[ \left( i g_x u_{x\mathbf g} + \frac{f_{xx \mathbf g}}{2} \right) \left( -i g_y u_{y\mathbf g}^* + \frac{f_{yy \mathbf g}^*}{2} \right) + \text{c.c.} \right], \\
     \frac{1}{A_\text{c}} \int_\text{cell} d^2 \mathbf r \, u_{(xy)}^2 & = \frac{1}{4} \sum_{\mathbf g} \left( i g_x u_{y\mathbf g} + i g_y u_{x\mathbf g} + f_{xy\mathbf g} \right) \left( -i g_x u_{y\mathbf g}^* - i g_y u_{x\mathbf g}^* + f_{xy\mathbf g}^* \right).
\end{align}
Putting everything together, we obtain
\begin{align}
    \mathcal H_\text{elas} & = \left( \frac{\lambda}{2} + \mu \right) \sum_{\mathbf g} \left( i g_x u_{x\mathbf g} + \frac{f_{xx \mathbf g}}{2} \right) \left( -i g_x u_{x\mathbf g}^* + \frac{f_{xx \mathbf g}^*}{2} \right) \\
    & + \left( \frac{\lambda}{2} + \mu \right) \sum_{\mathbf g} \left( i g_y u_{y\mathbf g} + \frac{f_{yy \mathbf g}}{2} \right) \left( -i g_y u_{y\mathbf g}^* + \frac{f_{yy \mathbf g}^*}{2} \right) \\
    & + \frac{\lambda}{2} \sum_{\mathbf g} \left[ \left( i g_x u_{x\mathbf g} + \frac{f_{xx\mathbf g}}{2} \right) \left( -i g_y u_{y\mathbf g}^* + \frac{f_{yy\mathbf g}^*}{2} \right) + \text{c.c.} \right] \\
    & + \frac{\mu}{2} \sum_{\mathbf g} \left( i g_x u_{y\mathbf g} + i g_y u_{x\mathbf g} + f_{xy\mathbf g} \right) \left( -i g_x u_{y\mathbf g}^* - i g_y u_{x\mathbf g}^* + f_{xy\mathbf g}^* \right) + \frac{\kappa}{2} \sum_{\mathbf g} | h_{\mathbf g} |^2 g^4.
\end{align}
Minimizing with respect to $u_{i\mathbf g}^*$ for $|\mathbf g| > 0$ yields the solutions for the Fourier components $u_{i\mathbf g}$ in terms of $f_{ij\mathbf g}$ (and thus $h_{\mathbf g}$). We find
\begin{align}
    \frac{\partial \mathcal H_\text{elas}}{\partial u_{x\mathbf g}^*} & = -i g_x \left[ \left( \lambda + 2\mu \right) \left( i g_x u_{x\mathbf g} + \frac{f_{xx \mathbf g}}{2} \right) + \lambda \left( i g_y u_{y\mathbf g} + \frac{f_{yy \mathbf g}}{2} \right) \right] - i \mu g_y \left( i g_x u_{y\mathbf g} + i g_y u_{x\mathbf g} + f_{xy\mathbf g} \right), \\
    \frac{\partial \mathcal H_\text{elas}}{\partial u_{y\mathbf g}^*} & = -i  g_y \left[ \left( \lambda + 2 \mu \right) \left( i g_y u_{y\mathbf g} + \frac{f_{yy \mathbf g}}{2} \right) + \lambda \left( i g_x u_{x\mathbf g} + \frac{f_{xx \mathbf g}}{2} \right) \right] - i \mu g_x \left( i g_x u_{y\mathbf g} + i g_y u_{x\mathbf g} + f_{xy\mathbf g} \right),
\end{align}
with roots
\begin{align}
    u_{x\mathbf g} & = \frac{i}{2 \left( \lambda + 2 \mu \right) g^4} \left\{ f_{xx\mathbf g} g_x \left[ g_x^2 \left( \lambda + 2 \mu \right) + g_y^2 \left( 3 \lambda + 4 \mu \right) \right] + \left( f_{yy\mathbf g} g_x - 2 f_{xy\mathbf g} g_y \right) \left[ g_x^2 \lambda - g_y^2 \left( \lambda + 2 \mu \right) \right]  \right\}, \label{eq:uxg} \\
    u_{y\mathbf g} & = \frac{i}{2 \left( \lambda + 2 \mu \right) g^4} \left\{ f_{yy\mathbf g} g_y \left[ g_y^2 \left( \lambda + 2 \mu \right) + g_x^2 \left( 3 \lambda + 4 \mu \right) \right] + \left( f_{xx\mathbf g} g_y - 2 f_{xy\mathbf g} g_x \right) \left[ g_y^2 \lambda - g_x^2 \left( \lambda + 2 \mu \right) \right]  \right\}, \label{eq:uyg}
\end{align}
where $g = |\mathbf g|$ and we find \autocite{guinea_PRB08,phong_PRL22}
\begin{align}
    u_{xx\mathbf g} + u_{yy\mathbf g} & = \frac{1-\nu}{2} \mathcal F_{\mathbf g}, \label{eq:trace} \\
    u_{xx\mathbf g} - u_{yy\mathbf g} & = \frac{1+\nu}{2} \frac{g_y^2 - g_x^2}{g^2} \mathcal F_{\mathbf g}, \label{eq:shear1} \\
    u_{xy\mathbf g} + u_{yx\mathbf g} & = -\frac{1+\nu}{2} \frac{2 g_x g_y}{g^2} \mathcal F_{\mathbf g}, \label{eq:shear2}
\end{align}
for nonzero $\mathbf g$ and $u_{ij\mathbf 0} = f_{ij\mathbf 0}/2$. These are the volumetric [Eq.\ \eqref{eq:trace}] and shear strains [Eqs.\ \eqref{eq:shear1} and \eqref{eq:shear2}]. In this work we use the value $\nu = 0.19$ for the Poisson ratio of graphene, obtained from density-functional theory calculations \autocite{carr_relaxation_2018}. Here we also defined,
\begin{equation} \label{eq:Fg}
    \mathcal F_{\mathbf g} = \frac{g_x^2 f_{yy\mathbf g} + g_y^2 f_{xx\mathbf g} - 2 g_x g_y f_{xy\mathbf g}}{g^2} = \frac{1}{g^2} \sum_{\mathbf g'} h_{\mathbf g-\mathbf g'} h_{\mathbf g'} \left( \mathbf g \times \mathbf g' \right)^2,
\end{equation}
which transforms as a scalar and is shown for a $C_{6v}$-symmetric height profile in Fig.\ \ref{fig:h}(c). Importantly, since $|\mathcal F_{\mathbf g}| \sim h^2 / L^2$ we expect continuum elasticity to be valid only for $h/L \ll 1$. This is always the case in the experiment. We further used the relations
\begin{equation}
    \mu = \frac{E}{2 ( 1 + \nu )}, \qquad \lambda = \frac{\nu E}{1 - \nu^2},  \qquad \nu = \frac{1}{1 + 2 \mu / \lambda},
\end{equation}
for isotropic linear elastic 2D materials, with $E$ the Young modulus and $\nu$ the Poisson ratio.

\paragraph{Relaxed strain fields}

\begin{figure}
    \centering
    \includegraphics[width=\linewidth]{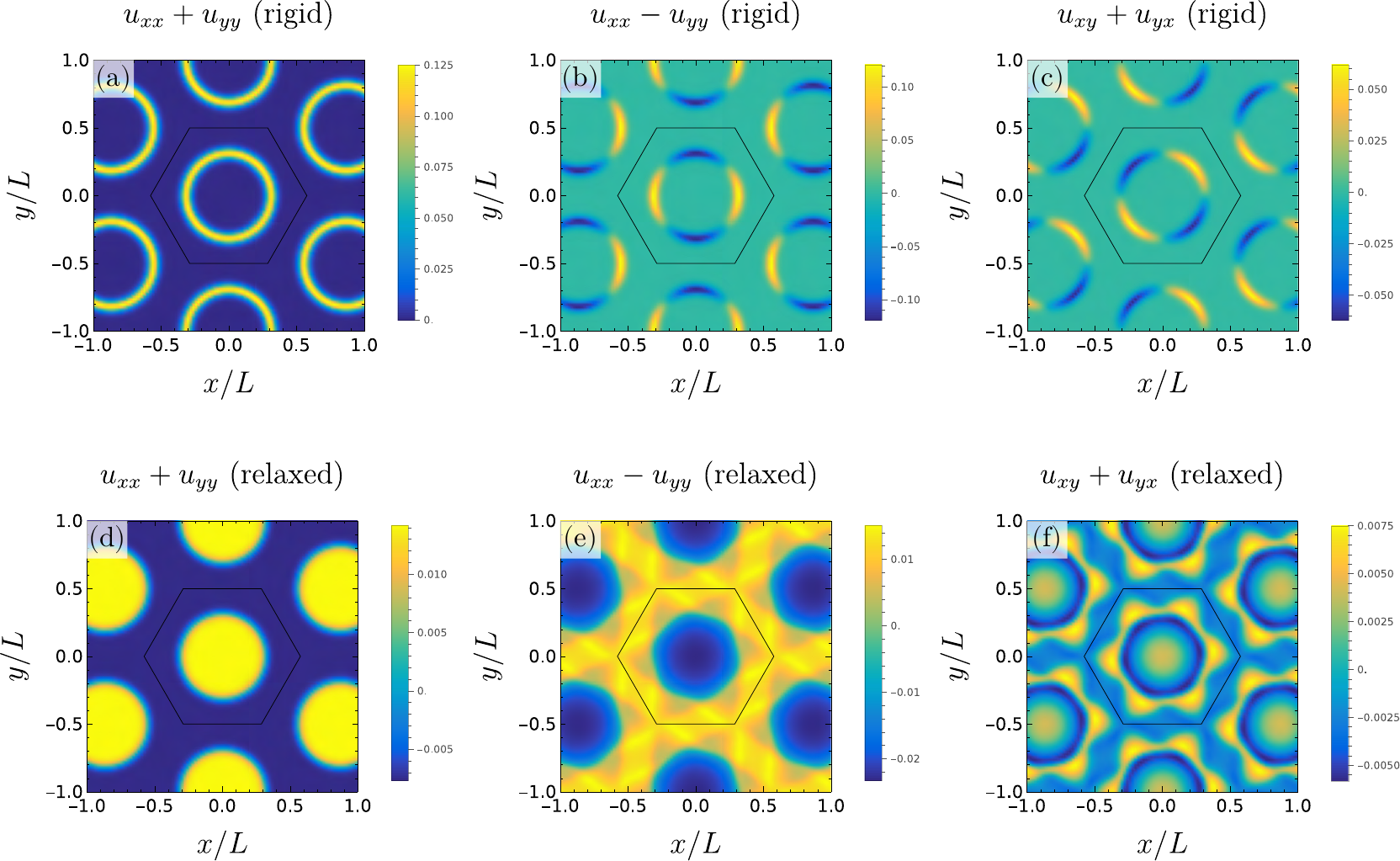}
    \caption{Rigid (a)--(c) and relaxed (d)--(f) strain fields due to a periodic height profile $h(\mathbf r)$ for $L = 400$~nm, $h_0 = 20$~nm, $r_0 / L = 0.3125$, and $w / L = 0.05$. We show the volumetric strain ($u_{xx} + u_{yy}$), corresponding to dilation (positive values) and compression (negative values), and the two shear strains ($u_{xx} - u_{yy}$ and $u_{xy} + u_{yx}$). While the trace of the strain tensor is a scalar, and thus mirrors the symmetry of the height profile, the shear strains do not transform as scalars.}
    \label{fig:strain}
\end{figure}

In Fig.\ \ref{fig:strain}, we compare the rigid and relaxed strain fields. While the rigid strain is on the order of 10\%, the relaxed strain is one order of magnitude smaller. Since this is the ground-state configuration of the elastic energy functional, the relaxed theory is inside the elastic regime. Consider, for example, starting from $h=0$ and thus $u_{ij} = 0$. Now increase the amplitude of the height profile from zero up to a value $h_0$. As long as the ground-state strain tensor remains in the elastic regime during this process, we can safely use the lowest-order expressions for the resulting pseudogauge field. Furthermore, we note that the trace of the strain tensor $(u_{xx} + u_{yy})$, which gives the volumetric strain (compression or dilation), enters the electronic theory as the deformation potential:
\begin{equation}
    V_\text{def}(\mathbf r) = \alpha \left( u_{xx} + u_{yy} \right),
\end{equation}
with $\alpha \approx -5.34$ eV where include both the on-site and next-nearest neighbor hopping contributions \autocite{fang_electronic_2018}. Experimentally, the deformation potential can be estimated using KPFM measurements. From Fig.\ \ref{fig:strain}(d), we find that the difference in the relaxed deformation potential inside and outside the holes $\Delta V_\text{def} \approx 0.022 |\alpha| \approx 117$~meV matches well with the measured potential difference of $200$~meV [see Fig.\ \ref{fig4}(b) of the main text]. This provides additional evidence that the strained graphene in the experiment is in the elastic regime, where in-plane lattice relaxation plays a crucial role for stabilizing the strain superlattice.

As an interesting aside, we note that $\mathcal F_{\mathbf g}$ is related to the Fourier components of the Gaussian curvature of the height profile \autocite{Wehling_2008}. Indeed, in the Monge gauge $h = h(x,y)$ the Gaussian curvature $K$ times the squared determinant of the metric $\mathfrak g_{ij} = \delta_{ij} + (\partial_i h)(\partial_j h)$ is given by the Hessian determinant:
\begin{align}
    K \left( \det \mathfrak g \right)^2 = (\partial_x^2 h) (\partial_y^2 h) - \left( \partial_x \partial_y h \right)^2 & = \sum_{\mathbf g,\mathbf g'} h_{\mathbf g} h_{\mathbf g'} \left[ \frac{g_x^2 ( g_y' )^2 + g_y^2 ( g_x' )^2}{2} - g_x g_y g_x' g_y' \right] e^{i ( \mathbf g + \mathbf g' ) \cdot \mathbf r} \\
    & = \frac{1}{2} \sum_{\mathbf g,\mathbf g'} h_{\mathbf g-\mathbf g'} h_{\mathbf g'} \left( \mathbf g \times \mathbf g' \right)^2 e^{i\mathbf g \cdot \mathbf r},
\end{align}
such that
\begin{equation}
    \left[ K \left( \det \mathfrak g \right)^2 \right]_{\mathbf g} = \frac{1}{2} \, g^2 \mathcal F_{\mathbf g}.
\end{equation}

Finally, the relaxed pseudogauge field becomes
\begin{equation} \label{eq:Ar}
    \mathbf A_\text{relax}(\mathbf r) = \frac{\hbar}{e} \frac{\beta \left( 1 + \nu \right)}{4a_0} R(3\theta) \sum_{\mathbf g \neq \mathbf 0} \frac{\mathcal F_{\mathbf g}}{g^2} \begin{pmatrix} g_x^2 - g_y^2 \\ -2g_xg_y \end{pmatrix} e^{i\mathbf g\cdot \mathbf r},
\end{equation}
where the sum excludes $\mathbf g = \mathbf 0$. The latter gives a constant term which does not contribute to the pseudomagnetic field. Indeed, a constant can always be removed by a gauge transformation, i.e.\ by changing the origin of momentum space in the electronic continuum Hamiltonian. So in practice, we start from $h_\text{single}(\mathbf r)$ from which we numerically compute the Fourier components $h_{\mathbf g}$ to obtain the periodic height profile. These also yield the $\mathcal F_{\mathbf g}$ using Eq.\ \eqref{eq:Fg} which then finally gives the relaxed pseudogauge field given in Eq.\ \eqref{eq:Ar}.
\begin{figure}
    \centering
    \includegraphics[width=.9\linewidth]{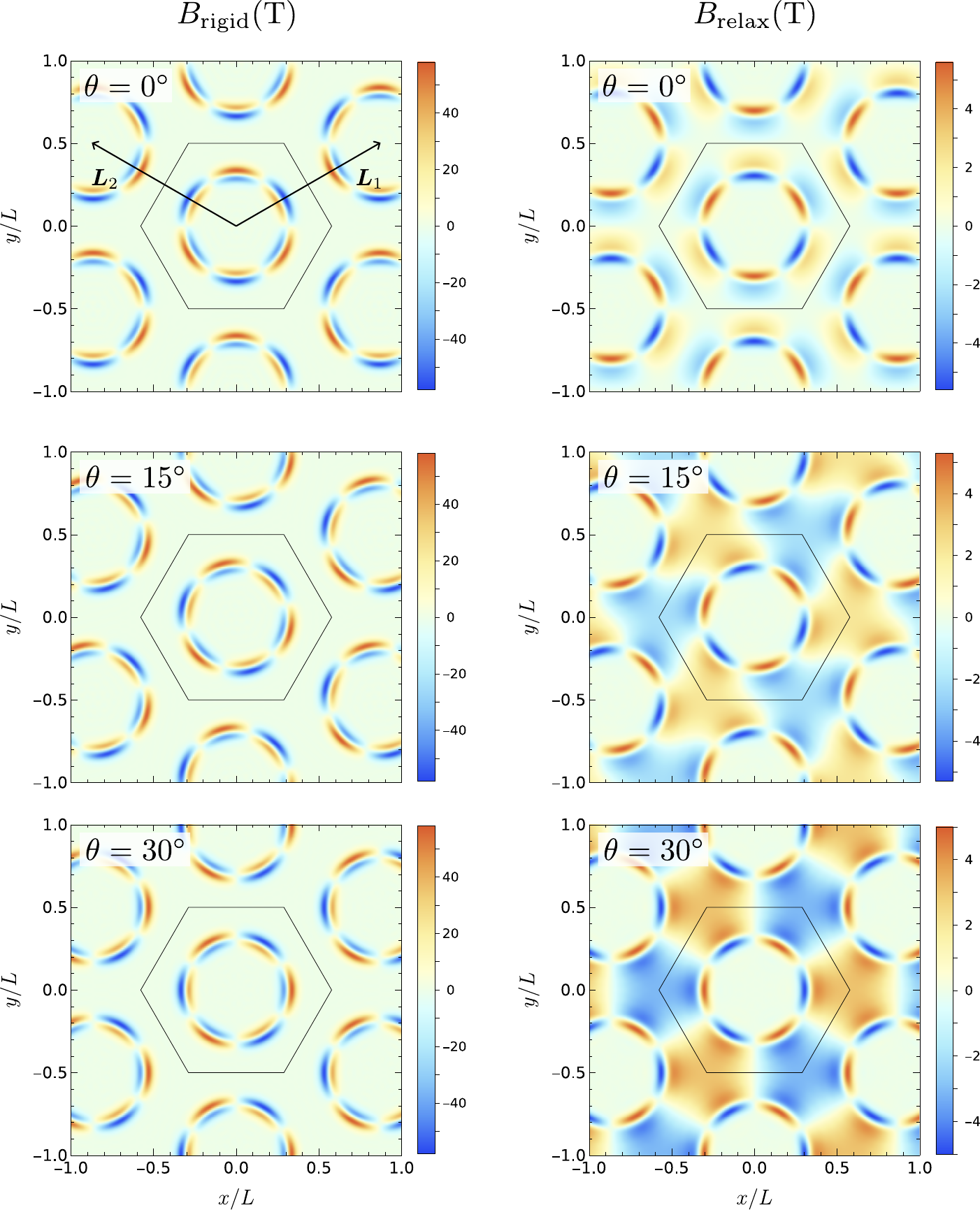}
    \caption{\textbf{Pseudomagnetic fields from a periodic $C_{6v}$ height modulation.} Rigid and relaxed PMFs in tesla for different orientations $\theta$ of graphene relative to the strain field. Here $\theta = 0$ ($\theta = \pi/6$) correspond to alignment of the zigzag (armchair) direction with the $x$ axis. The black hexagon gives the unit cell of the strain superlattice with primitive vectors $\mathbf L_{1,2}$. Shown for: $w/L = 0.05$, $r_0/L = 0.3125$, $h_0 = 20$ nm, $L = 400$ nm, $\beta = 3$, and $a_0 = 0.142$ nm.}
    \label{fig:pmf}
\end{figure}

The rigid (i.e.\ unrelaxed) and relaxed PMFs for the $C_{6v}$ height profile induced by a periodic array of nanoholes, are shown in Fig.\ \ref{fig:pmf} for different values of the orientation of the graphene lattice with respect to the height profile. Here, we use parameters fitted to the experimental device discussed in the main text. While lattice relaxation reduces the maximal strain [Fig.\ \ref{fig:strain}] as expected, elastic coupling between neighboring patterned holes extends the strain field to the regions in between the holes. Concomitantly, this leads to an overall reduction of the PMF by one order of magnitude, while also introducing a nonzero PMF to regions between adjacent holes. However, for zigzag alignment ($\theta = 0$ mod $\pi/3$), defined relative to the $x$ axis, the PMFs of neighboring holes tend to cancel. In contrast, for armchair alignment ($\theta = \pi/6$ mod $\pi/3$) the PMFs of neighboring holes add constructively, leading to a significant PMF between the holes. In this case, we see that electrons can be confined in hexagonal snake-like skipping orbits that form an emergent honeycomb lattice with counterpropagating orbits for different valleys and (emergent) sublattices.

\subsubsection{Topological flat bands in periodically strained bilayer graphene}

\noindent{\normalsize \textit{S3.1 Electronic continuum model}}\\
The low-energy electronic continuum Hamiltonian for Bernal bilayer graphene in the presence of homostrain \autocite{PhysRevB.108.125129} and an interlayer bias $V_z$ is given in lowest order by
\begin{equation} \label{eq:Heffapp}
    H = \sum_\tau \int d^2 \mathbf r \, \psi_\tau^\dag(\mathbf r) \left\{ \hbar v_F \left[ -i \nabla_{\mathbf r} + \frac{\tau e}{\hbar} \mathbf A(\mathbf r) \right] \cdot (\tau \sigma_x, \sigma_y) + \frac{V_z}{2} \lambda_z + \mathcal H_\perp \right\} \psi_\tau(\mathbf r),
\end{equation}
with interlayer coupling
\begin{equation}
    \mathcal H_\perp = \frac{\gamma}{2} \left( \lambda_x \sigma_x + \lambda_y \sigma_y \right),
\end{equation}
where we take $\hbar v_F = 3t_0 a_0/2$ with $t_0 = 3$ eV and $a_0 = 0.142$ nm, and $\gamma = 300$ meV. Here $\tau = \pm 1$ is the valley index corresponding to valley $K$ and $K'$, respectively, and $\sigma_i$ ($\lambda_i$) with $i=x,y,z$ are Pauli matrices acting in sublattice (layer) space, where juxtaposition of Pauli matrices corresponds to a Kronecker product.

We further introduced $\psi_\tau(\mathbf r) = \left[ \psi_{\tau A1}(\mathbf r), \psi_{\tau B1}(\mathbf r), \psi_{\tau A2}(\mathbf r), \psi_{\tau B2}(\mathbf r) \right]^\top$ which is a four-component fermion field operator that only contains long-wavelength momentum components near valley $\tau$. These satisfy the standard anticommutation relations $\{\psi_{\tau\sigma l}(\mathbf r), \psi_{\tau'\sigma'l'}^\dag(\mathbf r')\} = \delta_{\tau\tau'} \delta_{\sigma\sigma'} \delta_{ll'} \delta(\mathbf r-\mathbf r')$. We note that the shear component of the strain couples to the electrons as a pseudogauge field $\mathbf A(\mathbf r)$ giving rise to a pseudomagnetic field, while the volumetric part (trace of the strain tensor) results in a pseudoelectric field. In the previous section, we showed that the latter is radial and confined to the edges of the holes. Moreover, for electron densities $n \sim 10^{12}$ cm$^{-2}$ relevant to the experiment, the pseudoelectric field is screened and therefore we do not consider it further in this work \autocite{fogler_PRL08}.

In the presence of periodic scalar and gauge fields, the continuum Hamiltonian can be diagonalized by Fourier transformation (Bloch's theorem),
\begin{equation}
    \psi_\tau(\mathbf r) = \frac{1}{\sqrt{A}} \sum_{\mathbf k \in \text{SBZ}} \sum_{\mathbf g} e^{i (\mathbf k + \mathbf g ) \cdot \mathbf r} \, c_{\tau,\mathbf k+\mathbf g},
\end{equation}
where $A$ is the system size and the sum over $\mathbf k$ is restricted to the superlattice Brillouin zone (SBZ). Here we defined four-component creation (annihilation) operators $c_{\tau,\mathbf k+\mathbf g}^\dag$ ($c_{\tau,\mathbf k+\mathbf g}$) that create (destroy) a fermion in valley $\tau$ with momentum $\mathbf k+\mathbf g$. If we plug in the Fourier transform, the Hamiltonian becomes
\begin{equation}
    \begin{aligned}
        H & = \frac{1}{A} \sum_{\tau} \sum_{\mathbf k,\mathbf k'} \sum_{\mathbf g,\mathbf g'} \int d^2\mathbf r \, c_{\tau,\mathbf k+\mathbf g}^\dag e^{i(\mathbf k' + \mathbf g' - \mathbf k - \mathbf g) \cdot \mathbf r} \\
        & \qquad \times \left\{ \hbar v_F \left[ \mathbf k' + \mathbf g' + \frac{\tau e}{\hbar} \, \mathbf A(\mathbf r) \right] \cdot \left( \tau \sigma_x, \sigma_y \right) + \frac{V_z}{2} \lambda_z + \mathcal H_\perp \right\} c_{\tau,\mathbf k'+\mathbf g'},
    \end{aligned}
\end{equation}
To proceed, we note that for any function $f(\mathbf r)$ with the periodicity of the superlattice,
\begin{equation}
    \int d^2 \mathbf r \, e^{-i(\mathbf k+\mathbf g) \cdot \mathbf r} \, f(\mathbf r) \, e^{i(\mathbf k'+\mathbf g') \cdot \mathbf r} = A \delta_{\mathbf k\mathbf k'} f_{\mathbf g-\mathbf g'},
\end{equation}
where we used that $\mathbf k+\mathbf g$ with $\mathbf k$ in the SBZ and $\mathbf g$ a reciprocal vector of the superlattice, is a unique momentum decomposition. We further used $\int d^2 \mathbf r = \sum_{\mathbf R} \int_\text{cell} d^2\mathbf r$ with $\sum_{\mathbf R} e^{i \mathbf k \cdot \mathbf R} = N \delta_{\mathbf k\mathbf 0}$ where the sum runs over superlattice cells indexed by $\mathbf R$. We then obtain
\begin{equation}
    H = \sum_{\tau} \sum_{\mathbf k} \sum_{\mathbf g, \mathbf g'} c_{\tau,\mathbf k+\mathbf g}^\dag \left\{ \hbar v_F \left[ \left( \mathbf k+\mathbf g \right) \delta_{\mathbf g\mathbf g'} + \frac{\tau e}{\hbar} \, \mathbf A_{\mathbf g-\mathbf g'} \right] \cdot \left( \tau \sigma_x, \sigma_y \right) + \frac{V_z}{2} \lambda_z + \mathcal H_\perp \right\} c_{\tau,\mathbf k+\mathbf g'},
\end{equation}  
with Bloch Hamiltonian
\begin{equation}
    \mathcal H_{\mathbf g\mathbf g'}^{(\tau)}(\mathbf k) = \hbar v_F \left[ \left( \mathbf k+\mathbf g \right) \delta_{\mathbf g\mathbf g'} + \frac{\tau e}{\hbar} \, \mathbf A_{\mathbf g-\mathbf g'} \right] \cdot \left( \tau \sigma_x, \sigma_y \right) + \frac{V_z}{2} \lambda_z + \mathcal H_\perp,
\end{equation}
which is diagonalized numerically by taking a sufficient number of $\mathbf g$ vectors for convergence. All results shown in this work were obtained with a cutoff $|\mathbf g| < 8k_0$ where $k_0 = 4\pi / 3L$.
\begin{figure}
    \centering
    \includegraphics[width=.9\linewidth]{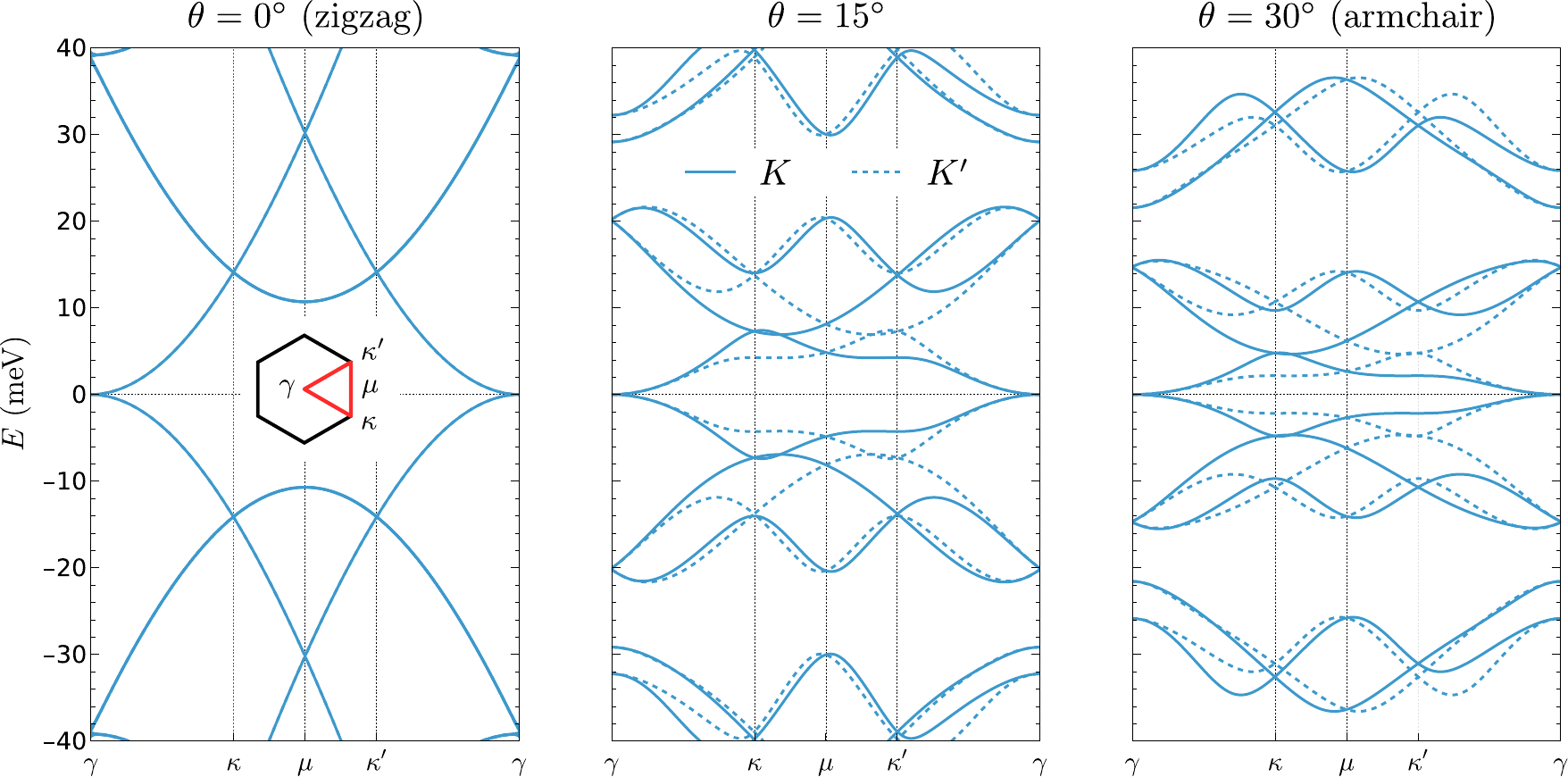}
    \caption{\textbf{Low-energy bands of periodically strained bilayer graphene without interlayer bias}. Bands for $V_z = 0$, $h_0 = 2$ nm, and $L = 40$ nm, such that $h_0 / L = 0.05$. We show the bands for different orientations $\theta$ along high-symmetry lines of the superlattice Brillouin zone as shown in the inset. Parameters for the nanohole are: $w/L = 0.05$ and $r_0/L = 0.3125$.}
    \label{fig:bands1}
\end{figure}
\begin{figure}
    \centering
    \includegraphics[width=.9\linewidth]{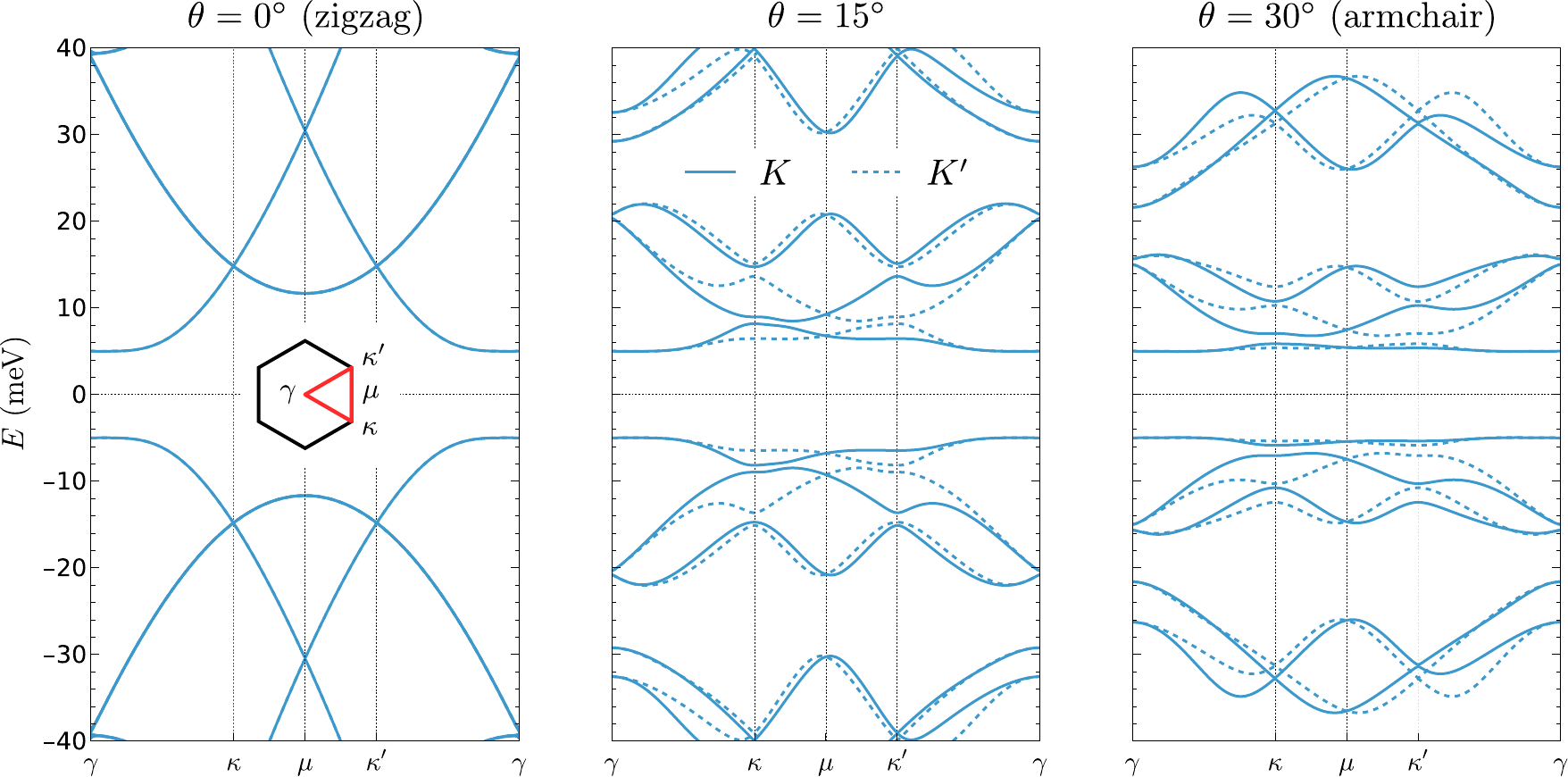}
    \caption{\textbf{Low-energy bands of periodically strained bilayer graphene with interlayer bias}. Bands for $V_z = 10$ meV, $h_0 = 2$ nm, and $L = 40$ nm, such that $h_0 / L = 0.05$. Shown along high-symmetry lines of the SBZ (inset). For $\theta = 30^\circ$, the valley Chern number sequence of the 6 bands near charge neutrality is $(0,-1,2,-2,1,0)$. Parameters: $w/L = 0.05$ and $r_0/L = 0.3125$.}
    \label{fig:bands2}
\end{figure}
This then yields minibands $E_{n,\mathbf k}^{(\tau)}$ with wave functions
\begin{equation}
    | \psi_{n,\mathbf k}^{(\tau)} \rangle = e^{i\mathbf k \cdot \hat{\mathbf r}} | u_{n,\mathbf k}^{(\tau)} \rangle, \qquad \langle \mathbf r | u_{n,\mathbf k}^{(\tau)} \rangle = u_{n,\mathbf k}^{(\tau)}(\mathbf r) = \frac{1}{\sqrt{A_\text{c}}} \sum_{\mathbf g} \Lambda_{n,\mathbf k + \mathbf g}^{(\tau)} e^{i\mathbf g\cdot \mathbf r},
\end{equation}
where $\Lambda_{n,\mathbf k + \mathbf g}^{(\tau)}$ are four-component vectors that contain the eigenstates of the Bloch Hamiltonian and which are obtained numerically. The energy bands for several orientations are shown in Figs.\ \ref{fig:bands1} and \ref{fig:bands2} in the absence of and with an interlayer bias, respectively. We see that the bands depend strongly on the orientation of the graphene lattice with respect to the corrugation. In fact, for zigzag alignment ($\theta = 0$ mod $\pi/3$) the strain superlattice barely modifies the energy spectrum.

Suppressing the valley index for clarity, the Bloch states satisfy the periodic gauge condition $| \psi_{n,\mathbf k+\mathbf g} \rangle = | \psi_{n,\mathbf k} \rangle$ with
\begin{equation}
    | u_{n,\mathbf k+\mathbf g} \rangle = e^{-i\mathbf g \cdot \hat{\mathbf r}} | u_{n,\mathbf k} \rangle,
\end{equation}
and normalization $\langle u_{n,\mathbf k} | u_{m,\mathbf k} \rangle_\text{cell} = \delta_{nm}$. For the calculation of quantum geometry, we need to evaluate overlaps between cell-periodic Bloch functions at neighboring $\mathbf k$ points:
\begin{align}
    \langle u_{n,\mathbf k} | u_{n',\mathbf k'} \rangle_\text{cell} & = \frac{1}{A_\mathrm{c}} \sum_{\mathbf g,\mathbf g'} \left( \Lambda_{n,\mathbf k+\mathbf g} \right)^\dag \Lambda_{n',\mathbf k'+\mathbf g'} \int_\text{cell} d^2\mathbf r \, e^{i(\mathbf g'-\mathbf g) \cdot \mathbf r} \\
    & = \sum_{\mathbf g} \left( \Lambda_{n,\mathbf k+\mathbf g} \right)^\dag \Lambda_{n',\mathbf k'+\mathbf g}.
\end{align}

\paragraph{Symmetries of the low-energy electronic theory}

Since the valleys are decoupled because the strain fields under consideration vary slowly with respect to the graphene lattice, we are mainly interested in the valley-preserving symmetries. We first consider the symmetry of the pseudogauge field and then discuss the magnetic symmetries in one valley of Bernal bilayer graphene under $C_{6v}$ homostrain with and without an interlayer bias.\\[4mm]
\noindent \textit{1.\ Symmetries of the pseudogauge field}\\
To determine how the pseudogauge field for a $C_{6v} = \left< \mathcal C_{6z}, \mathcal M_y \right>$ strain field transforms, we start from zigzag orientation ($\theta = 0$ mod $\pi/3$). In this case, the pseudogauge field obeys
\begin{align}
    \mathbf A_0(\mathcal C_{6z}\mathbf r) & = -\mathcal C_{6z} \mathbf A_0(\mathbf r), \label{eq:A01} \\
    \mathbf A_0(\mathcal C_{3z}\mathbf r) & = \mathcal C_{3z} \mathbf A_0(\mathbf r), \\
    \mathbf A_0(\mathcal M_x\mathbf r) & = -\mathcal M_x \mathbf A_0(\mathbf r), \label{eq:A03} \\
    \mathbf A_0(\mathcal M_y\mathbf r) & = \mathcal M_y \mathbf A_0(\mathbf r),
\end{align}
where $\mathcal M_x (x \mapsto -x)$ and $\mathcal M_y (y \mapsto -y)$ are mirror operations. The minus signs in Eqs.\ \eqref{eq:A01} and \eqref{eq:A03}can be understood from the fact that the corresponding operations also change the valley index, such that $\tau \mathbf A_0(\mathbf r)$ transforms as a vector under $C_{6v}$. Here we used Eq.\ \eqref{eq:Ar} and the fact that $\mathcal F_{\mathbf g}$ transforms as a scalar. For a general orientation,
\begin{equation}
    \mathbf A(\mathbf r) = R(3\theta) \mathbf A_0(\mathbf r),
\end{equation}
such that
\begin{align}
    \mathbf A(\mathcal C_{6z}\mathbf r) & = -\mathcal C_{6z} \mathbf A(\mathbf r), \\
    \mathbf A(\mathcal C_{3z}\mathbf r) & = \mathcal C_{3z} \mathbf A(\mathbf r), \\
    \mathbf A(\mathcal M_x\mathbf r) & = -\mathcal M_x(\theta) \mathbf A(\mathbf r), \\
    \mathbf A(\mathcal M_y\mathbf r) & = \mathcal M_y(\theta) \mathbf A(\mathbf r),
\end{align}
with $\mathcal M_{x,y}(\theta) = R(3\theta) \mathcal M_{x,y} R(-3\theta)$.

\noindent \textit{2.\ Magnetic point group}\\
Using the symmetry properties of the pseudogauge field, we determine the valley-preserving symmetries. We first discuss the case without an interlayer bias ($V_z = 0$) and we only consider zigzag and armchair alignment for simplicity. For the zigzag case ($\theta = 0$ mod $\pi/3$), the valley-preserving symmetries form the dichromatic group $D_{3d}(D_3)$ also denoted as $\bar 3'm'$ and which is generated by, for example, $\mathcal C_{3z} \mathcal P \mathcal T$ and $\mathcal M_x \mathcal T$ where $\mathcal P \, (\mathbf r \mapsto - \mathbf r, \lambda_x \sigma_x)$ is spatial inversion and $\mathcal T$ is spinless time-reversal symmetry. The same holds for armchair alignment ($\theta = \pi/6$ mod $\pi/3$) but now with generators $\mathcal C_{3z} \mathcal P \mathcal T$ and $\mathcal M_y \mathcal T$. In the presence of an interlayer bias, the group is reduced to $C_{3v}(C_3)$ or $3m'$. From the Dimmock-Wheeler sum rule, it follows that none of these yield nontrivial irreducible co-representations.

For a general orientation of the graphene with respect to the strain superlattice, we have the following symmetries that act on field operators as:

\begin{align}
    \mathcal C_{3z} \psi_\tau(\mathbf r) \mathcal C_{3z}^{-1} & = e^{-i \tau \pi \lambda_z / 3} e^{-i \tau \pi \sigma_z / 3} \psi_\tau(\mathcal C_{3z}^{-1} \mathbf r), \\
    \mathcal T \psi_\tau(\mathbf r) \mathcal T^{-1} & = \psi_{-\tau}(\mathbf r), \\
    \mathcal P \psi_\tau(\mathbf r) \mathcal P^{-1} & = \lambda_x \sigma_x \psi_{-\tau}(-\mathbf r),
\end{align}
with $\mathcal T^{-1}i\mathcal T=-i$ and $[H,\mathcal P] = 0$ only for $V_z = 0$. Note that $\mathcal T$ and $\mathcal P$ both interchange the valleys. In a single valley, we have
\begin{equation}
    \left( \mathcal P \mathcal T \right) \psi_\tau(\mathbf r) \left( \mathcal P \mathcal T \right)^{-1} = \lambda_x \sigma_x \psi_\tau(-\mathbf r).
\end{equation}
For $\theta = 0$ mod $\pi/3$ we also have
\begin{align}
    \mathcal M_x \psi_\tau(x,y) \mathcal M_x^{-1} & = \psi_{-\tau}(-x,y), \\
    \mathcal C_{2x} \psi_\tau(x,y) \mathcal C_{2x}^{-1} & = \lambda_x \sigma_x \psi_\tau(x,-y),
\end{align}
where $\mathcal C_{2x} = \mathcal M_y \mathcal M_z$ is only conserved for $V_z = 0$. Hence, for zigzag alignment, we find that $\mathcal M_x \mathcal T$ makes the energy bands (and Berry curvature) symmetric with respect to $k_y$. On the other hand, for $\theta = \pi/6$ mod $\pi/3$:
\begin{align}
    \mathcal M_y \psi_\tau(x,y) \mathcal M_y^{-1} & = \lambda_z \sigma_z \psi_{-\tau}(x,-y), \\
    \mathcal C_{2y} \psi_\tau(x,y) \mathcal C_{2y}^{-1} & = \lambda_y \sigma_y \psi_\tau(-x,y).
\end{align}
where $\mathcal C_{2y} = \mathcal M_x \mathcal M_z$ is only conserved for $V_z = 0$. For armchair alignment, $\mathcal M_y \mathcal T$ makes the energy and (Berry curvature) symmetric with respect to $k_x$.

\noindent \textit{3. Chiral symmetry}\\
The Hamiltonian also has model symmetries, which do not follow from symmetry alone but rather from the choice of model that omits certain allowed terms. We see that the energy bands are always symmetric around zero energy. This is enforced by a unitary chiral symmetry $[H,\mathcal C] = 0$ with
\begin{equation}
    \mathcal C \psi_\tau(\mathbf r) \mathcal C^{-1} = -i \lambda_x \sigma_y \psi_\tau^*(-\mathbf r),
\end{equation}
with $\mathcal C i \mathcal C^{-1} = i$ and where the phase factor is chosen for convenience. Here we defined $\psi_\tau^* = ( \psi_{\tau A1}^\dag, \psi_{\tau B1}^\dag, \psi_{\tau A2}^\dag, \psi_{\tau B2}^\dag )$. Chiral symmetry is conserved because $\mathbf A(\mathbf r) = \mathbf A(-\mathbf r)$ yielding an odd PMF profile, which is always the case regardless of the orientation, see Fig.\ \ref{fig:pmf}. A finite deformation potential $(u_{xx} + u_{yy}) \lambda_0 \sigma_0$ which is allowed by symmetry would break $\mathcal C$. We demonstrate that $\mathcal C$ commutes with the Hamiltonian by making use of
\begin{align}
    \psi_\tau^\top(\mathbf r) \lambda_z \psi_\tau^*(\mathbf r) & = -\psi_\tau^\dag(\mathbf r) \lambda_z \psi_\tau(\mathbf r), \\
    \psi_\tau^\top(\mathbf r) (\lambda_x \sigma_x + \lambda_y \sigma_y) \psi_\tau^*(\mathbf r) & = -\psi_\tau^\dag(\mathbf r) (\lambda_x \sigma_x + \lambda_y \sigma_y) \psi_\tau(\mathbf r), \\
    \psi_\tau^\top(\mathbf r) \sigma_x \psi_\tau^*(\mathbf r) & = -\psi_\tau^\dag(\mathbf r) \sigma_x \psi_\tau(\mathbf r), \\
    \psi_\tau^\top(\mathbf r) \sigma_y \psi_\tau^*(\mathbf r) & = + \psi_\tau^\dag(\mathbf r) \sigma_y \psi_\tau(\mathbf r),
\end{align}
and
\begin{equation}
    \{ \partial_i \psi_{\tau Al}(\mathbf r), \psi_{\tau Bl}(\mathbf r) \} = \lim_{\delta\rightarrow0} \{ \frac{\psi_{\tau Al}(\mathbf r+\delta \mathbf e_i) - \psi_{\tau Al}(\mathbf r)}{\delta}, \psi_{\tau Bl}(\mathbf r) \} = 0. 
\end{equation}
We then find that $\mathcal C H \mathcal C^{-1}$ gives
\begin{equation}
    \begin{aligned}
        & \int d^2 \mathbf r \, \psi_\tau^\top(-\mathbf r) \left[ -i \nabla + \frac{\tau e}{\hbar} \mathbf A(\mathbf r) \right] \cdot \left( -\tau \sigma_x, \sigma_y \right) \psi_\tau^*(-\mathbf r) \\
        & \hspace{3cm} = \int d^2 \mathbf r \, \psi_\tau^\dag(\mathbf r) \left[ -i \nabla + \frac{\tau e}{\hbar} \mathbf A(-\mathbf r) \right] \cdot \left( \tau \sigma_x, \sigma_y \right) \psi_\tau(\mathbf r).
    \end{aligned}
\end{equation}
In turn, this acts on the Bloch Hamiltonian as an anticommuting antiunitary operation $-i\lambda_x \sigma_y \mathcal K$ with $\mathcal K$ complex conjugation:
\begin{equation}
    \lambda_x \sigma_y [ \mathcal H^{(\tau)}(\mathbf k) ]^* \lambda_x \sigma_y = -\mathcal H^{(\tau)}(\mathbf k),
\end{equation}
which leaves $\mathbf k$ invariant and relies on the reality of $\mathbf A_{\mathbf g}$. Hence, the bands can be labeled as $E_{n,\mathbf k}^{(\tau)} = -E_{-n,\mathbf k}^{(\tau)}$ with $n = \pm 1, \pm 2, \ldots$. Moreover, it enforces opposite Berry curvatures (and thus opposite valley Chern numbers) for the valence and conduction bands: $\Omega_{n,\mathbf k}^{(\tau)} = -\Omega_{-n,\mathbf k}^{(\tau)}$ which are well defined only for isolated bands. The latter requires finite $V_z$ in practice.

While the chiral symmetry $\mathcal C$ is conserved even for finite $V_z$, there is an additional antiunitary chiral symmetry for $V_z = 0$:
\begin{equation}
    \tilde{\mathcal C} \psi_\tau(\mathbf r) \tilde{\mathcal C}^{-1} = \sigma_z \psi_\tau^*(\mathbf r),
\end{equation}
with $\tilde{\mathcal C} i \tilde{\mathcal C}^{-1} = -i$ and which holds irrespective of the pseudogauge field. Finally, note that combining these symmetries gives $\mathcal C \tilde{\mathcal C} = \mathcal P \mathcal T$. On the Bloch Hamiltonian, $\tilde{\mathcal C}$ acts as an anticommuting unitary and enforces equal Berry curvatures between valence and conduction band. Therefore, the Berry curvature has to vanish for $V_z = 0$ as expected from $\mathcal P \mathcal T$.\\

\noindent{\normalsize \textit{S3.2 Quantum geometry of flat bands}}\\
We consider the case of a single isolated band, but the method described below is easily generalized to the multiband case. To calculate the Berry curvature and quantum metric, we consider a square plaquette of area $\delta^2$ centered at $\mathbf k$ with corners: $\mathbf k_1 = \mathbf k + \tfrac{\delta}{2}(-1, -1)$, $\mathbf k_2 = \mathbf k + \tfrac{\delta}{2}(-1, 1)$, $\mathbf k_3 = \mathbf k + \tfrac{\delta}{2}(1, 1)$, and $\mathbf k_4 = \mathbf k + \tfrac{\delta}{2}(1, -1)$. We then consider the gauge-invariant product
 \begin{equation}
    \langle u_{\mathbf k_1} | u_{\mathbf k_2} \rangle \langle u_{\mathbf k_2} | u_{\mathbf k_3} \rangle \langle u_{\mathbf k_3} | u_{\mathbf k_4} \rangle \langle u_{\mathbf k_4} | u_{\mathbf k_1} \rangle = \prod_{m=1}^4 \langle u_{\mathbf k_m} | u_{\mathbf k_{m+1}} \rangle,
\end{equation}
where $\mathbf k_5 = \mathbf k_1$. Here we suppressed the band index. 

Using the shorthand $\partial_i = \partial/\partial k_i$, we expand the different factors up to second order in $\delta$ as,
\begin{align}
    & \langle u_{\mathbf k_m} | u_{\mathbf k_{m+1}} \rangle \\
    & = \left[ \langle u_{\mathbf k} | + \delta k_{m,i} \partial_i \langle u_{\mathbf k} | + \frac{1}{2} \delta k_{m,i} \delta k_{m,j} \partial_i \partial_j \langle u_{\mathbf k} | + \mathcal O(\delta^3) \right] \\
    & \times \left[ | u_{\mathbf k} \rangle + \delta k_{m+1,i} \partial_i | u_{\mathbf k} \rangle + \frac{1}{2} \delta k_{m+1,i} \delta k_{m+1,j} \partial_i \partial_j | u_{\mathbf k} \rangle + \mathcal O(\delta^3) \right] \\
    & = 1 + \left( \delta k_{m+1,i} - \delta k_{m,i} \right) \langle u_{\mathbf k} | \partial_i u_{\mathbf k} \rangle + \delta k_{m,i} \delta k_{m,j} \langle \partial_i u_{\mathbf k} | \partial_j u_{\mathbf k} \rangle + \frac{1}{2} \delta k_{m,i} \delta k_{m,j} \langle \partial_i \partial_j u_{\mathbf k} | u_{\mathbf k} \rangle \\
    & \quad + \frac{1}{2} \delta k_{m+1,i} \delta k_{m+1,j} \langle u_{\mathbf k} | \partial_i \partial_j u_{\mathbf k} \rangle + \mathcal O(\delta^3),
\end{align}
where we used $\langle \partial_i u_{\mathbf k} | u_{\mathbf k} \rangle = -\langle u_{\mathbf k} | \partial_i u_{\mathbf k} \rangle$. We obtain
\begin{align}
    \prod_{m=1}^4 \langle u_{\mathbf k_m} | u_{\mathbf k_{m+1}} \rangle & = 1 + \sum_{m=1}^4 \left( \delta k_{m+1,i} - \delta k_{m,i} \right) \langle u_{\mathbf k} | \partial_i u_{\mathbf k} \rangle +
   \sum_{m=1}^4 \delta k_{m,i} \delta k_{m+1,j} \langle \partial_i u_{\mathbf k} | \partial_j u_{\mathbf k} \rangle \\
    & + \frac{1}{2} \sum_{m=1}^4 \delta k_{m,i} \delta k_{m,j} \left( \langle \partial_i \partial_j u_{\mathbf k} | u_{\mathbf k} \rangle + \langle u_{\mathbf k} | \partial_i \partial_j u_{\mathbf k} \rangle \right) \\
    & + \left( \delta k_{2i} - \delta k_{1i} \right) \langle u_{\mathbf k} | \partial_i u_{\mathbf k} \rangle \left( \delta k_{3j} - \delta k_{2j} + \delta k_{4j} - \delta k_{3j} + \delta k_{1j} - \delta k_{4j} \right) \langle u_{\mathbf k} | \partial_j u_{\mathbf k} \rangle \\
    & + \left( \delta k_{3i} - \delta k_{2i} \right) \langle u_{\mathbf k} | \partial_i u_{\mathbf k} \rangle \left( \delta k_{4j} - \delta k_{3j} + \delta k_{1j} - \delta k_{4j} \right) \langle u_{\mathbf k} | \partial_j u_{\mathbf k} \rangle \\
    & + \left( \delta k_{4i} - \delta k_{3i} \right) \langle u_{\mathbf k} | \partial_i u_{\mathbf k} \rangle \left( \delta k_{1j} - \delta k_{4j} \right) \langle u_{\mathbf k} | \partial_j u_{\mathbf k} \rangle + \mathcal O(\delta^4) \\
    & = 1 + \sum_{m=1}^4 \delta k_{m,i} \delta k_{m+1,j} \langle \partial_i u_{\mathbf k} | \partial_j u_{\mathbf k} \rangle \\
    & - \frac{1}{2} \sum_{m=1}^4 \delta k_{m,i} \delta k_{m,j} \left( \langle \partial_i u_{\mathbf k} | \partial_j u_{\mathbf k} \rangle + \langle \partial_j u_{\mathbf k} | \partial_i u_{\mathbf k} \rangle \right) \\
    & + \left( \delta k_{2i} - \delta k_{1i} \right) \langle u_{\mathbf k} | \partial_i u_{\mathbf k} \rangle \left( \delta k_{1j} - \delta k_{2j} \right) \langle u_{\mathbf k} | \partial_j u_{\mathbf k} \rangle \\
    & + \left( \delta k_{3i} - \delta k_{2i} \right) \langle u_{\mathbf k} | \partial_i u_{\mathbf k} \rangle \left( \delta k_{1j} - \delta k_{3j} \right) \langle u_{\mathbf k} | \partial_j u_{\mathbf k} \rangle \\
    & + \left( \delta k_{4i} - \delta k_{3i} \right) \langle u_{\mathbf k} | \partial_i u_{\mathbf k} \rangle \left( \delta k_{1j} - \delta k_{4j} \right) \langle u_{\mathbf k} | \partial_j u_{\mathbf k} \rangle + \mathcal O(\delta^4),
\end{align}
where we used
\begin{equation}
    \langle \partial_i \partial_j u_{\mathbf k} | u_{\mathbf k} \rangle + \langle u_{\mathbf k} | \partial_i \partial_j u_{\mathbf k} \rangle = - \left( \langle \partial_i u_{\mathbf k} |  \partial_j u_{\mathbf k} \rangle + \langle \partial_j u_{\mathbf k} | \partial_i u_{\mathbf k} \rangle \right).
\end{equation}
Noting that
\begin{equation}
    \sum_{m=1}^4 \delta k_{m,i} \delta k_{m,j} = \delta^2 \begin{pmatrix} 1 & 0 \\ 0 & 1 \end{pmatrix}, \qquad 
    \sum_{m=1}^4 \delta k_{m,i} \delta k_{m+1,j} = \delta^2 \begin{pmatrix} 0 & -1 \\ 1 & 0 \end{pmatrix},
\end{equation}
and 
\begin{align}
    & \delta \mathbf k_2 - \delta \mathbf k_1 = \delta \mathbf k_3 - \delta \mathbf k_4 = \delta (0,1), \\
    & \delta \mathbf k_4 - \delta \mathbf k_1 = \delta \mathbf k_3 - \delta \mathbf k_2 = \delta (1,0), \\
    & \delta \mathbf k_3 - \delta \mathbf k_1 = \delta (1,1),
\end{align}
we find
\begin{align}
    \prod_{m=1}^4 \langle u_{\mathbf k_m} | u_{\mathbf k_{m+1}} \rangle & = 1 - \delta^2 \left( \langle \partial_{k_x} u_{\mathbf k} | \partial_{k_y} u_{\mathbf k} \rangle  - \langle \partial_{k_y} u_{\mathbf k} | \partial_{k_x} u_{\mathbf k} \rangle \right) \\
    & - \delta^2 \left( \langle \partial_{k_x} u_{\mathbf k} | \partial_{k_x} u_{\mathbf k} \rangle + \langle \partial_{k_y} u_{\mathbf k} | \partial_{k_y} u_{\mathbf k} \rangle + \langle u_{\mathbf k} | \partial_{k_x} u_{\mathbf k} \rangle^2 + \langle u_{\mathbf k} | \partial_{k_y} u_{\mathbf k} \rangle^2 \right) + \mathcal O(\delta^4) \\
    & = 1 - \delta^2 \left( \text{tr} \, g_{\mathbf k} - i \Omega_{\mathbf k} \right) + \mathcal O(\delta^4) \approx \exp \left[ -\delta^2 \left( \text{tr} \, g_{\mathbf k} - i \Omega_{\mathbf k} \right) \right].
\end{align}

Here
\begin{align}
    g_{\mathbf k}^{\mu\nu} & = \text{Re} \left( \langle \partial^\mu u_{\mathbf k} | \partial^\nu u_{\mathbf k} \rangle \right) + \langle u_{\mathbf k} | \partial^\mu u_{\mathbf k} \rangle \langle u_{\mathbf k} | \partial^\nu u_{\mathbf k} \rangle, \\
    \Omega_{\mathbf k} & = i \left( \langle \partial_{k_x} u_{\mathbf k} | \partial_{k_y} u_{\mathbf k} \rangle  - \langle \partial_{k_y} u_{\mathbf k} | \partial_{k_x} u_{\mathbf k} \rangle \right) = -2 \, \text{Im} \left( \langle \partial_{k_x} u_{\mathbf k} | \partial_{k_y} u_{\mathbf k} \rangle \right),
\end{align}
is the Fubini-Study quantum metric with $\text{tr} \, g_{\mathbf k} = g_{\mathbf k}^{xx} + g_{\mathbf k}^{yy}$ and the Berry curvature, respectively, with $\mu, \nu = x,y$. They form the real and imaginary components of the quantum geometric tensor:
\begin{equation}
    Q_{\mathbf k}^{\mu\nu} = \langle \partial^\mu u_{\mathbf k} | \left( 1 - | u_{\mathbf k} \rangle \langle u_{\mathbf k} | \right) | \partial^\nu u_{\mathbf k} \rangle = g_{\mathbf k}^{\mu\nu} - \frac{i}{2} \, \epsilon^{\mu\nu} \Omega_{\mathbf k},
\end{equation}
which is well-defined for an isolated band and where the projector ensures invariance under gauge transformations $| u_{\mathbf k} \rangle \mapsto e^{i\varphi_{\mathbf k}} | u_{\mathbf k} \rangle$. 
We thus have
\begin{equation}
    \Omega_{\mathbf k} = \lim_{\delta \rightarrow 0} \left( \frac{1}{\delta^2} \arg \prod_{m=1}^4 \langle u_{\mathbf k_m} | u_{\mathbf k_{m+1}} \rangle \right), \qquad \text{tr} \, g_{\mathbf k} = \lim_{\delta \rightarrow 0} \left[ \frac{1}{\delta^2} \, \text{Re} \left( 1 - \prod_{m=1}^4 \langle u_{\mathbf k_m} | u_{\mathbf k_{m+1}} \rangle \right) \right].
\end{equation}

For the multiband case, we simply let
\begin{equation}
    | u_{n,\mathbf k} \rangle \rightarrow U_{\mathcal M, \mathbf k} = \begin{pmatrix} | u_{1,\mathbf k} \rangle & | u_{2,\mathbf k} \rangle & \cdots & | u_{M,\mathbf k} \rangle \end{pmatrix},
\end{equation}
where $\mathcal M$ represents a manifold of $M$ bands $\left\{ E_{1,\mathbf k}, E_{2,\mathbf k}, \ldots, E_{M,\mathbf k} \right\}$ that are isolated from other bands. Finally, we define the Chern number:
\begin{equation}
    \mathcal C = \frac{1}{2\pi} \int_\text{SBZ} d^2 \mathbf k \, \Omega_{\mathbf k} \rightarrow \frac{1}{2\pi N} \sum_{\mathbf k \in \text{SBZ}} \Omega_{\mathbf k} A_\text{SBZ},
\end{equation}
where $N$ is the number of superlattice unit cells (which corresponds to the number of $\mathbf k$ points in our calculation) and $A_\text{SBZ} = (2\pi)^2 / A_\text{c}$ is the area of the superlattice Brillouin zone. A finite Chern number is allowed in a single valley as time-reversal symmetry is effectively broken. Note that the Berry curvature and quantum metric trace transform as
\begin{equation}
    \Omega(\mathbf k) = \pm \det(\mathcal S) \Omega(\pm \mathcal S \mathbf k), \qquad \text{tr} \, g(\mathbf k) = \text{tr} \, g(\pm \mathcal S \mathbf k),
\end{equation}
where the sign $+$ corresponds to a spatial symmetry $\mathcal S$ (constrained to the plane) and the sign $-$ corresponds to $\mathcal S \mathcal T$ when it is composed with time-reversal symmetry.
\begin{figure}
    \centering
    \includegraphics[width=.98\linewidth]{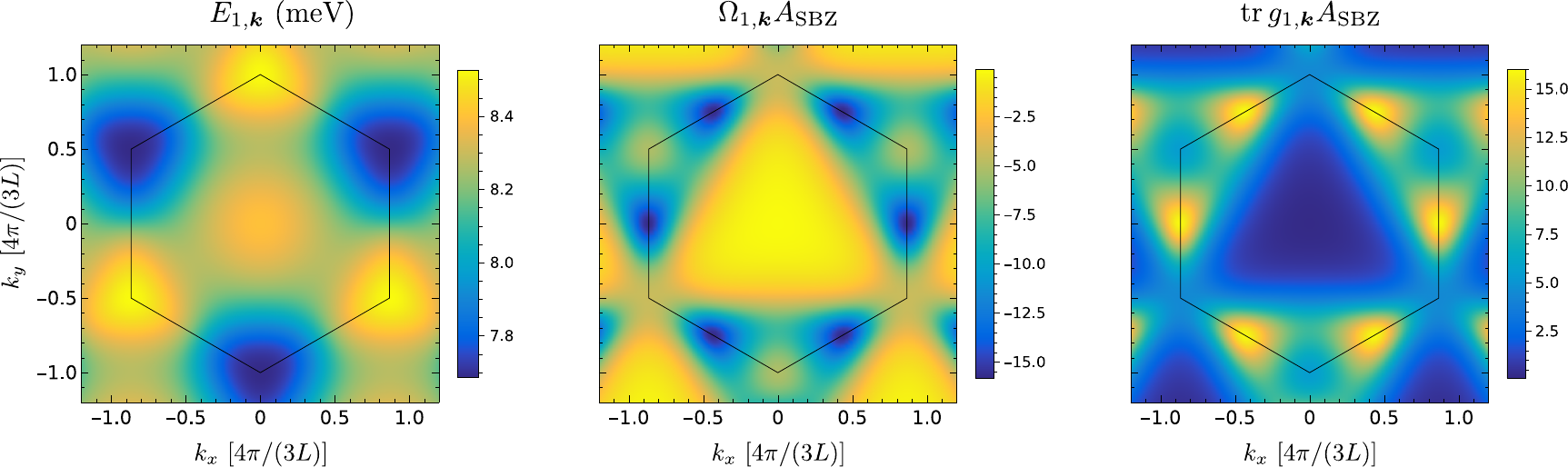}
    \caption{\textbf{Topological flatband in periodically strained bilayer graphene.} Energy (left panel), Berry curvature (middle panel), and quantum metric trace (right panel) of the first conduction band in $K$ with valley Chern number $-2$. Here $A_\text{SBZ} = (2\pi)^2 / A_\text{c}$ is the SBZ area and the parameters are: $\theta = 30^\circ$, $V_z = 16.8$ meV, $h_0 / L = 0.05$ nm, $w/L = 0.05$, and $r_0/L = 0.3125$.}
    \label{fig:bands3}
\end{figure}

\subsubsection{Current-dependent suppression of magnetic hysteresis}

\begin{figure}
    \centering
    \includegraphics[width=0.7\linewidth]{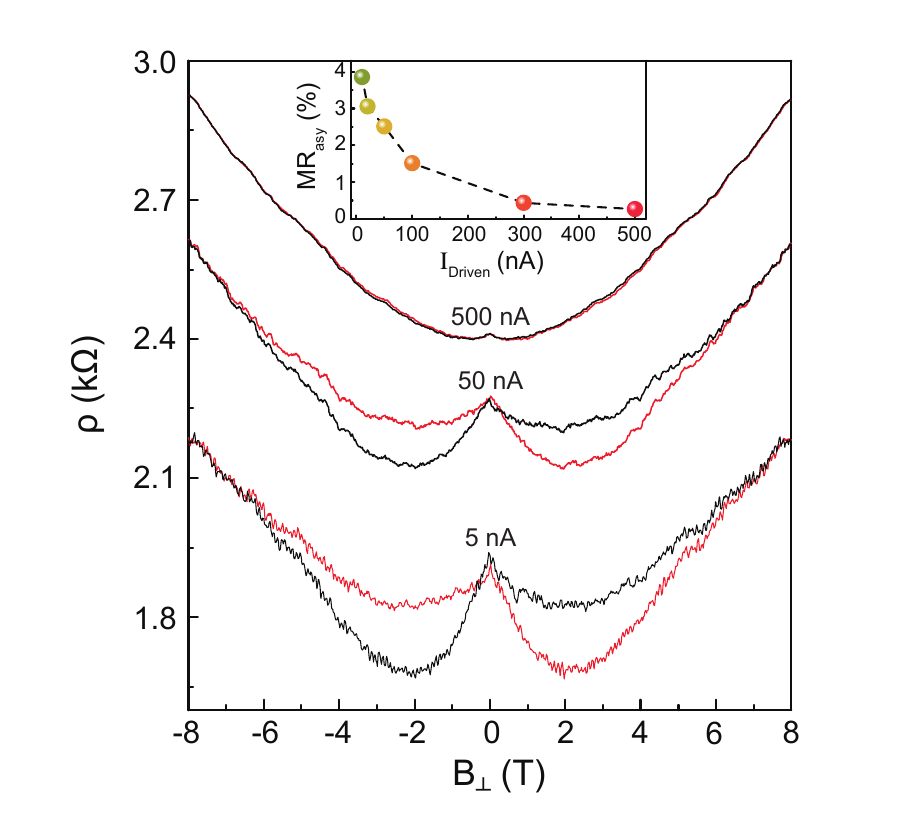}
    \caption{\textbf{Current dependence of magnetic hysteresis.} Magnetic hysteresis as a function of driving current at $T = 0.2$~K and a field-sweep rate of $0.2$~T/min. The inset shows the evolution of the hysteresis window with current, quantified by the magnetoresistivity asymmetry. The window decreases monotonically with increasing current, in contrast to the non-monotonic temperature dependence shown in Fig.~2e. }
    \label{figS4}
\end{figure}

We observe that the magnetic hysteresis is progressively suppressed with increasing drive current, as shown in Fig.~\ref{figS4}. The hysteresis window, quantified by the magnetoresistivity asymmetry $MR_{\mathrm{asy}}$, decreases monotonically with current, in clear contrast to the non-monotonic temperature dependence observed in Fig.~2e. This monotonic suppression cannot be attributed to simple thermal smearing, but instead reflects how electrical current perturbs the frustrated state, possibly through current-induced delocalization or non-equilibrium excitations.

\subsubsection{Monte Carlo simulations of the ruby-lattice spin-ice model}

To investigate the frustrated magnetism on the effective ruby superlattice, we perform the classical Monte Carlo simulation for the Hamiltonian:
\begin{align}
\begin{split}
\mathcal{H} &= J_1\sum_{\langle ij\rangle}\sigma_i\sigma_j + J_2\sum_{\langle\langle ij\rangle\rangle}\sigma_i\sigma_j-\alpha\mu_0H\sum_i\sigma_i,
\end{split}
\end{align}
where $J_1 > 0$ denotes the antiferromagnetic intra-triangle bonds, and $J_2 > 0$ denotes the antiferromagnetic inter-triangle bonds. The Ising variable $\sigma_i = +1$ ($-1$) represents a magnetic moment pointing out of (into) the triangular plaquette. In the Zeeman term, we introduce the factor $\alpha$ as the projection factor of the local moment to the external-field orientation, and the factor $\mu_0$ accounts for the magnetic moment. In the simulation, the energy scale of the Zeeman term is in terms of the reference scale $J_1$. The Monte Carlo data presented in the main text and the supplementary material are obtained from $L = 20$ systems (2400 spins) using the single-spin-flip Metropolis algorithm. Each curve in the simulation results is averaged over 100 independent runs. To capture the non-equilibrium hysteresis behavior under field sweeps caused by spin freezing, we apply thermalization steps only at the beginning of each simulation (at the largest field in magnitude), but not at each subsequent field-increasing or field-decreasing step.

In our simulation, the main observables are the magnetic charges, which are defined on triangular and rectangular plaquettes as:
\begin{align}
Q_{\triangle} =\sum_{i\in\triangle}\sigma_i, \quad Q_{\square} = \sum_{i\in\square}\sigma_i.
\end{align}
In our setup for the spin axes, $\sigma_i = +1$ ($-1$) corresponds to a magnetic moment with a positive (negative) $z$-component, so the magnetic charge on a plaquette represents the net number of spins aligned with the external field. The magnetic charge order parameter defined on the triangular plaquettes is given by
\begin{align}
Q_t = \frac{1}{2L^2}\left|\sum_{p}Q_{\triangle,p}\right|,
\end{align}
where $Q_{\triangle,p}$ denotes the magnetic charge on a single plaquette $p$. The denominator $2L^2$ is the total number of triangular plaquettes on an $L \times L$ ruby lattice, so that $Q_t = 3$ corresponds to the maximum value, indicating a charge-saturated state. For low and moderate field strengths, most $Q_{\triangle}$ values in the system are $\pm 1$, meaning that $Q_t$ quantifies the net triangular magnetic charge of the system.

Similar to kagome spin-ice systems, $Q_t = 0$ ($Q_t = +1$) indicates that the positive and negative charges on the triangular plaquettes are perfectly balanced (imbalanced). For the rectangular charge, because it can be charge neutral with $Q_\square = 0$, its role is analogous to the spin-ice rule on a tetrahedron in pyrochlore systems. Therefore, we take the absolute value of $Q_{\square}$ before summing over all rectangular plaquettes, yielding a quantity analogous to the magnetic monopole density in spin ice:
\begin{align}
D_r = \frac{1}{6L^2}\sum_p \left| Q_{\square,p} \right|.
\end{align}
Here, the denominator includes $3L^2$ from the number of rectangles on the lattice and an additional factor of $2$ to convert the charge magnitude into a density. For example, in a charge-saturated state where $Q_{\square} = +4$ on every plaquette, the density becomes $D_r = +2$, representing a $100\%$ occupation of double rectangular charges.

\begin{figure}
	\centering
	\includegraphics[width=1.0\textwidth]{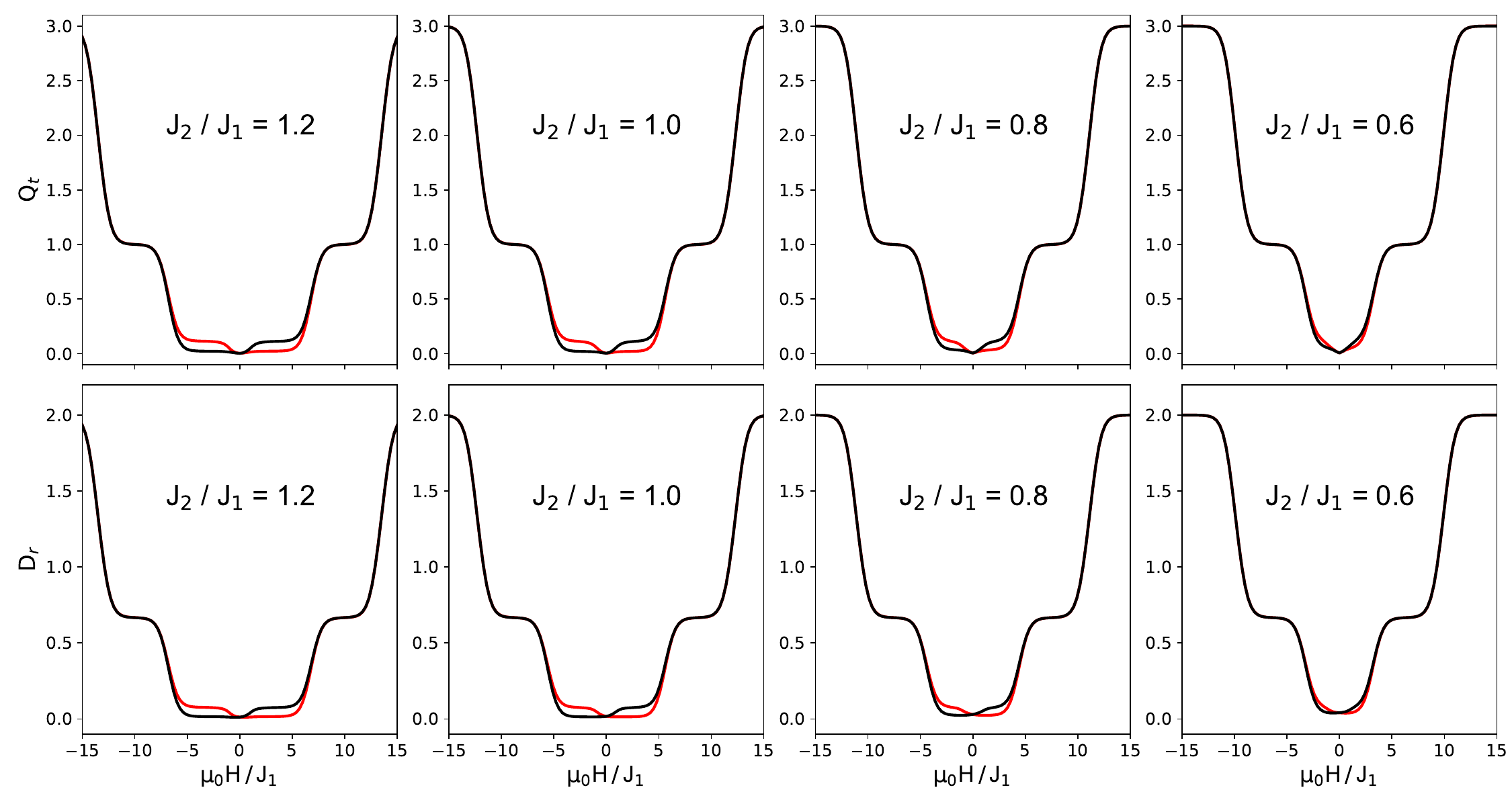}
	\caption{\textbf{Monte Carlo simulation of the spin-ice model on the ruby lattice.}
	The upper row shows the $J_2$ dependence of the triangular charge order parameter, while the lower row shows the $J_2$ dependence of the rectangular charge density. All the simulations are done under the temperature $k_B T/J_1 = 0.3$, which is low enough to demonstrate the hysteresis effect in the middle.}
	\label{fig:ruby1_T03}
\end{figure}

\begin{figure}
	\centering
	\includegraphics[width=1\textwidth]{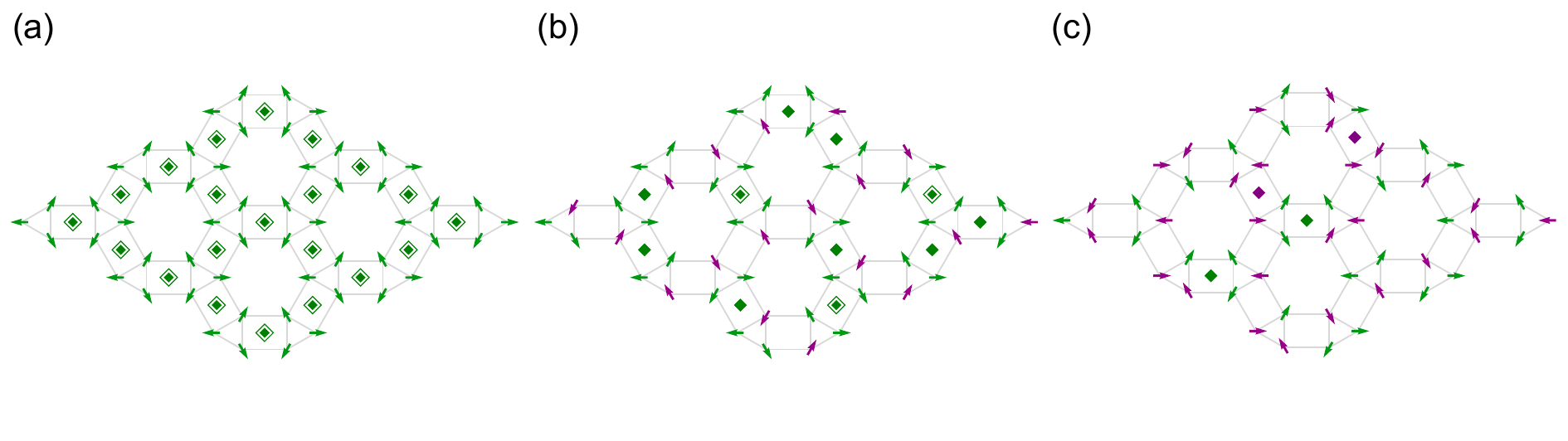}
	\caption{\textbf{Typical magnetic-charge configurations in different field regions.}
		(\textbf{A}) Charge-saturated (CS) state. This is the ground state for $H>H_{c,1}$, where the ice rule is broken on every triangular plaquette. The magnetic charges are uniform across the system, with $Q_{\triangle} = +3$ and $Q_{\square}= +4$.  (\textbf{B}) Triangular-charge-ordered (TCO) phase. This happens at $H_{c,2}<H<H_{c,1}$, where the Zeeman field induces perfect imbalance for triangular charge ($Q_{\triangle}= +1$) and finite density of the rectangular charge ($Q_{\square}=+2$ and $Q_{\square} = +4$). Note that the spins remain disordered. (\textbf{C}) Triangular-charge-disordered (TCD) phase. This is the typical configuration at $H = 0$. The absence of Zeeman field makes the positive and negative magnetic charges populate equally, such as $Q_{\triangle}=\pm 1$ for triangular plaquettes and $Q_{\square}= \pm 2$ for rectangular plaquettes.}
	\label{fig:ruby2_config}
\end{figure}

In Figure \ref{fig:ruby1_T03}, we present the simulation results for $Q_t$ and $D_r$ under various values of $J_2$. The saturated values $Q_t = +3$ and $D_r = +2$ at very strong magnetic fields indicate the all-spin-out state for the triangular plaquettes, called the charge-saturated (CS) state, as shown in Figure \ref{fig:ruby2_config}A. 
At moderate field strengths, there exists a plateau for $Q_t = +1$ and $D_r = +2/3$, where the system has mostly the 2-out-1-in spin configuration on the triangles. This is the triangular-charge-ordered (TCO) phase. Note that the spins and rectangular charges remain disordered (Figure \ref{fig:ruby2_config}B). Around $H = 0$, even the triangular charges become disordered, giving the triangular-charge-disordered (TCD) phase (Figure \ref{fig:ruby2_config}C).

\begin{figure}
	\centering
	\includegraphics[width=1\textwidth]{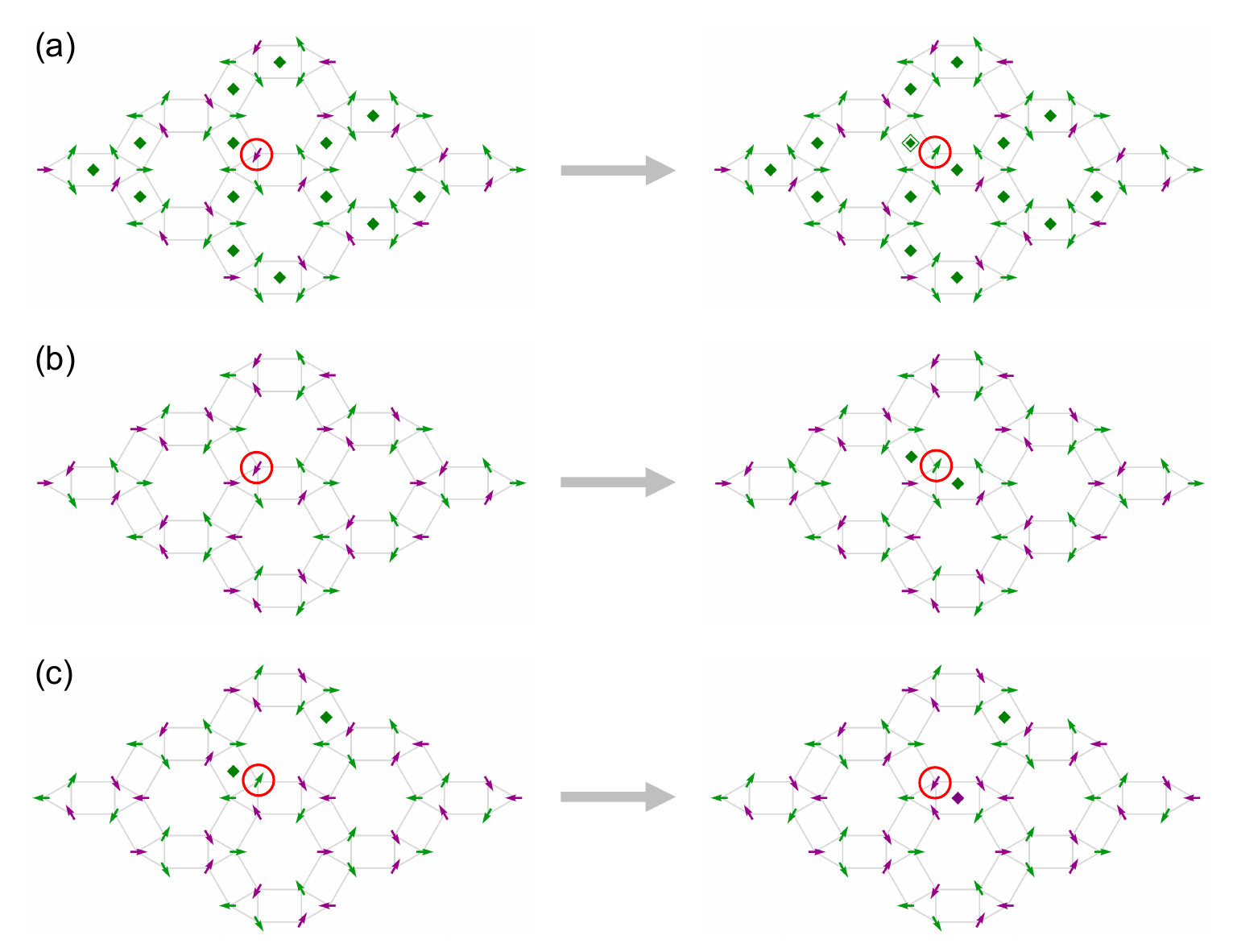}
	\caption{\textbf{Local magnetic charge dynamics through single spin flip.}
		(\textbf{A}) Rectangular-charge pair creation on the triangular-charge-ordered phase. This process breaks the ice-rule on a triangle and sets the critical field $H_{c,1}$. (\textbf{B}) Rectangular-charge pair creation on the triangular-charge-disordered phase. This process does not break the triangular ice-rule but still creates two rectangular charges. This sets the critical field $H_{c,2}$. (\textbf{C}) Rectangular-charge migration. This process moves a single charge to the neighboring plaquette and flips its sign. Because of the sign flip, the charge migration is impeded when $H\neq 0$. }
	\label{fig:ruby3_process}
\end{figure}

Between the TCO phase and the CS state, the critical field strength for this transition can be estimated from the energy cost of a single-spin flip associated with the pair creation of rectangular charges (Figure \ref{fig:ruby3_process}A):
\begin{align}
\Delta E = 4J_1 + 4J_2 - 2\alpha \mu_0 H \to 0, \quad \mu_0 H_{c,1} = \frac{2}{\alpha}(J_1 + J_2).
\end{align}
For instance, when $J_1 = 1.0$ and $J_2 = 0.8$, the critical field is $\mu_0 H_{c,1} = 10.8$. 

\begin{figure}
	\centering
	\includegraphics[width=0.6\textwidth]{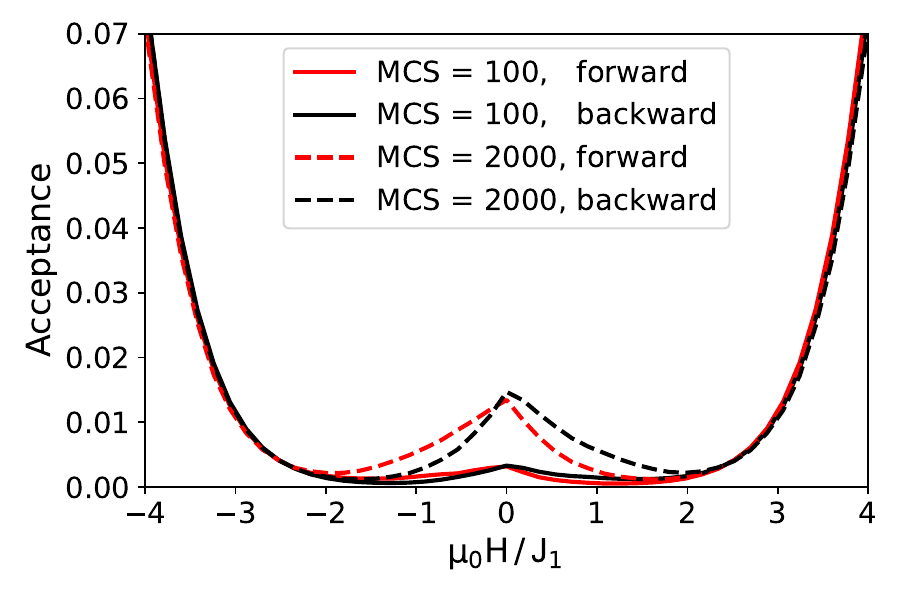}
	\caption{\textbf{Monte Carlo acceptance rate under different field-sweep rates.} The acceptance rate of the single-spin flip updates is averaged over the number of MCS at each field. The solid lines represent the faster sweep while the dashed lines represent the slower sweep. The temperature is fixed at $k_B T / J_1 = 0.3$.
		 }
	\label{fig:ruby_acceptance}
\end{figure}

Below this critical field $H_{c,1}$, an intermediate plateau appears, where the system exhibits a high density of rectangular charges in the TCO phase. This phase remains robust until the second critical field is reached, whose energy cost can be estimated from the pair creation of single charges without breaking the triangular ice rule (Figure \ref{fig:ruby3_process}B):
\begin{align}
\Delta E = 4J_2 - 2\alpha \mu_0 H \to 0, \quad \mu_0 H_{c,2} = \frac{2}{\alpha} J_2.
\end{align}
For $J_2 = 0.8$, the critical field is $\mu_0 H_{c,2} = 4.8$. Around this field, the system exhibits a high single-spin-flip acceptance rate and strong charge fluctuations. Below this point, the second-neighbor interaction $J_2$ becomes the dominant bottleneck for local spin dynamics, and the remaining sparse single charges can barely migrate across the lattice to further annihilate each other. To better understand the underlying dynamics, we calculate the spin-flip acceptance rate (Figure \ref{fig:ruby_acceptance}), which tracks how easily magnetic moments can rearrange to move or create magnetic charge defects. The rate rises sharply near the critical field where charge defects proliferate and break the local constraint, and peaks near zero field, where they can move across relatively small energy barriers. Moreover, slower sweeps allow trapped defects to relax, reducing both the peak and hysteresis. This mechanism underlies the metastability responsible for the hysteresis discussed extensively in the main text.

When the field sweep passes through zero field, the spin and charge dynamics are slightly revived through the charge-migration process, as shown in Figure \ref{fig:ruby3_process}C. It is straightforward to see that the critical field for charge migration is zero, since the number of anti-aligned $J_1$ and $J_2$ pairs remains unchanged before and after the process. This mechanism enables charges to migrate on the lattice around $\mu_0 H = 0$, further reducing the rectangular charge density through pair annihilation when two charges meet. Note that during each charge-migration step, the sign of the rectangular charge is flipped. This implies that the perfect balance of the rectangular charges $Q_{\square}=\pm 2$ is only favored at $H = 0$. Away from the zero field, some of the remaining sparse charges prefer to move one more step and flip their signs to be energetically aligned with the Zeeman field. 

\begin{figure}
	\centering
	\includegraphics[width=1\textwidth]{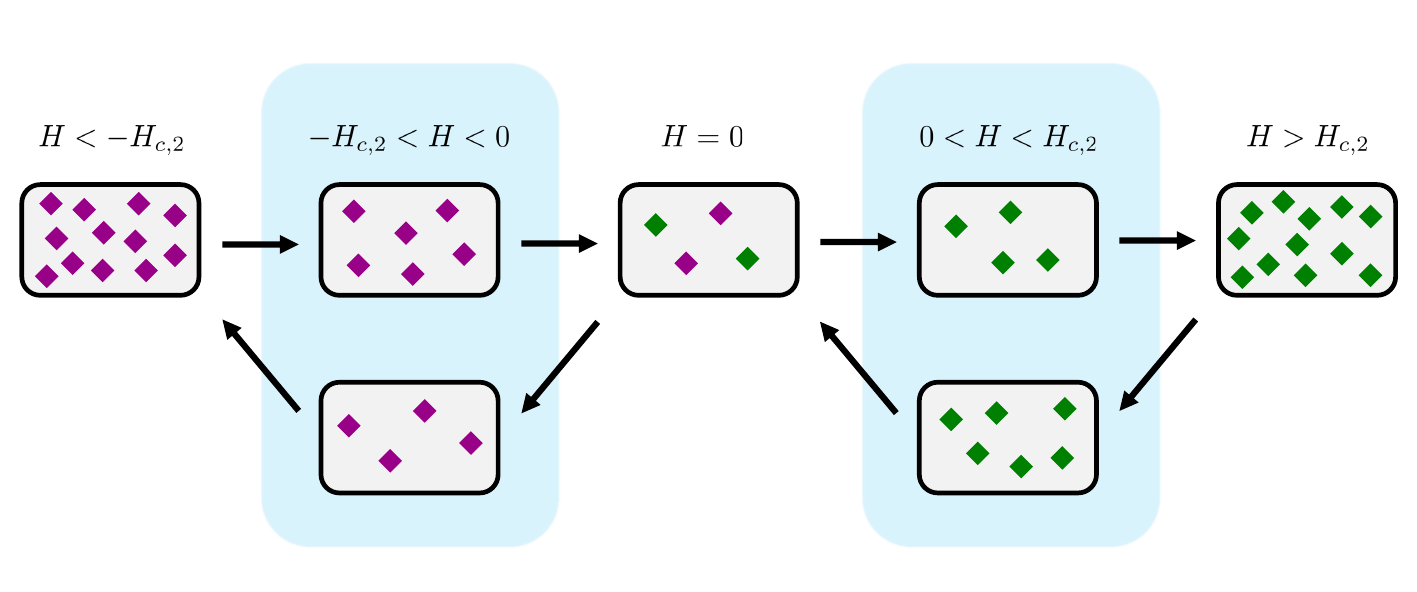}
	\caption{\textbf{Schematic illustration of hysteresis in field sweeps.} This figure illustrates the field-dependent dynamics of rectangular charges within the range $|H| < H_{c,2}$, highlighting the origin of the hysteresis. Right (left) arrows indicate the forward (backward) field sweep. The blue-shaded regions mark the hysteretic regime, where the system exhibits distinct charge order and charge density between forward and backward sweeps.}
	\label{fig:ruby4_schematic}
\end{figure}

During the backward field sweep, as the external field becomes negative, the remaining charges preferentially stay as $Q_{\square} = -2$, gradually slowing down and becoming immobilized until the critical field for the pair creation of negative charges is reached. The hysteresis region can therefore be understood as follows: within this region ($|H| < H_{c,2}$), rectangular charges are sparsely trapped in both sweep directions, but the branch that has passed through zero field contains fewer charges due to additional migration and annihilation processes. This mechanism is summarized in Figure \ref{fig:ruby4_schematic}.

\begin{figure}
	\centering
	\includegraphics[width=1\textwidth]{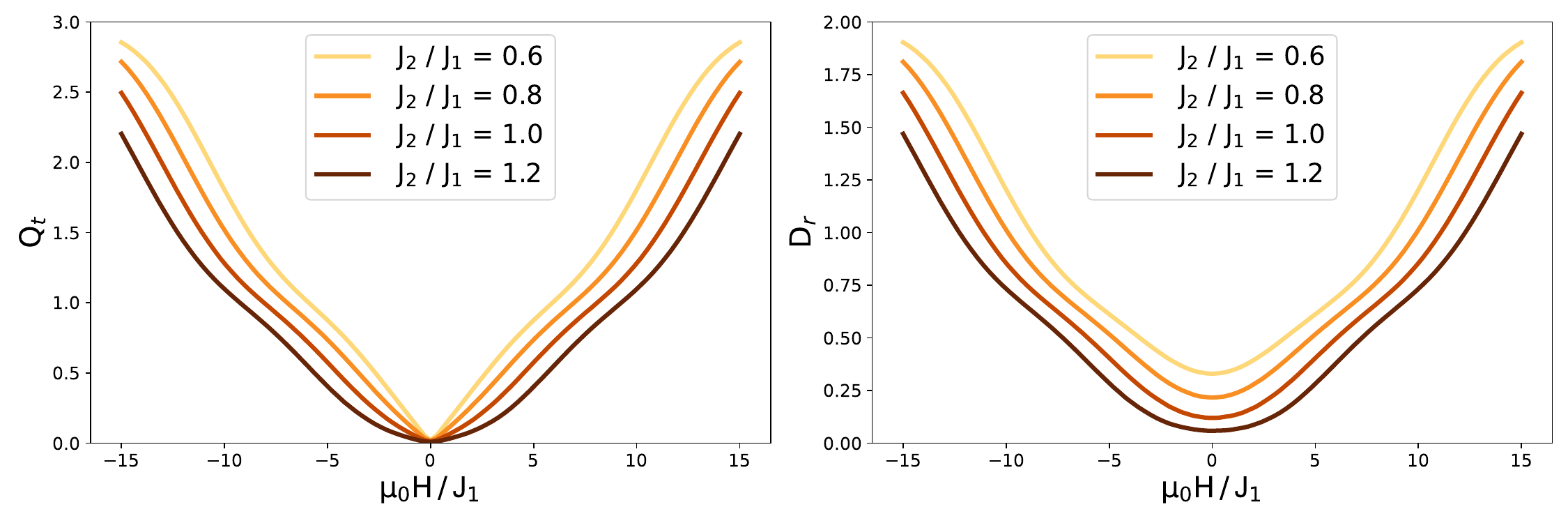}
	\caption{\textbf{Monte Carlo simulation without the hysteresis.} Comparison of triangular-charge order and rectangular-charge density at $T/J_1 = 1.0$ where the forward and backward sweeps coincide. At $\mu_0H = 0$, $Q_t = 0$ indicates the triangular-charge-disordered phase where $Q_{\triangle}=\pm1$ are balanced, while $D_r > 0 $ reveals the thermally proliferated rectangular charges.
    }
	\label{fig:ruby5_T10}
\end{figure}

At high temperatures, such as $k_B T / J_1 = 1.0$, thermal fluctuations are strong enough to equilibrate the system during both forward and backward field sweeps (Figure \ref{fig:ruby5_T10}). However, the distinct roles of triangular and rectangular charges can be identified by comparing $Q_t$ and $D_r$. At zero field, the triangular charges are on average balanced between $Q_{\triangle} = +1$ and $Q_{\triangle} = -1$, resulting in a vanishing charge order. This feature is robust against variations in temperature and coupling strength $J_2$. In contrast, the rectangular charge density $D_r$, which counts both $Q_{\square} = +2$ and $Q_{\square} = -2$ in absolute value, captures the presence of rectangular charge fluctuations even at zero field. In this case, the local energy barrier for creating a pair of rectangular charges from the charge vacuum is simply $4J_2$, so that a smaller $J_2$ leads to a higher $D_r$.

Finally, we comment on the energy scale and effective magnetic moment by relating the characteristic temperature and magnetic field strength in the Monte Carlo simulation to the experiment. First, we compare the temperature scales of the hysteresis loop onset, which is about $T^{*}_{\rm exp} = 2$~(K) and $T^{*}_{\rm MC} = 0.5 J_1$, which gives the estimate $J_1 \sim 4$~(K) and $J_2 \sim 3.2$~(K). Violating the 2-in-1-out constraint on a triangular plaquette costs $4J_1$, whereas creating a pair of rectangular charge defects costs $4J_2$. The corresponding defect energies, $4J_1\sim16$~K and $4J_2\sim13$~K, are comparable to the temperature range over which the static MFM texture disappears.

Second, by comparing the Zeeman-energy scale for the hysteresis-loop closure at the critical field $B^{*}_{\perp} \sim 5.5$~(T) of the experiment and $\mu_0 H_{c,2} = 4.8 J_1 = 19.2$~(K) in the magnetic model, we get the effective moment along the perpendicular field direction
\begin{align}
\mu_{\rm eff,\perp} B^{*}_{\perp} = \mu_0 H, \qquad \mu_{\rm eff,\perp} = 5.179~\mu_B, 
\end{align}
and thus the effective moment is estimated as $\mu_{\rm eff} = \sqrt{3}\mu_{\rm eff,\perp} \sim 9.0~\mu_B$. Note that this estimate relies on the effective magnetic model used in the simulation and not solely based on the experimental result, but the effective moment scale is comparable to other magnetic systems of graphene superlattices, such as twisted bilayer graphene ~\autocite{Li2020_TBG,Tschirhart2021_TBG,Sharpe2021_TBG}.

\subsubsection{Field-dependent MFM and AFM topography}

To examine the magnetic-field evolution of the MFM textures while minimizing possible topographic contributions, we prepared an hBN-capped strain-patterned graphene device, as illustrated in Fig.~S16a. The capped structure provides a relatively flat and uniform surface for the MFM measurement. The optical image in Fig.~S16b confirms a clean and bubble-free surface over the patterned region.
At zero field, no static magnetic texture is resolved within the MFM observation timescale (Fig.~S16c). This may reflect a weaker or more dynamic magnetic response near zero field. An intermediate magnetic field partially lifts the degeneracy and stabilizes metastable textures (Fig.~S16d--f). The emergence of these field-stabilized textures is qualitatively consistent with our calculations: the spin-flip acceptance rate is large near zero field and near the transition to the polarized state, but is significantly reduced at intermediate fields, allowing metastable textures to remain sufficiently static to be resolved by MFM.

Here, we also provide the AFM topographies corresponding to the MFM images shown in Figs.~4d and 4g of the main text (Fig.~S17a,c). The magnetic textures show spatial features distinct from the nanohole topography, indicating that the MFM contrast is not dominated by surface morphology. The long blue feature in Fig.~S17d coincides with the hBN wrinkle visible in Fig.~S17c and is therefore attributed to the underlying wrinkle rather than an intrinsic magnetic domain.

\begin{figure}[H]
    \centering
    \includegraphics[width=0.9\textwidth]{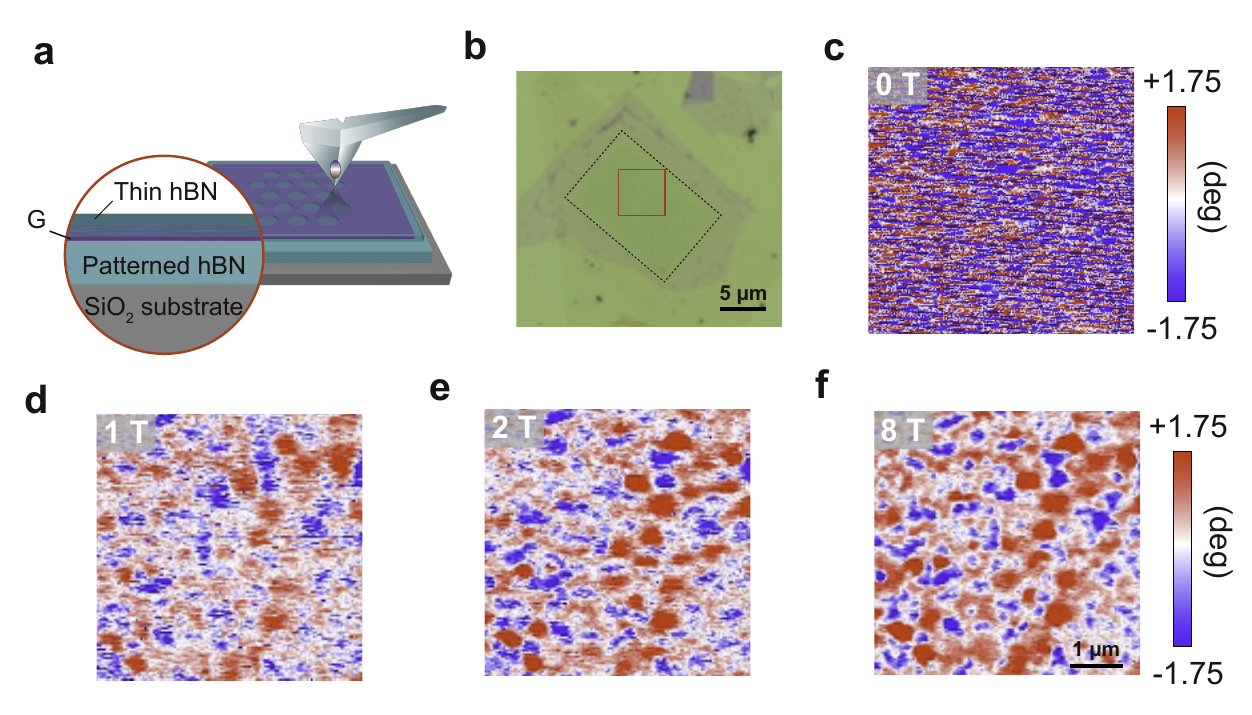}
    \caption{
    \textbf{Evolution of MFM textures with magnetic field in an hBN-capped strain-patterned graphene device.} \textbf{a}, Schematic of the hBN-capped strain-patterned graphene device. \textbf{b}, Optical microscope image of the device, with the strain-patterned region outlined by the black dashed rectangle. \textbf{c--f}, Cryogenic MFM images acquired at $1.6$~K under increasing out-of-plane magnetic fields. No static magnetic texture is resolved at zero field, whereas distinct magnetic textures emerge and progressively evolve under finite fields. The MFM images were acquired from the region marked by the red square in \textbf{b}.
       }
	\label{FigS8}
\end{figure}

\begin{figure}[H]
    \centering
    \includegraphics[width=0.9\textwidth]{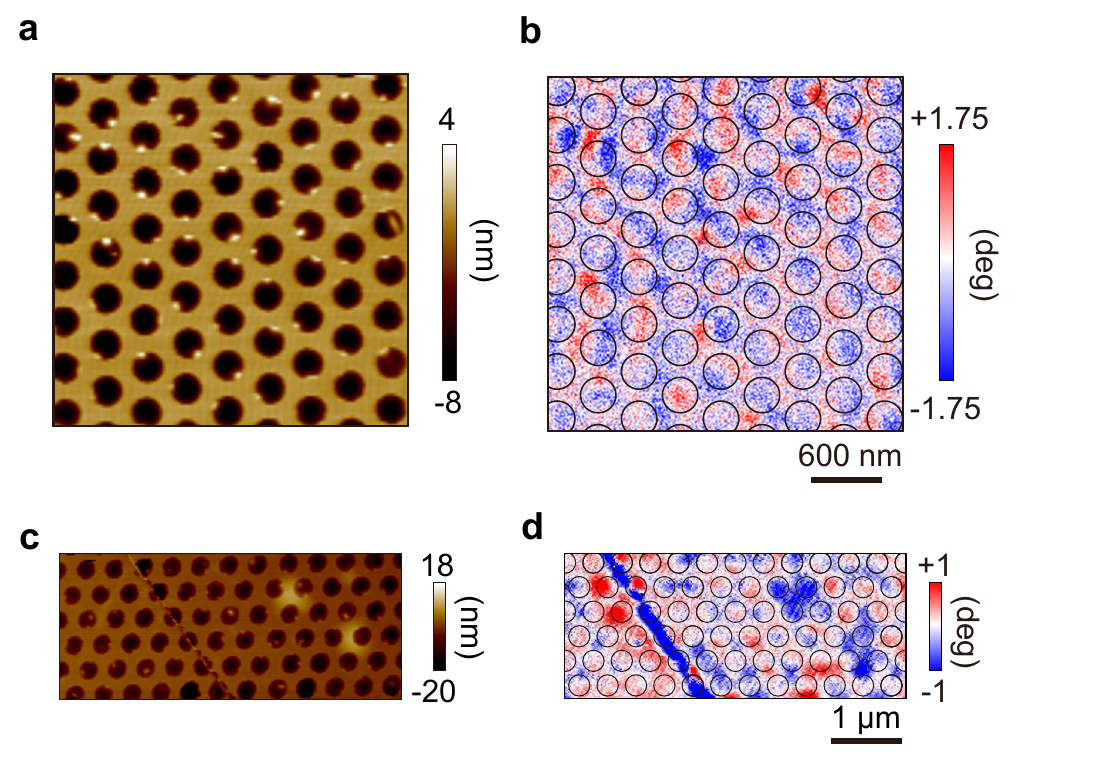}
    \caption{\textbf{AFM topography and MFM images.} 
    \textbf{a},\textbf{b}, AFM topography (a) and the corresponding MFM image (b) of the same region; the same MFM image is shown in Fig.~4d of the main text. \textbf{c},\textbf{d}, AFM topography (c) and the corresponding MFM image (d); the same MFM image is shown in Fig.~4g of the main text. The circular outlines in b and d mark the corresponding nanohole positions.
        }
	\label{EXT9}
\end{figure}

\subsubsection{Cooling-history irreversibility in strain-patterned graphene}

To examine whether the observed phenomena could arise from a disorder-driven glassy state, we compared the magnetotransport behavior following zero-field cooling (ZFC) and field cooling (FC), as shown in Fig.~S18. The nearly identical hysteresis curves show no clear spin-glass-like irreversibility. A pronounced dependence on the cooling history would generally be expected for conventional spin-glass freezing, because the ZFC and FC protocols can prepare different frozen configurations. Together with the corrugated-device control in Extended Data Fig.~5, these results indicate that the response is closely associated with the frustrated ruby geometry rather than a consequence of disorder or fabrication.

\begin{figure}[H]
    \centering
    \includegraphics[width=0.7\linewidth]{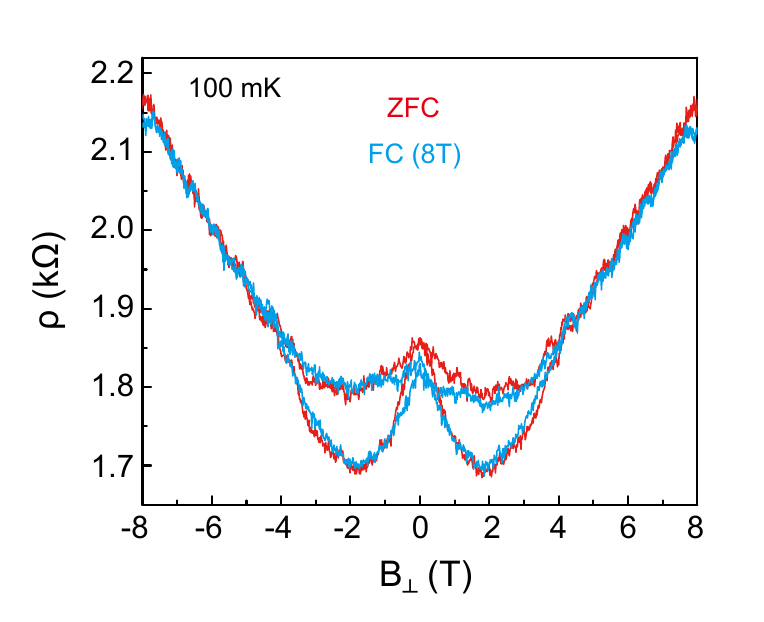}
    \caption{\textbf{Comparison of zero-field-cooled and field-cooled magnetotransport.}Magnetotransport measured at $100$~mK after zero-field cooling (ZFC) and field cooling (FC) at $8$~T, shown by the red and blue curves, respectively. The nearly identical magnetoresistivity responses show no clear irreversibility characteristic of a conventional spin glass.}
    \label{figS6}
\end{figure}

\subsubsection{Transverse transport response }
In our previous study of 1D corrugated bilayer graphene \cite{ho_NE21}, we showed that nonuniform strain can generate a pseudo-planar Hall effect through warped band dispersion, giving rise to a finite transverse response even without breaking time-reversal symmetry. Any possible anomalous Hall contribution associated with the magnetic response may therefore be superimposed on this strain-induced transverse background and cannot be unambiguously separated. We do not observe a clear Hall plateau or other feature that can be uniquely attributed to an anomalous Hall response. Nevertheless, the pronounced hysteresis and its nonmonotonic temperature dependence, closely following the longitudinal magnetotransport, provide an additional transport signature of the history-dependent magnetic response. 

\begin{figure}[H]
    \centering
    \includegraphics[width=0.7\linewidth]{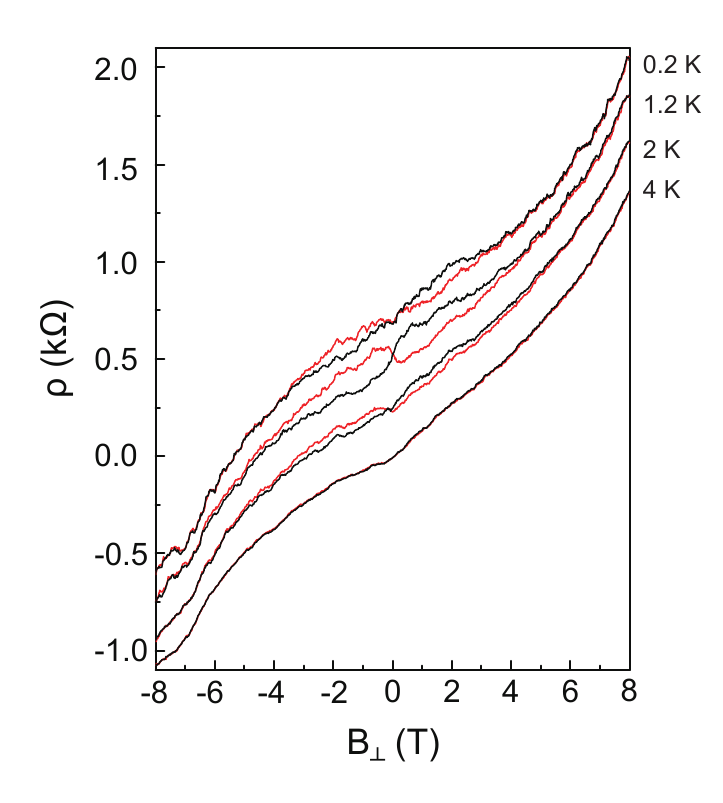}
    \caption{Transverse magnetotransport of strain-patterned graphene in the same trilayer device and under the same measurement conditions as in Fig. 2 of the main text, under an out-of-plane magnetic field. Clear hysteresis is observed, and the window size also shows non-monotonic temperature dependence. Traces are offset for clarity.}
    \label{figS7}
\end{figure}

\newpage

\printbibliography[segment=\therefsegment, heading=subbibliography, title={Supplementary References}]

\end{refsegment}

\end{document}